\documentclass[rmp,aps,twocolumn,nofootinbib]{revtex4}

\usepackage{amssymb,amsfonts}
\usepackage{graphicx}
\usepackage{epsfig}
\usepackage{epstopdf}
\usepackage[all]{xy}
\usepackage{bm}
\usepackage{amsthm}

\newcommand{\be}{\begin{equation}}
\newcommand{\ee}{\end{equation}}
\newcommand{\bea}{\begin{eqnarray}}
\newcommand{\nn}{\nonumber}
\newcommand{\eea}{\end{eqnarray}} 

\newcommand{\gn}{\bar{\nabla}}

\def\inbar{\,\vrule height1.5ex width.4pt depth0pt}
\def\IR{\relax{\rm I\kern-.18em R}}
\def\IC{\relax\hbox{$\inbar\kern-.3em{\rm C}$}}

\begin{document}
\title{$f(R)$ theories of gravity}

\author{Thomas P. Sotiriou\footnote{Present address: Department of Applied Mathematics and Theoretical Physics, Centre for Mathematical Sciences, University of Cambridge, Wilberforce Road, Cambridge, CB3 0WA, UK.\\
T.Sotiriou@damtp.cam.ac.uk}}
\affiliation{Center for Fundamental Physics,  University of Maryland, College Park, MD 20742-4111, USA}

\author{Valerio Faraoni\footnote{vfaraoni@ubishops.ca}}
\affiliation{Physics Department, Bishop's University, 2600 College St., Sherbrooke, Qu\`{e}bec, Canada 
J1M~1Z7}

\begin{abstract} 
Modified gravity theories have received 
increased attention lately due to combined motivation coming 
from high-energy physics, cosmology and astrophysics. 
Among numerous alternatives to Einstein's theory of gravity, 
theories which include higher order curvature invariants, and 
specifically the particular class of $f(R)$ theories, have a 
long history. In the last five years there has been a new stimulus 
for their study, leading to a number of interesting results. We 
review here $f(R)$ theories of gravity in an attempt to 
comprehensively present their most important aspects and cover 
the largest possible portion of the relevant literature. All 
known formalisms are presented --- metric, Palatini and 
metric-affine --- and the following topics are discussed: 
motivation; actions, field equations and theoretical aspects; 
equivalence with other theories; cosmological aspects and 
constraints; viability criteria; astrophysical applications. 
\end{abstract}

\maketitle
\tableofcontents

\section{Introduction}
\label{sec:intro}

\subsection{Historical}
  
As we are approaching the closing of a century after the 
introduction of General Relativity (GR) in 1915, questions 
related to its limitations are becoming more and more pertinent. 
However, before coming to the contemporary reasons for 
challenging a theory as successful as Einstein's theory, it is 
worth mentioning that it took only four years from its 
introduction for people to start questioning its unique status 
among gravitation theories. Indeed, it was just 1919 when 
Weyl, and 1923 when Eddington (the very man that three 
years earlier had provided the first experimental verification 
of GR 
by measuring light bending during a solar eclipse) started 
considering modifications of the theory by including higher 
order invariants in its action \cite{weyl,eddin}.

These early attempts were triggered mainly by scientific 
curiosity and a will to question and, therefore, understand the 
then newly proposed theory. It is quite straightforward 
to realize that complicating the action and, consequently, the 
field equations with no apparent theoretical or experimental 
motivation is not very appealing. However, the motivation was 
soon to come.

Beginning in the 1960's, there appeared indications that 
complicating the gravitational action might indeed have its 
merits. GR is not renormalizable and, therefore, 
can not be conventionally quantized. In 1962, Utiyama and De 
Witt 
showed that renormalization at one-loop demands that the 
Einstein--Hilbert action  be supplemented by higher order 
curvature terms \cite{Utiyama:1962sn}. Later on, Stelle showed 
that higher order actions are indeed renormalizable (but not 
unitary) \cite{Stelle:1976gc}. More recent results show that 
when quantum corrections or string theory are taken into 
account, the effective low energy gravitational action admits 
higher order curvature invariants \cite{quant1, quant2, quant3}.

Such considerations stimulated the interest of the scientific 
community in higher-order theories of gravity, {\em 
i.e.}, modifications of the Einstein--Hilbert action in order to include 
higher-order curvature invariants with respect to the Ricci 
scalar [see \cite{Schmidt:2006jt} for a historical review and a 
 list of references to early work]. However, the 
relevance of such terms in the action was considered to be 
restricted to very strong gravity regimes and they were 
expected to be strongly suppressed by small couplings, as one 
would expect when simple effective field theory considerations 
are taken into account. Therefore, corrections to GR 
were considered to be  important only at scales 
close to the Planck scale 
and, consequently, in the early universe or near black hole 
singularities  --- and indeed there are 
relevant studies, such as the well-known curvature-driven 
inflation scenario \cite{Starobinsky:1980te}  and attempts to 
avoid cosmological and black hole singularities 
\cite{Brandenberger:1995hd, Trodden:1993dm, 
Brandenberger:1993ef,Brandenberger:1993mw, Brandenberger:1992sy, 
Mukhanov:1991zn, Shahid-Saless:1990aaa}. However, it was not 
expected that such corrections could affect the gravitational 
phenomenology 
at low energies, and consequently large scales such as, for 
instance, the late universe.

\subsection{Contemporary Motivation}

More recently, new evidence coming from astrophysics and 
cosmology has revealed a quite unexpected picture of the 
universe. Our latest datasets coming from different sources, 
such as the Cosmic Microwave Background Radiation (CMBR) and 
supernovae surveys, seem to indicate that the energy budget of 
the universe is the following: 4\% ordinary baryonic matter, 
20\% {\em dark matter} and 76\% {\em dark energy} 
\cite{Spergel:2006hy,Riess:2004nr,Astier:2005qq,Eisenstein:2005su}. 
The term dark matter refers to an unkown form of matter, which 
has the clustering properties of ordinary matter but has not 
 yet been detected in the laboratory. The term dark energy is 
reserved for an unknown form of energy which not only has not 
been detected directly, but also does not cluster as ordinary matter 
does. More  rigorously, one could use the various energy 
conditions  \cite{wald}   to distinguish dark matter and dark energy: Ordinary 
matter and dark matter satisfy the Strong Energy Condition, 
whereas Dark Energy does not. Additionally, dark energy seems 
to resemble in high detail a cosmological 
constant. Due to its dominance over matter (ordinary and dark) 
at present times, the expansion of the universe seems to be an 
accelerated one, contrary to past expectations.\footnote{Recall  
that, from GR in the absence of the cosmological constant and 
under the standard cosmological assumptions (spatial 
homogeneity and   
isotropy {\em etc.}), one obtains the 
second Friedmann equation 
\begin{equation} 
\frac{\ddot{a}}{a}=-\frac{4\pi\,G}{3}\left( \rho+3P \right), 
\end{equation} 
where $a$ is the scale factor, $G$ is the gravitational constant 
and $\rho$ and $P$ are the energy density and the pressure of 
the cosmological fluid, respectively. Therefore, if the Strong 
Energy Condition $\rho+3 P \geq 0$ is satisfied, there can be no 
acceleration (gravity is attractive).}

Note that this late time speed-up comes to be added to an early 
time accelerated epoch as predicted by the inflationary paradigm 
\cite{Guth:1980zm,infl1,infl2}. The inflationary epoch is needed 
to address the so-called {\em horizon},  {\em flatness} and {\em 
monopole 
problems} \cite{Misner:1967uu,weinberg,infl1,infl2}, as well as 
to provide the mechanism that generates primordial 
inhomogeneities acting as seeds for the formation of large scale 
structures  
\cite{Mukhanov:2003xw}. Recall also that, in between these two 
periods of acceleration, there should be a period of decelerated 
expansion, so that the more conventional cosmological eras of 
radiation domination and matter domination can take place. 
Indeed, there are stringent observational bounds on the 
abundances of light elements, such as deuterium, helium and 
lithium, which require that Big Bang Nucleosynthesis (BBN), the 
production of nuclei other than hydrogen, takes place during 
radiation domination \cite{Burles:2000zk,Carroll:2001bv}. On the 
other hand, a matter-dominated era is required for structure 
formation to take place.

Puzzling observations do not stop here. Dark matter does 
not only make its appearance in cosmological data but also in 
astrophysical observations. The ``missing mass'' question had 
already been posed in 1933 for galaxy clusters 
\cite{Zwicky:1933gu} and in 1959 for individual galaxies 
\cite{KahnWol} and a satisfactory final answer has been pending 
ever since 
\cite{Rubin1,Rubin:1980zd, 
Bosma:1978,Persic:1995ru,Ellis:2002qd,Moore:2001fc} 
.

One, therefore, has to admit that our current picture of the 
evolution and the matter/energy content of the universe is at 
least surprising and definitely calls for an explanation. The 
simplest model which adequately fits the data creating this 
picture is the so called concordance model or $\Lambda$CDM 
($\Lambda$-Cold Dark Matter), supplemented by some inflationary  
scenario,  usually based on some scalar field called inflaton. 
Besides not explaining the origin of the inflaton or the 
nature of dark matter by itself, the $\Lambda$CDM model is 
burdened with the well known cosmological constant problems 
\cite{Weinberg:1988cp,Carroll:2000fy}: the magnitude 
problem, according to which the observed value of the 
cosmological constant is extravagantly small to be attributed to 
the vacuum energy of matter fields, and the coincidence problem, 
which can be summed up in the question: since there is just an 
extremely short period of time in the evolution of the universe 
in which the energy density of the cosmological constant is 
comparable with that of matter, why is this happening today that we are present to observe it?

These problems make the $\Lambda$CDM model more of an empirical 
fit to the data whose theoretical motivation can be regarded as
quite poor. Consequently, there have been several attempts to 
either directly motivate the presence of a cosmological constant 
or to propose dynamical alternatives to dark energy.  
Unfortunately, none of these attempts are problem-free. For 
instance, the so-called anthropic reasoning for the magnitude of 
$\Lambda$ \cite{carter1, BarrowTipler}, even when placed into 
the firmer grounds through the idea of the ``anthropic or string 
landscape'' \cite{Susskind:2003kw}, still makes many physicists 
feel uncomfortable due to its probabilistic nature. On the other 
hand, simple scenarios for dynamical dark energy, such as 
quintessence 
\cite{Peebles:1987ek,Ratra:1987rm, 
Wetterich:1987fm,Ostriker:1995su,Caldwell:1997ii,Carroll:1998zi, 
Bahcall:1999xn,Wang:1999fa} 
do not seem to be as well motivated theoretically as one would 
desire.\footnote{We are referring here to the fact that, not 
only the mass 
of the scalar turns out to be many orders of magnitude smaller 
than any of the masses of the scalar fields usually encountered 
in particle physics, but also to the inability to motivate the 
absence of any coupling of the scalar field to matter (there is 
no mechanism or symmetry preventing this) \cite{Carroll:2001xs}.}

Another perspective towards resolving the issues 
described above, which might appear as more radical to some, is 
the following: gravity is by far the dominant interaction at 
cosmological scales and, therefore, it is the force 
governing the evolution of the universe. Could it be that our 
description of the gravitational interaction at the relevant 
scales is not sufficiently adequate and stands at the root of all or 
some of these problems? Should we consider modifying our theory 
of gravitation and if so, would this help in avoiding dark 
components and answering the cosmological and astrophysical 
riddles?

It is rather pointless to argue whether such a perspective 
would be better or worse than any of the other solutions already 
proposed. It is definitely a different way to address the same 
problems and, as long as these problems do not find a plausible, 
well accepted and simple, solution, it is worth pursuing all 
alternatives. Additionally, questioning the gravitational theory 
itself definitely has its merits: it helps us to obtain a deeper 
understanding of the relevant issues and of the gravitational 
interaction, it has high chances to lead to new physics and it 
has worked in the past. Recall that the precession of Mercury's 
orbit was at first attributed to some unobserved (``dark'') planet orbiting inside Mercury's orbit, 
but it actually took the passage from Newtonian gravity to 
GR to be explained.

\subsection{$f(R)$ theories as toy theories}

Even if one decides that modifying gravity is the way to go, 
this is not an easy task. To begin with, there are numerous 
ways to deviate from GR. Setting aside the early 
attempts to generalize Einstein's theory, most of which have been 
shown to be non-viable \cite{willbook}, and the most well known 
alternative to GR, scalar-tensor theory \cite{Brans:1961sx, 
Wagoner:1970vr,valeriobook, Dicke:1961gz, Bergmann:1968ve, 
Nordtvedt:1970uv}, 
there are still numerous proposals for modified gravity in 
contemporary literature. Typical examples are DGP 
(Dvali-Gabadadze-Porrati) gravity \cite{Dvali:2000hr}, 
brane-world gravity \cite{Maartens:2003tw}, TeVeS 
(Tensor-Vector-Scalar) \cite{Bekenstein:2004ne} and 
Einstein-Aether theory \cite{Jacobson:2000xp}. The subject of 
this review is a different class of theories, $f(R)$ theories of 
gravity. These theories come about by a straightforward 
generalization of the Lagrangian in the Einstein--Hilbert 
action, 
\begin{equation}
 S_{\rm EH}=\frac{1}{2\kappa} \int d^4 x 
\sqrt{-g} 
\, R, 
\ee 
where $\kappa\equiv 8\pi G$, $G$ is the gravitational constant, 
$g$ is the 
determinant of the metric and $R$ is the Ricci scalar 
($c=\hbar=1$), to become a general function of $R$, {\em i.e.}, 
\begin{equation} \label{3}
S=\frac{1}{2\kappa} \int d^4 x \sqrt{-g} \, f(R). 
\ee

Before going further into the discussion of the details and the 
history of  such actions --- this will happen in the forthcoming 
section ---
some remarks are in order. We have already mentioned the motivation 
coming from high-energy physics for adding higher order 
invariants to the gravitational action, as well as a general 
motivation coming from cosmology and astrophysics for seeking 
generalizations of GR. There are, however, still two 
questions that might be troubling the reader. The first one is: 
Why specifically $f(R)$ actions and not more general ones, which 
include other higher order invariants, such as 
$R_{\mu\nu}R^{\mu\nu}$?

The answer to this question is twofold. First of all, there is 
simplicity: $f(R)$ actions are sufficiently general  to 
encapsulate 
some of the basic characteristics of higher-order gravity, but 
at the same time they are simple enough to be easy to 
handle. 
For instance, viewing $f$ as a series expansion, {\em i.e.},  
\begin{equation} 
\label{series} 
f(R)=\ldots+\frac{\alpha_2}{R^2} 
+\frac{\alpha_1}{R}-2\Lambda+R+\frac{R^2}{\beta_2} 
+\frac{R^3}{\beta_3}+\ldots, 
\ee 
where the $\alpha_i$ and $\beta_i$ coefficients have the 
appropriate dimensions, we see that the action includes a number 
of phenomenologically interesting terms. In brief, $f(R)$ 
theories make excellent candidates for toy-theories---tools from 
which  one gains some insight in such gravity 
modifications. Second, there are serious reasons to believe 
that $f(R)$ theories are unique among higher-order gravity 
theories, in the sense that they seem to be the only ones which 
can avoid the long known and fatal Ostrogradski instability  
\cite{Woodard:2006nt}.

The second question calling for an answer is related to a 
possible loophole that one may have already spotted in the 
motivation presented: How can high-energy modifications of the 
gravitational action have anything to do with late-time 
cosmological phenomenology? Wouldn't effective field theory 
considerations require that the coefficients in 
eq.~(\ref{series}) be such, as to make any corrections to the 
standard Einstein--Hilbert term important only near the Planck 
scale?

Conservatively thinking, the answer would be positive. However, one also 
has to stress two other serious factors: first, there is a 
large ambiguity on how gravity really works at small scales or 
high energies. Indeed there are certain results already in the 
literature claiming that terms responsible for late time 
gravitational phenomenology might be predicted by some more 
fundamental theory, such as string theory [see, for instance, 
\cite{Nojiri:2003rz}]. On the other hand, one should not forget 
that the observationally measured value of the cosmological 
constant corresponds to some energy scale. Effective field 
theory or any other high-energy theory consideration has  
thus far failed to predict or explain it. Yet, it stands  
as an 
experimental fact and putting the number in the right context 
can be crucial in explaining its value. Therefore, in any 
phenomenological approach, its seems inevitable that some 
parameter will appear to be unnaturally small at first (the mass 
of a scalar, a coefficient of some expansions, {\em etc.}~according 
to the approach). The real question is whether this 
initial 
``unnaturalness" still has room to be explained.

In other words, in all sincerity, the motivation for infrared 
modifications of gravity in general and $f(R)$ gravity in 
particular is, to some extent, hand-waving. However, the 
importance of the issues leading to this motivation and our 
inability to find other, more straightforward and maybe better 
motivated, successful ways to address them combined with the 
significant room for speculation which our quantum gravity candidates 
leave, have triggered an increase of 
interest in modified gravity that is probably reasonable.

To conclude, when all of the above is taken into account, 
$f(R)$ gravity should neither be over- nor 
under-estimated. It is an interesting and relatively simple 
alternative to 
GR, from the study of which some useful conclusions have 
been derived already. However, it is still a toy-theory, as 
already mentioned; an easy-to-handle deviation from Einstein's 
theory mostly to be used in order to understand the principles 
and limitations of modified gravity. Similar considerations 
apply to modifying gravity in general:  we are probably far from 
concluding whether it is the answer to our problems at the 
moment. However, in some sense, such an approach is bound to be 
fruitful since, even if it only leads to the conclusion that GR 
is the only correct theory of gravitation, it will still have 
helped us to both understand GR better and secure our faith 
in it.

\section{Actions and field equations}

As can be found in many textbooks ---  see, for example 
\cite{mtw,wald} --- there  are actually two variational 
principles that one can apply to the  Einstein--Hilbert action 
in 
order to derive Einstein's equations:  the standard metric 
variation and a less standard variation  dubbed Palatini 
variation [even though it was Einstein and not Palatini who introduced it \cite{ferfrancreina}]. In the latter the metric and the connection are 
assumed to be independent variables and one varies the action 
with respect to both of them (we will see how   this variation 
leads to Einstein's equations shortly), under  the important 
assumption that the matter action does not depend  on the 
connection. The choice of the variational principle  is usually 
referred to as a formalism, so one can use the terms  metric 
(or second order) formalism and Palatini (or first order)  
formalism. However, even though  both 
variational principles lead to the same field equation  for an 
action whose Lagrangian is linear in $R$, this is no  longer 
true for a more general action. Therefore, it is  intuitive that 
there will be two version of $f(R)$ gravity, according  to which 
variational principle or formalism is used. 
Indeed this is the case: $f(R)$ gravity in the metric formalism 
is called {\em metric $f(R)$ gravity} and $f(R)$ gravity in the 
Palatini formalism is called {\em Palatini $f(R)$ gravity} \cite{Buchdahl:1983zz}.

Finally, there is actually even a third version of $f(R)$ 
gravity: {\em metric-affine $f(R)$ gravity} \cite{Sotiriou:2006qn,Sotiriou:2006mu}. This comes about if 
one 
uses the Palatini variation but abandons the assumption that 
the matter action is independent of the connection. Clearly, 
metric affine $f(R)$ gravity is the most general of these 
theories and reduces to metric or Palatini $f(R)$ gravity if 
further assumptions are made. In this section we will present 
the actions and field equations of all three versions of $f(R)$ 
gravity and point out their difference. We will also clarify 
the physical meaning behind the assumptions that discriminate 
them.

For an introduction to metric $f(R)$ gravity see also 
\cite{Nojiri:2006ri}, for a shorter review of metric and 
Palatini $f(R)$ gravity see \cite{Capozziello:2007ec} and for 
an extensive analysis of all versions of $f(R)$ gravity and 
other alternative theories of gravity see \cite{Sotiriou:2007}.

\subsection{Metric formalism}
\label{sec:actionmetric}

Beginning from the action~(\ref{3}) and adding a matter term 
$S_M$, the total action for $f(R)$ gravity takes the form
\be
\label{metaction}
S_{met}=\frac{1}{2\kappa}\int d^4 x \sqrt{-g} \, f(R) 
+S_M(g_{\mu\nu},\psi),
\ee
where $\psi$ collectively denotes the matter fields.
Variation with respect to the metric gives,  after some 
manipulations and modulo surface terms
 \be
\label{metf}
 f'(R)R_{\mu\nu}-\frac{1}{2}f(R)g_{\mu\nu}- 
\left[\nabla_\mu\nabla_\nu -g_{\mu\nu}\Box\right] f'(R)= 
\kappa \,T_{\mu\nu}, 
\ee
where, as usual, 
\begin{equation}
\label{set}
T_{\mu\nu}=\frac{-2}{\sqrt{-g}}\, \frac{\delta
S_M }{\delta g^{\mu\nu} }  ,
\end{equation}
a prime denotes differentiation with respect to the argument, 
$\nabla_\mu$ is the covariant derivative associated with the 
 Levi-Civita connection of the metric, and $\Box\equiv 
 \nabla^\mu\nabla_\mu$. Metric $f(R)$ gravity was first 
rigorously studied in \cite{Buchdahl:1983zz}.\footnote{Specific  
attention to 
higher-dimensional $f(R)$ gravity was paid in 
\cite{Saidov:2006xr, Saidov:2006xp, Gunther:2002aa, 
Gunther:2003zn, Gunther:2004ht}.}

It has to be stressed that there is a mathematical jump in 
deriving eq.~(\ref{metf}) from the action (\ref{metaction}) 
having to do with the surface terms that appear in the 
variation: as in the case of the Einstein--Hilbert action, the 
surface terms do not vanish just by fixing the metric on the 
 boundary. For the Einstein--Hilbert action, however, these 
terms 
gather into a total variation of a quantity. Therefore, it is 
possible to add a total divergence to the action in order to 
``heal" it and arrive to a well-defined variational principle 
(this is the well known Gibbons--Hawking--York surface term 
\cite{York:1972sj,PhysRevD.15.2752}). Unfortunately, the  
surface  terms in the variation of the action~(\ref{3}) do not 
consist of a total variation of some quantity (the reader is 
urged to 
calculate the variation in order to verify this fact) and it is 
not possible to ``heal'' the action  by just subtracting some 
surface term before performing the 
variation.

The way out comes from the fact that the action includes higher
order derivatives of the metric and, therefore, it should be 
possible to fix
more degrees of freedom on the boundary than those of the metric
itself. There is no unique prescription for such a fixing in the
literature so far. Note also that the choice of fixing is not void of physical meaning, since it will be relevant for the Hamiltonian
formulation of the theory. However, the field equations $(\ref{metf})$ would be unaffected by the
fixing chosen and from a purely classical perspective, such as the one followed here, the field
equations are all that one needs [see~\cite{Sotiriou:2007} for a 
more detailed discussion on these issues]. 

Setting aside the complications of the variation we can now 
focus on the field equations~(\ref{metf}). These are  obviously 
fourth order partial differential
equations in the metric, since $ R$ already includes  second 
derivatives of the latter. For an action which is linear in $R$, 
the fourth 
order terms --- the last two on the left hand side --- vanish 
and the theory reduces to GR. 

Notice also that the trace of eq.~(\ref{metf})
 \be
\label{metftrace}
 f'(R)R-2f(R) + 3\Box f'  =\kappa \,T,
\ee
 where $T=g^{\mu\nu}T_{\mu\nu}$, relates $R$ with $T$ differentially
and not algebraically as in GR, where $R=-\kappa \,T$.
This is already an indication that the field equations of $f(R)$
theories will admit a larger variety of solutions than 
Einstein's theory. As an 
example, we mention here that the Jebsen-Birkhoff's theorem, 
stating that the
Schwarzschild solution is the unique spherically symmetric vacuum
solution, no longer holds in metric $f(R)$ gravity.
Without going into  details, let us stress that $T=0$ no longer
implies that $R=0$, or is even constant.

Eq.~(\ref{metftrace}) will turn out to be very useful in 
studying various aspects of $f(R)$ gravity, notably its 
stability and  weak-field limit. For the moment, let us use it 
to make some remarks about maximally symmetric solutions. Recall 
that maximally
symmetric solutions lead to a constant Ricci scalar. For $R={\rm
constant}$ and $T_{\mu\nu}=0$, eq.~(\ref{metftrace}) reduces to
 \be
\label{metftr}
f'(R)R-2f(R)=0,
\ee
 which, for a given $f$, is an algebraic equation in $R$. If $R=0$ is
a root of this equation and one takes this root, then eq.~(\ref{metf})
reduces to $R_{\mu\nu}=0$ and the maximally symmetric solution is
Minkowski spacetime. On the other hand, if the root of
eq.~(\ref{metftr}) is $R=C$, where $C$ is a constant, then
eq.~(\ref{metf}) reduces to $ R_{\mu\nu}= g_{\mu\nu} C/4 $ and 
the
maximally symmetric solution is de  Sitter or anti-de Sitter 
space  
depending
on the sign of $C$, just as in GR with a cosmological
constant.

Another issue that should be stressed is that of energy 
conservation. In metric $f(R)$ gravity the matter is  minimally 
coupled to the metric. One can, therefore, use the usual arguments based on the invariance of the action under diffeomorphisms of the spacetime manifold 
[coordinate transformations  $x^{\mu}\rightarrow 
x'^{\mu}=x^{\mu}+ 
\xi^{\mu} $ followed by a pullback,  
with the field $\xi^{\mu} $ vanishing on the boundary of the 
spacetime region considered, leave the physics unchanged, see 
\cite{wald}] to show that $T_{\mu\nu}$ is 
divergence-free. The 
same can be done at the level of the field equations: a ``brute 
force'' calculation reveals that the left hand side of  
 eq.~(\ref{metf}) is divergence-free (generalized Bianchi 
identity)  implying that $\nabla_\mu T^{\mu\nu}=0$ 
\cite{Koivisto:2005yk}.\footnote{Energy-momentum complexes in 
the  spherically symmetric case have been computed in 
\cite{Multamaki:2007wb}.}

Finally, let us note that it is possible to write the field 
equations in the form of  Einstein 
equations with an effective stress-energy tensor composed of   
curvature terms moved to the right hand side. This approach 
is questionable in principle (the theory is not Einstein's 
theory and it is artificial to force upon it an 
interpretation in terms of Einstein equations) but, in 
practice, it has  been proved to be useful in scalar-tensor 
gravity.  Specifically, eq.~(\ref{metf}) can be 
written as 
\begin{eqnarray}
G_{\mu\nu} & \equiv & R_{\mu\nu}-\frac{1}{2} \, g_{\mu\nu}  
R\nn\\&=& \frac{\kappa \, T_{\mu\nu}}{f'(R)}+g_{\mu\nu} \, \frac{ \left[ 
f(R)-Rf'(R) \right]}{2f'(R)} \nonumber \\
& &+ \frac{ \left[ 
\nabla_{\mu}\nabla_{\nu} f'(R) 
-g_{\mu\nu} \Box f'(R) \right]}{f'(R)} 
\end{eqnarray}
or
\be \label{Tabeff}
G_{\mu\nu}=\frac{\kappa}{f'(R)} \left(  T_{\mu\nu} +  
T_{\mu\nu}^{(eff)}  \right) ,
\ee 
where the quantity $G_{eff}\equiv G/f'(R)$ can be regarded as  
the effective gravitational coupling strength in analogy to 
what is done in scalar-tensor gravity --- positivity of 
$G_{eff}$ (equivalent to the requirement that the graviton is 
not a ghost) imposes that $f'(R)>0$. Moreover, 
\begin{eqnarray} 
 T_{\mu\nu}^{(eff)} &\equiv &\frac{1}{\kappa} \Big[  
\frac{f(R)-Rf'(R)}{2} \,  g_{\mu\nu} + \nabla_{\mu}\nabla_{\nu}  
f'(R) \nonumber \\
&& \qquad -g_{\mu\nu} \Box f'(R) \Big] 
\label{effectivetab}
\end{eqnarray}
is an effective stress-energy tensor which  does 
not  have the canonical form quadratic in the first derivatives 
of the  field $f'(R)$, but contains terms linear in the 
second 
derivatives. The effective energy density derived from it is 
not positive-definite and none of the energy conditions holds. 
Again, this situation is  analogous to that occurring in  
scalar-tensor gravity.  The effective 
stress-energy tensor (\ref{effectivetab}) can be put in the form 
of a perfect fluid energy-momentum tensor, which will turn out 
to be useful in Sec.~\ref{sec:chapIV}.

\subsection{Palatini formalism}
\label{sec:palatinifield}

We have already mentioned that the Einstein equations can be
derived using, instead of the standard metric variation of the
Einstein--Hilbert action, the Palatini formalism, {\em i.e.}, an
independent variation with respect to the metric and an independent
connection (Palatini variation). The action is formally the 
same but now the Riemann tensor and the Ricci tensor are constructed with the independent
connection. Note that the metric is not needed to obtain the 
latter from 
the former. For clarity of notation, we denote the Ricci tensor 
constructed with this independent connection as ${\cal
R}_{\mu\nu}$ and the corresponding Ricci scalar\footnote{The 
term ``$f(R)$ gravity'' is used generically for a theory in 
which 
the  action is some function of some Ricci scalar, not 
necessarily $R$.} is 
${\cal R}=g^{\mu\nu}{\cal R}_{\mu\nu}$. The action now takes 
the form
\be
\label{palaction}
 S_{pal}=\frac{1}{2\kappa }\int d^4 x \sqrt{-g} \, f({\cal R}) 
+S_M(g_{\mu\nu}, \psi).
\ee
GR will come about, as we will see shortly, when $f({\cal R})={\cal R}$.
Note that the matter action $S_M$ is assumed to depend only on 
the metric and the matter fields and not on the independent 
connection. This assumption is crucial for the derivation of 
Einstein's equations from the linear version of the 
action~(\ref{palaction})
and is the main feature of the Palatini formalism. 

It has already been mentioned that this assumption has consequences for the physical
meaning of the independent connection \cite{Sotiriou:2006sr,Sotiriou:2006qn,Sotiriou:2006hs}. Let us elaborate on 
this: recall that an affine connection usually defines parallel 
transport and the covariant derivative.  On the other hand, the 
matter action $S_M$  is supposed to be a generally covariant 
scalar which includes  derivatives of the matter fields. 
Therefore, these derivatives ought to be covariant derivatives 
for a general matter field. Exceptions exist, such as a scalar 
field, for which a covariant and a partial derivative coincide, 
and the electromagnetic field, for which one can write a 
covariant action without the use of the covariant derivative [it 
is the exterior derivative that is actually needed, see next 
section and \cite{Sotiriou:2006qn}]. However, $S_M$ should include all possible fields. 
Therefore, assuming that $S_M$ is independent of the connection 
can imply one of two things \cite{Sotiriou:2006sr}: either we 
are restricting ourselves to specific fields,    or we 
are  implicitly assuming that it is the Levi-Civita connection 
of the  metric that actually defines parallel transport. Since 
the first  option is implausibly limiting for a gravitational 
theory, we  are left with the conclusion that the independent 
connection $\Gamma^\lambda_{\phantom{a}\mu\nu}$  does
not define parallel transport or the covariant derivative and the geometry is actually
pseudo-Riemannian.  The covariant derivative is actually defined by the Levi-Civita connection of the metric $\{^\lambda_{\phantom{a}\mu\nu}\}$. 

This also implies that Palatini $f(R)$ gravity is a {\em metric 
theory} in the sense that it satisfies the metric postulates 
\cite{willbook}. Let us clarify this: matter is   
minimally coupled to the metric and not coupled to any other  
 fields. Once again, as in GR or metric $f(R)$ gravity, one 
could 
use  diffeomorphism invariance to show that the stress energy 
tensor  is conserved by the covariant derivative defined with 
the  Levi-Civita connection of the metric, {\em 
 i.e.},~$\nabla_\mu T^{\mu\nu}=0$ (but $\bar{\nabla}_{\mu} 
 T^{\mu\nu}\neq 0$). This can also be shown by using the field 
 equations, which we will present shortly, in order  to 
 calculate the divergence of $T_{\mu\nu}$ with respect to the 
 Levi-Civita connection of the metric and show that it vanishes 
 \cite{Barraco:1998eq,Koivisto:2005yk}.\footnote{Energy 
 supertensors and pseudotensors in Palatini $f(R) $ gravity were 
 studied in 
 \cite{Barraco:1998eq, Borowiec:1993ne, Ferraris:1992dx, 
 Borowiec:1996kg} and alternative energy 
 definitions were  given in \cite{Deser:2007vs, Deser:2003up, 
Deser:2002jk, Deser:2002rt}.} Clearly then,  
Palatini $f(R)$ gravity is a metric theory according to the 
definition of \cite{willbook} (not to be confused with  
the term ``metric" in ``metric $f(R)$ gravity", which simply 
refers to the fact that one only varies the action with respect 
to the metric). Conventionally thinking, as a consequence of the covariant conservation of the 
matter energy-momentum tensor, test particles should follow geodesics 
of the metric  in Palatini $f(R)$ gravity. This 
can  be seen by considering a dust fluid with $T_{\mu\nu}=\rho 
\, 
u_{\mu}  u_{\nu} $ and  projecting the conservation equation 
$\nabla^{\beta} 
T_{\mu\beta}=0$ onto the fluid four-velocity $u^{\beta}$. 
Similarly, theories that satisfy the metric postulates are 
supposed to satisfy the Einstein Equivalence Principle     as 
well \cite{willbook}. Unfortunately, things are more complicated 
here and, therefore, we set this issue aside for the moment. We 
will return to it and attempt to fully clarify it in 
Secs.~\ref{sec:conflict} and~\ref{sec:surfsing}. For now,    let 
us proceed with our discussion of the field equations.

Varying the action~(\ref{palaction}) independently with respect 
to  the metric and the connection, respectively, and using the 
formula
\bea
\label{varR}
\delta {\cal R}_{\mu\nu}&=&\gn_\lambda \delta 
\Gamma^\lambda_{\phantom{a}\mu\nu}-\gn_{\nu}\delta 
\Gamma^\lambda_{\phantom{a}\mu\lambda}.
\eea
yields
\bea
\label{palf1}
&&f'({\cal R}) {\cal R}_{(\mu\nu)}-\frac{1}{2}f({\cal 
R})g_{\mu\nu}=\kappa \, T_{\mu\nu},\\
\label{palf2} 
&&-\gn_\lambda\left(\sqrt{-g}f'({\cal 
R} )g^{\mu\nu}\right) \nn\\
& &  \qquad+\gn_\sigma\left(\sqrt{-g}f'({\cal 
R}) g^{\sigma(\mu}\right)\delta^{\nu)}_\lambda=0,
\eea 
where $T_{\mu\nu}$ is defined in the usual way as in 
eq.~(\ref{set}),  $\bar{\nabla}_\mu$ denotes the covariant 
derivative defined with the independent connection  
$\Gamma^\lambda_{\phantom{a}\mu\nu}$, and $(\mu\nu)$ and 
$[\mu\nu]$ denote symmetrization or  
anti-symmetrization over 
the indices $\mu$ and $\nu$, respectively.  Taking the trace of 
eq.~(\ref{palf2}), it can be easily shown that
\be
\gn_\sigma\left(\sqrt{-g}f'({\cal R})g^{\sigma\mu}\right)=0,
\ee
 which implies that we can bring the field equations into the more 
economical form
\bea
\label{palf12}
f'({\cal R}) {\cal R}_{(\mu\nu)}-\frac{1}{2}f({\cal 
R})g_{\mu\nu}&=&\kappa \,G\, T_{\mu\nu},\\
 \label{palf22} 
\gn_\lambda\left(\sqrt{-g}f'({\cal R})g^{\mu\nu}\right)&=&0,
\eea
It is now easy to see how the Palatini formalism leads to GR 
when $f({\cal R})={\cal R}$; in this case $f'({\cal R})=1$ and 
eq.~(\ref{palf22}) becomes the definition of the Levi-Civita 
connection for the initially independent connection 
$\Gamma^\lambda_{\phantom{a}\mu\nu}$. Then, ${\cal 
R}_{\mu\nu}=R_{\mu\nu}$, ${\cal R}=R$ and eq.~(\ref{palf12}) 
yields Einstein's equations. This reproduces the result that 
can be found in textbooks \cite{mtw,wald}. Note that in the 
Palatini formalism for GR, the fact that the connection turns 
out to be the Levi-Civita one is a dynamical feature instead of 
an {\em a priori} assumption.

It is  now evident that generalizing the action to be a general 
function of ${\cal R}$ in the Palatini formalism is just as 
natural as it is to generalize the Einstein--Hilbert action in 
the metric formalism.\footnote{See, however, 
\cite{Sotiriou:2007} for  further analysis of the $f(R)$ 
action and how it can be derived from first principles in the 
two formalisms.}  Remarkably, even though the two formalisms 
give the same results for linear actions, they lead to  
different results for more general actions 
\cite{Buchdahl:1983zz, Exirifard:2007da,Burton:1997sj, 
Burton:1997pe, Querella:1998ke,Shahid-Saless:1987}.

Finally,  let us present some useful manipulations of the field 
equations. Taking the trace of eq.~(\ref{palf12}) yields
\be
\label{paltrace}
f'({\cal R}) {\cal R}-2f({\cal R})=\kappa\,T.
\ee
 As in the metric case, this equation will prove very useful 
later on. For a given  $f$, it is an algebraic equation in 
${\cal R}$. For all cases in  which $T=0$, including vacuum 
and electrovacuum, ${\cal  R}$ will therefore be a constant and 
a root of the equation
 \be
\label{paltracev}
f'({\cal R}) {\cal R}-2f({\cal R})=0.
\ee
We will not consider cases for which this equation has  no roots 
since it can be shown that the field equations are then 
inconsistent
\cite{Ferraris:1992dx}. Therefore, choices of $f$ that lead to 
this behaviour
should simply be avoided. Eq.~(\ref{paltracev}) can also be
identically satisfied if $f({\cal R})\propto {\cal R}^2$. This very
particular choice for $f$ leads to a conformally invariant theory
\cite{Ferraris:1992dx}. As is apparent from eq.~(\ref{paltrace}), if $f({\cal
R})\propto {\cal R}^2$ then only conformally invariant matter, for
which $T=0$ identically, can be coupled to gravity. Matter is not
generically conformally invariant though, and so  this 
particular choice of $ f $ is not suitable for a low energy 
theory of gravity. We will, therefore, not consider it further 
[see \cite{Sotiriou:2006hs} for a discussion].

Next, we consider eq.~(\ref{palf22}). Let us define a
metric conformal to $g_{\mu\nu}$ as 
 \be
 \label{hgconf}
h_{\mu\nu} \equiv f'({\cal R})g_{\mu\nu}.
\ee
It can easily be shown that\footnote{This calculation holds in 
four    dimensions. When the number of dimensions $D$ is 
different from 4 then, instead of using eq.~(\ref{hgconf}), 
the  conformal metric $h_{\mu\nu}$ should be introduced as 
$ h_{\mu\nu} \equiv [f'({\cal R})]^{2/(D-2)}g_{\mu\nu}$ in order 
for eq.~(\ref{midstep}) to still hold.}
\begin{equation}
\label{midstep}
\sqrt{-h} \, h^{\mu\nu}=\sqrt{-g} \, f'({\cal R}) g^{\mu\nu}.
\end{equation}

Then, eq.~(\ref{palf22}) becomes the definition of the 
Levi-Civita connection of $h_{\mu\nu}$ and can 
be solved algebraically to give
 \be
\Gamma^\lambda_{\phantom{a}\mu\nu}= \frac{1}{2}
h^{\lambda\sigma}\left(\partial_\mu h_{\nu\sigma} +\partial_\nu 
h_{\mu\sigma}-\partial_\sigma h_{\mu\nu}\right), 
\ee
or, equivalently, in terms of $g_{\mu\nu}$,
\bea
\label{gammagmn}
\Gamma^\lambda_{\phantom{a}\mu\nu}&=&\frac{1}{2}\frac{1}{f'({\cal 
R})}g^{\lambda\sigma}\Big[\partial_\mu \left(f'({\cal 
R})g_{\nu\sigma}\right)+\partial_\nu \left(f'({\cal 
R})g_{\mu\sigma}\right)\nn\\
&&-\partial_\sigma \left(f'({\cal 
R})g_{\mu\nu}\right)\Big],
\eea
Given that eq.~(\ref{paltrace}) relates ${\cal R}$ 
algebraically with
$T$, and since we have an explicit expression for
$\Gamma^\lambda_{\phantom{a}\mu\nu}$ in terms of ${\cal R}$ and
$g^{\mu\nu}$, we can  in principle eliminate the independent 
connection
from the field equations and express them only in terms of the metric
and the matter fields. Actually, the fact that we can 
algebraically express $\Gamma^\lambda_{\phantom{a}\mu\nu}$ in 
terms of the latter two already indicates that this connections 
act as some sort of auxiliary field. We will explore this 
further in Sec.~\ref{sec:equiv}. For the moment, let us  take 
into account how the Ricci
tensor transforms under conformal transformations and write
 \bea
\label{confrel1}
{\cal R}_{\mu\nu}=R_{\mu\nu}&+&\frac{3}{2}\frac{1}{(f'({\cal 
R}))^2}\left(\nabla_\mu f'({\cal R})\right)\left(\nabla_\nu 
f'({\cal R})\right)-\nn\\
& -&\frac{1}{f'({\cal 
R})}\left(\nabla_\mu 
\nabla_\nu-\frac{1}{2}g_{\mu\nu}\Box\right)f'({\cal R}).
\eea
Contraction with $g^{\mu\nu}$ yields
\bea
\label{confrel2}
{\cal R}=R&+&\frac{3}{2(f'({\cal R}))^2}\left(\nabla_\mu f'({\cal 
R})\right)\left(\nabla^\mu f'({\cal 
R})\right)\nn\\&+&\frac{3}{f'({\cal R})}\Box f'({\cal R}).
\eea
 Note the difference between ${\cal R}$ and the Ricci scalar of
$h_{\mu\nu}$ due to the fact that $g_{\mu\nu}$ is used here for the
contraction of ${\cal R}_{\mu\nu}$.

Replacing eqs.~(\ref{confrel1}) and (\ref{confrel2}) in
eq.~(\ref{palf12}), and after some easy manipulations, one 
obtains
\bea
\label{eq:field}
G_{\mu \nu} &= &\frac{\kappa}{f'}T_{\mu \nu}- \frac{1}{2} 
g_{\mu \nu} \left({\cal R} - \frac{f}{f'} \right) \\
&&+ \frac{1}{f'} \left(
			\nabla_{\mu} \nabla_{\nu}
			- g_{\mu \nu} \Box
		\right) f'-\nn\\
& &- \frac{3}{2}\frac{1}{f'^2} \left[
			(\nabla_{\mu}f')(\nabla_{\nu}f')
			- \frac{1}{2}g_{\mu \nu} (\nabla f')^2
		\right]. \nn
\eea
 Notice that, assuming that we know the root of 
eq.~(\ref{paltrace}), ${\cal R}={\cal R}(T)$, we have 
completely eliminated  the independent connection from this 
equation. Therefore, we have 
successfully reduced
the number of field equations to one and at the same time both 
sides of
eq.~(\ref{eq:field})  depend only on the metric and the matter 
fields. In a sense, the theory has been brought to the form of  
GR with a modified source.

We can now straightforwardly deduce the following:
\begin{itemize}
 \item When $f({\cal R})={\cal R}$, the theory reduces to GR, as 
discussed previously.
 \item For matter fields with $T=0$, due to 
eq.~(\ref{paltracev}),
${\cal R}$ and consequently $f({\cal R})$ and $f'({\cal R})$ are
constants and the theory reduces to GR with a
cosmological constant and a modified coupling constant $G/f'$. If we
denote the value of ${\cal R}$ when $T=0$ as ${\cal R}_0$, then the
value of the cosmological constant is
 \be
 \label{cosmcon}
\frac{1}{2}\left({\cal R}_0 - \frac{f({\cal R}_0)}{f'({\cal 
R}_0)} \right)=\frac{{\cal R}_0}{4},
\ee
 where we have used eq.~(\ref{paltracev}). Besides vacuum, $T=0$ also
for electromagnetic fields, radiation, and any other conformally
invariant type of matter.
\item In the general case $T\neq 0$, the modified source on the right
hand side of eq.~(\ref{eq:field}) includes derivatives of the 
stress-energy tensor, unlike in
GR. These are  implicit in the last two terms of 
eq.~(\ref{eq:field}), since
$f'$ is in practice a function of $T$, given that\footnote{Note  
that, apart from special
cases such as a perfect fluid, $T_{\mu\nu}$ and consequently $T$
already include first derivatives of the matter fields, given that the
matter action has such a dependence. This implies that the right hand
side of eq.~(\ref{eq:field}) will include at least second derivatives
of the matter fields, and possibly up to third derivatives.}  
$f'=f'({\cal R})$ and ${\cal R}={\cal R}(T)$.
 \end{itemize}

The serious implications of this last observation will become 
clear in Sec.~\ref{sec:astro}.

\subsection{Metric-affine formalism}

As we already pointed out, the Palatini formalism of $f(R)$ 
gravity relies on the crucial assumption that the matter action 
does not depend on the independent connection. We also argued 
that this assumption relegates this connection to the role 
of some sort of 
auxiliary field and the connection carrying the usual 
geometrical meaning ---  parallel transport and definition of 
the covariant derivative    --- remains the Levi-Civita    
 connection of the metric. All of these statements will be 
supported further 
in  the forthcoming sections, but for the moment let us consider 
what  would be the outcome if we decided to be faithful to the 
geometrical interpretation of the independent connection  
 $\Gamma^{\lambda}_{\phantom{a}\mu\nu}$: this would imply that 
we  would define the covariant derivatives of the matter fields 
with  this connection and, therefore, we would have 
$S_M = S_M(g_{\mu\nu},\Gamma^\lambda_{\mu\nu},\psi)$. The action 
of this  theory, dubbed metric-affine $f(R)$ gravity 
\cite{Sotiriou:2006qn}, would 
then be 
[note the  difference with respect to the 
action~(\ref{palaction})]
\be
\label{maaction}
S_{ma}=\frac{1}{ 2\kappa }\int d^4 x \sqrt{-g} f({\cal R}) 
+S_M(g_{\mu\nu}, \Gamma^{\lambda}_{\phantom{a}\mu\nu}, \psi).
\ee

\subsubsection{Preliminaries}

Before going further and deriving field equations from this 
action  certain issues need to be clarified. First, 
since now the matter action depends on the connection, we 
should define a quantity representing the 
variation  of $S_M$ with respect to  the connection, which 
mimics  the 
definition of the stress-energy tensor. We  call this quantity 
the {\em hypermomentum} and is defined as \cite{hehl}
\be
\label{defD}
\Delta_{\lambda}^{\phantom{a}\mu\nu} 
\equiv-\frac{2}{\sqrt{-g}}\frac{\delta {S}_M}{\delta  
\Gamma^\lambda_{\phantom{a}\mu\nu}}. 
\ee

 Additionally, since the connection is now promoted to the role 
of a  
 completely independent field, it is interesting to consider not 
 placing any restrictions to it. Therefore, besides dropping the 
 assumption that the connection is related to the metric, we 
 will also drop the assumption that the connection is symmetric. 
 It is useful to define the following quantities: the 
 non-metricity tensor
\be
\label{nonmet}
Q_{\mu\nu\lambda}\equiv-\gn_\mu g_{\nu\lambda},
\ee
which measures the failure of the connection to covariantly 
conserve the metric,  the trace of the non-metricity tensor with 
respect to its last two 
(symmetric) indices, which is called the Weyl vector,
\be
\label{weyl}
Q_\mu\equiv \frac{1}{4}Q_{\mu\nu}^{\phantom{a}\phantom{b}\nu} ,
\ee
and  the Cartan  torsion tensor
\be
\label{cartan}
S_{\mu\nu}^{\phantom{ab}\lambda}\equiv 
\Gamma^{\lambda}_{\phantom{a}[\mu\nu]},
\ee
which is the antisymmetric part of the connection.

By allowing a non-vanishing Cartan torsion tensor we are 
allowing the theory to naturally include torsion. Even though 
this brings complications, it has been considered by 
some to be an advantage for a gravity theory since some matter 
fields, such as  Dirac fields, can be coupled to gravity in a way 
which  might be considered more natural \cite{Hehl:1994ue}: one 
might expect that at some intermediate or high energy
regime,  the spin of particles might interact with the geometry 
(in the  same sense that macroscopic angular momentum interacts 
with geometry) and
torsion can naturally arise. Theories with torsion have a long 
history, probably starting with 
the Einstein--Cartan(--Sciama--Kibble)   
 theory 
\cite{cartan1,cartan2, 
cartan3,Sciama:1964wt,Kibble:1961ba,Hehl:1976kj}.  In this 
theory, as well as in other theories with an independent  
connection, some part of the connection is  still related to the 
metric ({\em e.g.},~the non-metricity is set to zero). In our  
case, the 
connection is left completely unconstrained and is  to be 
determined by the field equations. Metric-affine gravity with  
 the linear version of the action (\ref{maaction}) was initially 
proposed in~\cite{hehl} and 	the generalization to $f({\cal 
R})$ actions  was considered in 
\cite{Sotiriou:2006mu,Sotiriou:2006qn}.

 Unfortunately, leaving the connection completely unconstrained 
 comes with a complication. Let us consider the projective 
transformation
\be
\label{proj}
\Gamma^{\lambda}_{\phantom{a}\mu\nu}\rightarrow 
\Gamma^{\lambda}_{\phantom{a}\mu\nu}+{\delta^\lambda}_\mu\xi_\nu, 
\ee
where $\xi_\nu$ is an arbitrary covariant vector field. One can
easily show that the {\cal R}icci tensor will correspondingly
transform like
 \be
\label{projRicci}
{\cal R}_{\mu\nu}\rightarrow {\cal 
R}_{\mu\nu}-2\partial_{[\mu}\xi_{\nu]}.
\ee
 However, given that the metric is symmetric, this implies that the 
curvature scalar does not change
\be
{\cal R}\rightarrow {\cal R},
\ee
 {\em i.e.},~${\cal R}$ is invariant under projective 
transformations. 
Hence the Einstein--Hilbert action or any other action built from a
function of ${\cal R}$, such as the one used here, is projective
invariant in metric-affine gravity.  However, the matter action is not
generically projective invariant and this would be the cause of an
inconsistency in the field equations.

 One could try to avoid this problem by generalizing the 
gravitational action in order to break projective invariance. 
 This can be done in several ways, such as allowing for the 
 metric to be non-symmetric as well, adding higher order 
curvature invariants or terms including the Cartan torsion 
tensor  [see \cite{Sotiriou:2007,Sotiriou:2006qn}  for a more 
detailed discussion].  However, if one wants to stay within the 
framework of $f(R)$ gravity, which is our subject here,  then 
there is only one way to cure this problem: to somehow  
constrain the connection. In fact, it is evident  from 
eq.~(\ref{proj}) that, if the connection were symmetric,  
projective invariance would be broken.  However, one does not 
have to take such a  drastic measure.

To understand this issue further,
we should re-examine the meaning of projective invariance. This is
very similar to gauge invariance in electromagnetism (EM). It 
tells us
that the corresponding field, in this case the connections
$\Gamma^\lambda_{\phantom{a}\mu\nu}$, can be determined from  
the field
equations up to a projective transformation [eq.~(\ref{proj})].
Breaking this invariance can therefore come by fixing some degrees of
freedom of the field, similarly to gauge fixing. The number of degrees
of freedom which we need to fix is obviously the number of the
components of the four-vector used for the transformation, {\em
i.e.},~simply four. In practice, this means that we should start 
by
assuming that the connection is not the most general which one can
construct, but satisfies some constraints.

Since the degrees of freedom that we need to fix are four and 
seem  to be related to the non-symmetric part of the connection,  
the most obvious prescription is to demand that  $  
 S_\mu=S_{\sigma\mu}^{\phantom{ab}\sigma}$ be equal to zero, 
which 
 was first suggested in \cite{Sandberg:1975db} for a linear 
 action and shown to work also for an $f({\cal R})$  action in 
 \cite{Sotiriou:2006qn}.\footnote{The proposal of \cite{hehl} to 
 fix part of the non-metricity, namely the Weyl vector $Q_\mu$,  
 in order to break projective invariance works only when 
 $f({\cal R})={\cal R}$ \cite{Sotiriou:2007,Sotiriou:2006qn}.}
Note that this does not mean that 
$\Gamma^{\phantom{ab}\sigma}_{\mu\sigma}$
should vanish, but merely that
$\Gamma^{\phantom{ab}\sigma}_{\mu\sigma} 
=\Gamma^{\phantom{ab}\sigma}_{\sigma\mu}$.
 Imposing this constraint can easily be done by adding a 
Lagrange multiplier $B^\mu$. 
The additional term in the action will be
\be
\label{lm2}
S_{LM}=\int d^4 x \sqrt{-g} \, B^\mu S_{\mu}.
\ee
The action~(\ref{maaction}) with the addition of the term in 
eq.~(\ref{lm2})  is, therefore, the action of the  most general 
metric-affine $f(R)$ theory of gravity.

\subsubsection{Field Equations}

We are now ready to vary the action and obtain field equations. 
Due to space limitations, we will not present the various steps 
of the variation    here. Instead we merely give the 
formula
\bea
\label{varR2}
\delta {\cal R}_{\mu\nu}&=&\gn_\lambda \delta 
\Gamma^\lambda_{\phantom{a}\mu\nu}-\gn_{\nu}\delta 
\Gamma^\lambda_{\phantom{a}\mu\lambda} 
+2\Gamma^\sigma_{\phantom{a}[\nu\lambda]} 
\delta\Gamma^\lambda_{\phantom{a}\mu\sigma}, \eea
 which is useful to those wanting to repeat the 
variation as an exercise, and we also stress our definitions for 
the covariant derivative
 \be
\gn_\mu A^\nu_{\phantom{a}\sigma}= \partial_\mu 
A^\nu_{\phantom{a}\sigma}+\Gamma^\nu_{\phantom{a}\mu\alpha}  
A^\alpha_{\phantom{a}\sigma}-\Gamma^\alpha_{\phantom{a}\mu\sigma}  
A^\nu_{\phantom{a}\alpha}.
\ee
and for the Ricci tensor of an independent connection
\bea
\label{ricci}
{\cal R}_{\mu\nu}&=&{\cal 
R}^\lambda_{\phantom{a}\mu\lambda\nu}\\&=&\partial_\lambda 
\Gamma^\lambda_{\phantom{a}\mu\nu}-\partial_\nu 
\Gamma^\lambda_{\phantom{a}\mu\lambda} 
+\Gamma^\lambda_{\phantom{a}\sigma\lambda} 
\Gamma^\sigma_{\phantom{a}\mu\nu}- 
\Gamma^\lambda_{\phantom{a}\sigma\nu} 
\Gamma^{\sigma}_{\phantom{a}\mu\lambda}.\nn 
\eea
The outcome of varying independently with respect to the 
metric, the connection and the Lagrange multiplier is, 
respectively,
\bea
\label{field1t1}
& &f'({\cal R})  {\cal R}_{(\mu\nu)}-\frac{1}{2}f({\cal 
R})g_{\mu\nu}=\kappa T_{\mu\nu},\\
 \label{field2t1}
& &\!\frac{1}{ 
\sqrt{-g}}\bigg[\!\!-\!\!\gn_\lambda\! \left(\sqrt{-g}f'({\cal 
R})g^{\mu\nu}\right)\!+\!\gn_\sigma\! \left(\sqrt{-g}f'({\cal 
R})g^{\mu\sigma}\right){\delta^\nu}_\lambda\bigg]\nn \\
&&\quad+2f'({\cal  
R})\left(g^{\mu\nu} 
S^{\phantom{ab}\sigma}_{\lambda\sigma} 
-g^{\mu\rho}S^{\phantom{ab}\sigma}_{\rho\sigma}{\delta^\nu}_\lambda 
+g^{\mu\sigma}S^{\phantom{ab}\nu}_{\sigma\lambda}\right) 
\nn\\
&&\qquad\qquad\qquad 
=\kappa(\Delta_{\lambda}^{\phantom{a}\mu\nu} 
-B^{[\mu}{\delta^{\nu]}}_{\lambda}),\\
& & S_{\mu\sigma}^{\phantom{ab}\sigma}=0. \label{4444}
\eea
Taking the trace of eq.~(\ref{field2t1}) over the 
indices $\mu$ and 
$\lambda$ and using eq.~(\ref{4444})  yields
\be
B^\mu=\frac{2}{3}\Delta_{\sigma}^{\phantom{a}\sigma\mu}.
\ee
Therefore, the final form of the field equations is
\bea
\label{field1t}
& &f'({\cal R})  {\cal R}_{(\mu\nu)}-\frac{1}{2}f({\cal 
R})g_{\mu\nu}=\kappa T_{\mu\nu},\\
\label{field2t}
& &\frac{1}{\sqrt{-g}}\bigg[\!\!-\!\!\gn_\lambda\!\left( 
\sqrt{-g}f'({\cal R})g^{\mu\nu}\right)\! 
+\!\gn_\sigma\!\left(\sqrt{-g}f'({\cal R}) 
g^{\mu\sigma}\right){\delta^\nu}_\lambda\bigg]\nn\\
&&\quad+2f'({\cal R})g^{\mu\sigma} 
S^{\phantom{ab}\nu}_{\sigma\lambda} 
=\kappa(\Delta_{\lambda}^{\phantom{a}\mu\nu} 
-\frac{2}{3}\Delta_{\sigma}^{\phantom{a} 
\sigma[\nu}{\delta^{\mu]}}_{\lambda}), \\
\label{field3t}
& & S_{\mu\sigma}^{\phantom{ab}\sigma}=0.
\eea
Next, we examine the role of 
$ \Delta_{\lambda}^{\phantom{a}\mu\nu}$. By splitting 
eq.~(\ref{field2t})  into a symmetric and an antisymmetric part 
and performing contractions  and manipulations it can be shown 
that \cite{Sotiriou:2006qn}
\be
\label{torsion}
\Delta_{\lambda}^{\phantom{a}[\mu\nu]}=0 \Rightarrow 
S_{\mu\nu}^{\phantom{ab}\lambda}=0.
\ee
This straightforwardly implies two things: a) Any torsion is introduced by matter fields for which  $\Delta_{\lambda}^{\phantom{a}[\mu\nu]}$ is non-vanishing; b) torsion is not propagating, since it is given algebraically in terms of the matter fields through $\Delta_{\lambda}^{\phantom{a}[\mu\nu]}$. It can, therefore, only be detected in the presence of such matter fields. In the absence of the latter, spacetime will have no torsion. 

In a similar fashion, one can use the symmetrized version of  
eq.~(\ref{field2t}) to show that the symmetric part of the  
hypermomentum $\Delta_{\lambda}^{\phantom{a}(\mu\nu)}$   is 
algebraically related to the non-metricity $Q_{\mu\nu\lambda}$.  
Therefore, matter fields  with non-vanishing 
$\Delta_{\lambda}^{\phantom{a}(\mu\nu)}$  will introduce 
non-metricity. However, in this case things  are slightly more 
complicated because part of the  non-metricity is also due to 
the 
functional form of the Lagrangian itself [see  \cite{Sotiriou:2006qn}].

We will not perform a detailed study of different matter fields and their role in metric-affine gravity. We refer the reader to the more exhaustive analysis of \cite{Sotiriou:2006qn} for details and we restrict ourselves to the following 
remarks: Obviously, there are certain types of matter fields for which $\Delta_{\lambda}^{\phantom{a}\mu\nu}=0$. Characteristic examples are
\begin{itemize}
\item A scalar field, since in this case the covariant derivative can be replaced with a partial derivative. Therefore, the connection does not enter the matter action.
\item The electromagnetic field (and gauge fields in general), since the electromagnetic field tensor $F_{\mu\nu}$ is defined in a covariant manner using the exterior derivative. This definition remains unaffected when torsion is included [this can be related to gauge invariance, see \cite{Sotiriou:2006qn} for a discussion].
\end{itemize}
On the contrary, particles with spin,  such as Dirac fields,  
 generically have a non-vanishing hypermomentum and will, 
therefore, introduce torsion.
A more complicated case is that of a perfect fluid with 
vanishing 
vorticity.  If we set torsion aside, or if we consider a fluid 
describing particles that would initially not introduce any 
torsion  then, as for a usual perfect fluid in GR, the matter 
action can  be   written  in terms of three scalars: the energy 
density, the  pressure, and the velocity potential 
\cite{Stone:1999gi,schakel}. Therefore such a fluid will lead to 
a vanishing $\Delta_{\lambda}^{\phantom{a}\mu\nu}$. However, 
complications arise when torsion is taken into account:  Even 
though it can be argued that the spins of the individual 
particles composing the fluids will be randomly oriented, and 
therefore the expectation value for the spin should add up to 
zero, fluctuations around this value will affect spacetime   
 \cite{Hehl:1976kj,Sotiriou:2006qn}. Of course, such effects 
will 
 be largely suppressed, especially in situations in which the 
energy density is small, such as late time cosmology. 

It should be evident by now that, due to eq.~(\ref{torsion}),  
the field equations of metric $f(R)$ gravity  reduce to 
eqs.~(\ref{palf1}) and (\ref{palf2}) and, ultimately,  to the 
field equations of Palatini $f(R)$ gravity~(\ref{palf12})  
and~(\ref{palf22}), for all cases in which  
 $\Delta_{\lambda}^{\phantom{a}\mu\nu}=0$. Consequently, in 
 vacuo, where also $T_{\mu\nu}=0$, they will reduce to the 
Einstein equations with an  effective cosmological constant 
given by eq.~(\ref{cosmcon}), as discussed at the end  of 
Sec.~\ref{sec:palatinifield} for Palatini $f(R)$ gravity. 

In conclusion, metric-affine $f(R)$ gravity appears to be the 
most general case of $f(R)$ gravity. It includes enriched 
phenomenology, such as matter-induced non-metricity and torsion. 
It is worth stressing that torsion comes quite naturally, since 
it is actually introduced by particles with spin (excluding 
gauge fields). Remarkably, the theory reduces to GR in vacuo or 
for conformally invariant types of matter, such as the 
electromagnetic field, and departs from GR in the same way that 
Palatini $f(R)$ gravity does  for most matter fields that are 
usually studied as sources of gravity. However, at the same 
time,  
it exhibits new phenomenology in less studied cases, such as in 
the presence of Dirac fields, which include torsion and 
non-metricity. Finally, let us repeat once more that  Palatini 
$f(R)$ gravity, despite appearances, is really a metric theory 
according to the definition of \cite{willbook} 
(and the geometry is {\em a priori} 
pseudo-Riemannian).\footnote{As mentioned in 
Sec.~\ref{sec:palatinifield}, although the metric postulates are 
manifestly satisfied, there are ambiguities regarding the 
physical interpretation of this property and its relation with 
the Einstein Equivalence Principle (see Sec.~\ref{sec:astro}).}
On  the contrary,  
metric-affine $f(R)$ gravity is not a metric theory (hence the 
name). Consequently, it should also be clear that $T^{\mu\nu}$ 
is not divergence-free                  with respect to the 
covariant derivative defined with the  Levi-Civita connection 
(nor with $\bar{\nabla}_\mu$ actually).  However, the physical 
meaning of this last statement is questionable  and deserves 
further analysis, since in metric-affine gravity  $T_{\mu\nu}$ 
does not really carry the usual meaning of a  stress-energy 
tensor (for instance, it does not reduce to  the special 
relativistic tensor at an appropriate limit  and at the same 
time there is also another quantity, the  hypermomentum, which 
describes matter characteristics).


\section{Equivalence with Brans--Dicke theory and classification 
of theories }
\label{sec:equiv}

In the same way that one can make variable redefinitions in 
classical mechanics in order to bring an equation describing a  
system to  a more attractive, or easy to handle, form (and in a 
 very similar way to changing coordinate systems), one can also 
 perform field redefinitions in a field theory, in order to 
rewrite the action or the field equations. 

There is no unique prescription for redefining the fields of a 
theory. One can introduce auxiliary fields, perform  
renormalizations or conformal transformations, or even  simply 
redefine fields to one's convenience. 

It is important to mention that, at least within a classical
perspective such as the one followed here, two theories are 
considered to be dynamically equivalent if, under a suitable 
redefinition of the
gravitational and matter fields, one can make their field equations
coincide. The  same statement can be made at the level of the 
action. Dynamically equivalent theories give exactly the same 
results when  
describing a dynamical system which falls within the purview of  
these theories. There are clear advantages in exploring the 
dynamical
equivalence between theories:  we can use results already 
derived for one theory in the study of another, equivalent, 
theory. 

The term ``dynamical equivalence'' can be considered misleading 
in classical gravity. Within a classical perspective, a theory 
is fully
described by a set of field equations. When we are referring to
gravitation theories, these equations  describe the 
dynamics of gravitating systems. Therefore, two dynamically 
equivalent theories
are actually just different representations of the same
theory (which also makes it clear  that all allowed 
representations can be used on an equal footing).

The issue of distinguishing between truly different theories and
different representations of the same theory (or dynamically
equivalent theories) is an intricate one. It has  serious 
implications
and has been the cause of many misconceptions in the past, especially
when conformal transformations are used in order to redefine the
fields ({\em e.g.,~}the Jordan and Einstein frames in 
scalar-tensor
theory). It goes beyond the scope of this review to present a 
detailed analysis of this issue. We refer the reader to the 
literature, and specifically to \cite{Sotiriou:2007zu} and 
references therein for a detailed discussion. Here, we     
 simply mention that, given that they are handled carefully, 
field redefinitions and different representations of the same 
theory are perfectly legitimate and constitute very useful tools 
for understanding gravitational theories. 

In what follows, we review the equivalence between metric and 
Palatini $f(R)$ gravity with specific theories within the 
Brans--Dicke class with a potential. It is shown that these 
versions of $f(R)$ gravity are nothing      but different 
representations of Brans--Dicke theory with  Brans--Dicke 
parameter $\omega_0=0$ and $\omega_0=-3/2$,  respectively. We 
comment on this equivalence and on whether preference to a specific representation should be an issue. Finally, we use this equivalence to perform a classification of $f(R)$ gravity.

\subsection{Metric formalism}
\label{sec:equivalencemetric}

It has been noticed quite early that metric quadratic gravity 
can be cast into the form of a Brans--Dicke theory  and it did 
not take long for these results to be extended to more 
general actions which are functions of the Ricci scalar of the 
metric     \cite{Teyssandier:1983zz, Wands:1993uu, 
Barrow:1988xh,Barrow:1988xi} [see also 
\cite{Flanagan:2003iw} and 
\cite{Cecotti:1987sa, Wands:1993uu} for the extension to 
theories of the type 
$f(R, 
\Box^k R)$ with $k\geq 1$ of interest in supergravity]. This 
 equivalence has been re-examined recently due to the increased 
interest in metric $f(R)$ gravity 
 \cite{Chiba:2003ir,Flanagan:2004bz,Sotiriou:2006hs}. Let us 
present this equivalence in some detail.

We will work at the level of the action but the same approach 
can be used
to work directly at the level of the field equations. We begin with
metric $f(R)$ gravity. For the convenience of the reader, we  
rewrite
here the action (\ref{metaction}):
\be
S_{met}=\frac{1}{ 2\kappa }\int d^4 x \sqrt{-g} \, f(R) 
+S_M(g_{\mu\nu},\psi).
\ee
One can introduce a new field $\chi$ and write the dynamically 
equivalent action 
\bea
\label{metactionH}
S_{met}&=&\frac{1}{ 2\kappa }\int d^4 x \sqrt{-g} 
\left[ f(\chi)+f'(\chi)(R-\chi)\right] 
+\nn\\&&\qquad\qquad\qquad+S_M(g_{\mu\nu},\psi).
\eea
 Variation with respect to $\chi$ leads to the equation
 \be
 \label{1600}
 f''( \chi )(R-\chi)=0.
 \ee
 Therefore,
  $\chi=R$ if
$f''(\chi)\neq 0$, which reproduces the 
action~(\ref{metaction}).\footnote{The action is sometimes 
called  ``$R$-regular'' by mathematical physicists if  
$f''(R)\neq 0$  [{\em e.g.}, \cite{Magnano:1993bd}].}
Redefining the field $\chi$ by $\phi=f'(\chi)$ and setting
 \be
\label{defV}
V(\phi)=\chi(\phi)\phi-f(\chi(\phi)),
\ee
 the action takes the form
\be
\label{metactionH2}
S_{met}=\frac{1}{ 2\kappa }\int d^4 x \sqrt{-g} \left[ \phi 
R-V(\phi)\right] +S_M(g_{\mu\nu},\psi).
\ee
This is the Jordan frame representation of the  action of a 
Brans--Dicke theory with
Brans--Dicke parameter $\omega_0=0$. An $\omega_0=0$ 
Brans--Dicke 
theory [sometimes called ``massive 
dilaton gravity'' \cite{Wands:1993uu}] was originally 
proposed by~\cite{O'Hanlon:1972ya} in order to 
generate a Yukawa term in the Newtonian limit   and has 
been occasionally considered in the 
literature \cite{Anderson:1971dm, 
Fujii:1982ms, Deser:1970hs, Barber:2003ik, 
Davidson:2004hh, Dabrowski:2005yn, O'Hanlon:1972a, 
O'Hanlon:1972b}.  It should be stressed that the scalar 
degree of  freedom $\phi=f'(\chi)$ is quite different from a 
matter field; 
for example, like all nonminimally coupled scalars, it can 
violate all of the energy conditions 
\cite{valeriobook}.

The  field equations corresponding to the action 
(\ref{metactionH2}) are
\bea
\label{bdmetf1}
G_{\mu\nu}&=&\frac{\kappa}{\phi}  \, T_{\mu\nu} -\frac{1}{2\phi} 
\, 
g_{\mu\nu} V(\phi) \nonumber \\
&& +\frac{1}{\phi}\left( 
\nabla_{\mu}\!\nabla_{\nu} 
\phi -g_{\mu\nu} \Box \phi \right) ,\;\\
\label{RV} 
 R&=&V'(\phi).
\eea
These field equations could have been derived directly from 
eq.~(\ref{metf}) using the same field redefinitions that were 
mentioned above for the action. By taking the trace of 
eq.~(\ref{bdmetf1})  in order to replace $R$ in eq.~(\ref{RV}), 
one gets
\be
\label{metbdfphi}
3\Box \phi +2V(\phi) -\phi\, \frac{dV}{d\phi}=\kappa 
T .
\ee
This last equation determines the dynamics of $\phi$ for given 
matter sources.

The condition $f''\neq 0$ for the scalar-tensor theory to be 
equivalent to the original $f(R)$ gravity theory can be 
seen as the condition that the change of variable 
$ \phi = f'(R)$ needed to express the 
theory as a Brans--Dicke one~(\ref{metactionH2})  be invertible, 
{\em 
i.e.}, $ 
d\phi/dR=f''\neq 0$. This is a sufficient but not 
necessary condition for invertibility: it is only necessary 
that $f'(R)$ be continuous and  one-to-one  
\cite{Olmo:2006eh}. By looking at 
eq.~(\ref{1600}), it is seen that $f''\neq 0$ implies $\phi=f'(R)$ 
and the equivalence of the actions (\ref{3}) and 
(\ref{metactionH}). When $f''$ is not defined, or it vanishes, 
the equality $ \phi=f'(R) $ and 
the equivalence between the two theories can not be 
guaranteed (although this it is not {\em a priori} excluded by 
$f''=0$).

 Finally, let us mention that, as usual in Brans--Dicke theory  
and more general scalar-tensor theories, one can 
perform a  conformal transformation and rewrite the 
action~(\ref{metactionH2}) in what is called the Einstein frame  
(as opposed to the Jordan frame). Specifically, by 
performing the conformal transformation
\be
\label{bdconf}
g_{\mu\nu}\rightarrow \tilde{g}_{\mu\nu}=f'(R) \, g_{\mu\nu}  
\equiv \phi \, g_{\mu\nu}
\ee
and the scalar field redefinition $\phi=f'(R) \rightarrow 
\tilde{\phi} $ with
\be
d\tilde{\phi} =\sqrt{\frac{2\omega_0 +3}{2\kappa} } \, 
\frac{d\phi}{\phi} ,
\ee
a scalar-tensor theory is mapped into the Einstein frame in 
which 
the ``new'' scalar field $\tilde{\phi}$ couples minimally to the 
Ricci curvature and has canonical kinetic energy, as described 
by the gravitational action
\be
 S^{(g)}=\int d^4x \, \sqrt{-\tilde{g}} \, \left[ 
\frac{\tilde{R}}{2\kappa}-\frac{1}{2} \, 
\partial^{\alpha}\tilde{\phi}\partial_{\alpha}\tilde{\phi} -U( 
\tilde{\phi} )  \right] .
\ee
For the $\omega_0=0$ equivalent of metric $f(R)$ gravity we have
\be
\phi \equiv f'(R)=\mbox{e}^{\sqrt{ \frac{2\kappa}{3}} \, 
\tilde{\phi}} ,
\ee
\be
U( \tilde{\phi} )=\frac{R f'(R)-f(R)}{2\kappa \left( f'(R) 
\right)^2} ,
\ee
where $R=R(\tilde{\phi}) $, and the complete action is
\bea
\label{metactionein}
S'_{met}&=&\int  d^4x \, \sqrt{-\tilde{g}} \, 
\left[\frac{\tilde{R}}{2\kappa }-\frac{1}{2} \, 
\partial^\alpha\tilde{\phi}\partial_\alpha\tilde{\phi} -U( \tilde{\phi} ) 
\right] +\nn\\
&&\qquad\qquad 
\qquad+S_M(e^{-\sqrt{2\kappa/3}\,\tilde{\phi}} 
\tilde{g}_{\mu\nu},\psi) . \eea
A direct transformation to the Einstein frame, without the 
intermediate passage from the Jordan frame, has been discovered 
in  \cite{Whitt:1984pd,Barrow:1988xh}. 

We stress once more  that the 
actions~(\ref{metaction}), (\ref{metactionH2}), 
and~(\ref{metactionein}) are nothing but different 
representations of the same theory.\footnote{This 
has been an issue of debate and confusion, see for example 
the references in 
\cite{Faraoni:2006fx}.}  Additionally, there is 
nothing exceptional about the Jordan or the  Einstein frame of 
the Brans--Dicke representation, and one can actually find 
infinitely  many conformal frames 
\cite{Flanagan:2004bz,Sotiriou:2007zu}.

\subsection{Palatini formalism}
\label{sec:equivalencePalatini}

Palatini $f(R)$ gravity can also be cast  in the form of a 
Brans--Dicke theory with a potential  
 \cite{Flanagan:2003rb,Olmo:2005jd,Sotiriou:2006hs}. As a matter 
of  fact, beginning from the Palatini $f(R)$ action, which we 
repeat here for the reader's convenience
 \be
\label{palaction2}
S_{pal} = \frac{1}{2\kappa}\int d^4 x \sqrt{-g} \, f({\cal R}) 
+S_M(g_{\mu\nu}, \psi),
\ee
and following exactly the same steps as before, {\em 
i.e.}, introducing  a scalar field $\chi$ which we later 
redefine in terms of $\phi$, yields 
\be
\label{palactionH2}
S_{pal}=\frac{1}{2\kappa}\int d^4 x \sqrt{-g} \left[ \phi {\cal 
R}-V(\phi)\right] +S_M(g_{\mu\nu}, \psi).
\ee
 Even though the gravitational part of this action is formally 
the same as that of the action~(\ref{metactionH2}), this action 
is not a Brans--Dicke one with $\omega_0=0$: ${\cal R}$ is 
not 
the Ricci scalar of the metric $ g_{\mu\nu}$.  However, we have 
already seen that the field equation~(\ref{palf12}) can be 
solved algebraically for the independent connection  yielding 
eq.~(\ref{gammagmn}). This implies that we can replace the 
connection  in the action without affecting the dynamics of the 
theory (the  independent connection is practically an auxiliary 
field).  Alternatively, we can directly use 
eq.~(\ref{confrel2}),  which relates $R$ and ${\cal R}$. 
Therefore, the action~(\ref{palactionH2}) can be rewritten, 
modulo surface terms, as
\bea
\label{palactionH2d0}
S_{pal}&=&\frac{1}{2\kappa}\int d^4 x \sqrt{-g} \left(\phi 
 R+\frac{3}{2\phi}\partial_\mu \phi \partial^\mu 
\phi-V(\phi)\right) \nn\\
&&\qquad\qquad\qquad+S_M(g_{\mu\nu}, \psi).
\eea
This is the action of a
Brans--Dicke theory with Brans--Dicke parameter $\omega_0=-3/2$.  
The corresponding field equations  obtained from the 
action~(\ref{palactionH2d0}) through variation with respect to 
the metric and the scalar are 
\bea
\label{stf1}
G_{\mu\nu}&=&\frac{\kappa}{\phi} T_{\mu\nu} -\frac{3}{2\phi^2}  
\Big(\nabla_\mu \phi \nabla_\nu \phi 
 -\frac{1}{2}g_{\mu\nu} \nabla^\lambda \phi \nabla_\lambda \phi 
\Big)+\nn\\
&&\qquad+\frac{1}{\phi}   (\nabla_\mu\nabla_\nu \phi-g_{\mu\nu} 
\Box \phi)-\frac{V}{2\phi}g_{\mu\nu},\\
\label{bdf2'}
\Box\phi&=& 
\frac{\phi}{3}(R-V')+\frac{1}{2\phi}\nabla^\mu\phi 
\nabla_\mu\phi.
\eea
Once again, we can use the trace of eq.~(\ref{stf1}) in order 
to eliminate $R$ in eq.~(\ref{bdf2'}) and relate $\phi$ 
directly to the matter sources. The outcome is
\be
\label{bdf2}
 2V-\phi V'=\kappa\, T.
\ee

Finally, one can also perform the conformal 
transformation~(\ref{bdconf}) in order to rewrite the 
action~(\ref{palactionH2d0}) in the Einstein frame. The result is
\be
\label{palactionein}
S'_{pal}=\int d^4x \, \sqrt{-\tilde{g}}  \, \left[\frac{\tilde{R}}{2\kappa}-U( 
\phi ) 
\right] +S_M(\phi^{-1}\tilde{g}_{\mu\nu},\psi) ,
\ee
where $U(\phi)=V(\phi)/(2\kappa \,\phi^2)$. Note that  here we 
have not used any redefinition for the scalar.

To  conclude, we have established that Palatini $f(R)$ gravity 
can  be cast into the form of an $\omega_0=-3/2$ Brans--Dicke 
theory with a potential. 

\subsection{Classification}
\label{sec:class}

The scope of this section is to present a classification of 
the different versions of $f(R)$ gravity. However, before doing 
so, some remarks are in order.

Let us, first of all, use the Brans--Dicke representation of 
both metric and Palatini $f(R)$ gravity  to comment on the 
dynamics of these theories. This  representation makes it 
transparent that metric $f(R)$  gravity has just one extra 
scalar degree of freedom with  respect to GR. The absence of a 
kinetic term for the scalar in  the action~(\ref{metactionH2}) 
or in  eq.~(\ref{RV}) should not mislead us to think that this 
degree of freedom does not carry dynamics.  As can be seen by 
eq.~(\ref{metbdfphi}), $\phi$ is dynamically related to the 
matter fields and, therefore, it is a dynamical degree of 
freedom. Of course, one should also not fail to mention that 
eq.~(\ref{RV}) does constrain the dynamics of $\phi$. In this 
sense metric $f(R)$ gravity and $\omega_0=0$ Brans--Dicke theory 
differs from the general Brans--Dicke theories and constitutes a 
special case. On the other hand, in the $\omega_0=-3/2$ case       
 which corresponds to Palatini $f(R)$ gravity, the scalar   
$\phi$ appears to have dynamics in the 
action~(\ref{palactionH2d0}) or in eq.~(\ref{bdf2'}). However, 
once again this is misleading since, as  is clear from 
eq.~(\ref{bdf2}), $\phi$ is in this case  algebraically related 
to the matter and, therefore,  carries no dynamics of its own 
[indeed the field eqs.~(\ref{stf1}) and  (\ref{bdf2}) could be 
combined to give eq.~(\ref{eq:field}),  eliminating 
$\phi$ completely].  As a remark, let us state   that the 
equivalence 
between  Palatini $f(R)$ gravity and $\omega_0=-3/2$ 
Brans--Dicke theory and the clarifications just made highlight 
two issues already mentioned: the  fact that Palatini $f(R)$ 
gravity is a metric theory according to the definition of 
\cite{willbook}, and the  fact that the independent 
connection is actually some sort of auxiliary field.

The fact that the dynamics of $\phi$ are not transparent at the 
level of the action in both cases should not come as a big  
 surprise: $\phi$ is coupled to the derivatives of the metric 
(through the coupling  with $R$) and, therefore, partial 
integrations to ``free"  $\delta \phi$ or $\delta g^{\mu\nu}$ 
during the variation are  bound to generate dynamical terms even 
if they are not initially  present in the action. The 
$\omega_0=-3/2$ case is even  more intricate because the 
dynamical terms generated  through this procedure exactly cancel 
the existing one in the action. 

 We already saw an example of how different representations of 
 the theory can highlight some of its characteristics and be 
 very useful for our understanding of it. The equivalence 
 between $f(R)$ gravity and Brans--Dicke theory will turn out to 
be very useful in the forthcoming sections.

Until now we have not discussed any possible equivalence 
between Brans--Dicke theory and metric-affine $f(R)$ gravity.  
However, it is quite straightforward to  see that there cannot 
be any. Metric-affine $f(R)$ gravity is not a metric theory 
and, consequently, it can not be cast into  the form of one, 
such 
as Brans--Dicke theory. For the sake of clarity,  let us 
state that one 
could still start from the action~(\ref{maaction})  and follow 
the  steps of the previous section to bring  its gravitational 
part  into the form of the action~(\ref{palactionH2d0}). 
However,  the matter action would have an explicit dependence 
from  the connection. Additionally, one would not be able to use 
eq.~(\ref{confrel2})  to eliminate ${\cal R}$ in favour of $R$ 
since this only holds in Palatini $f(R)$ gravity.

 In conclusion, metric-affine $f(R)$ gravity is the most general 
 case of $f(R)$ gravity. Imposing further assumptions can lead 
 to both metric or Palatini $f(R)$ gravity, which can be cast 
 into the form of $\omega_0=0$ and $\omega_0=-3/2$ Brans--Dicke 
theories with a potential. In both cases, restricting the 
functional form of 
the action leads  to GR. These 
results are summarized in the schematic diagram of 
Fig.~\ref{fig1}.
\begin{widetext}
\centering
\begin{tabular}{|c|}\hline  $\xymatrix{& \textrm{$f(R)$ GRAVITY} \ar[dl]|{\textrm{{\small $\Gamma^{\lambda}_{\phantom{a}\mu\nu}$ and $g_{\mu\nu}$ independent}}}\ar[ddr]|{\textrm{{\small $\Gamma^{\lambda}_{\phantom{a}\mu\nu}=
\left\{^\lambda_{\phantom{a}\mu\nu}\right\}
$}}} & \\ \textrm{METRIC-AFFINE $f(R)$} \ar[d]^{{\small \textrm{$S_M=S_M(g_{\mu\nu},\psi)$}}}& &  \\
 \textrm{PALATINI $f(R)$}\ar@2{<->}[dd]|{\textrm{{\small $f''(R)\neq 0$}}}\ar[dr]|{\textrm{{\small $f(R)=R$}}}&  & \textrm{METRIC $f(R)$} \ar@2{<->}[dd]|{\textrm{{\small $f''(R)\neq 0$}}}\ar[dl]|{\textrm{{\small $f(R)=R$}}} \\
&  \textrm{GR} &\\
\textrm{BRANS--DICKE, $\omega_0=-\frac{3}{2}$}& &\textrm{BRANS--DICKE, $\omega_0=0$}} 
$\\\hline\end{tabular}
\begin{figure}[h]
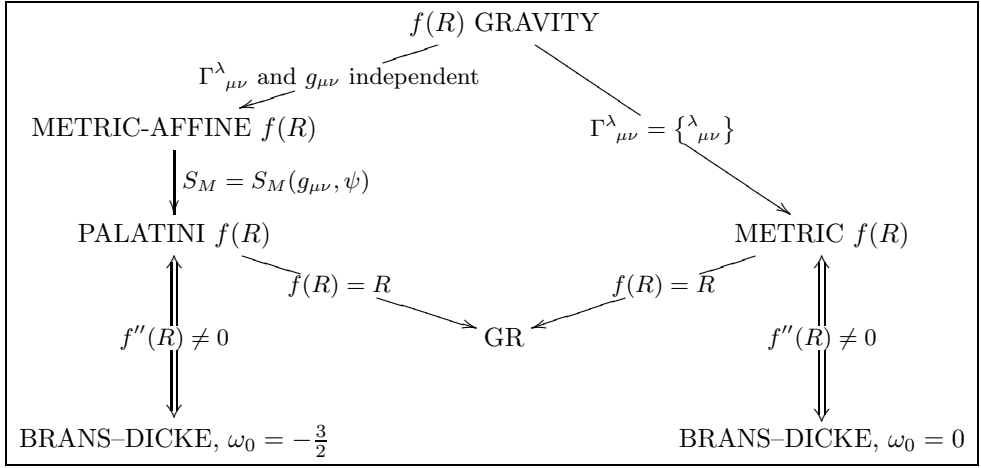

\centering
\caption{\label{fig1}Classification of $f(R)$ theories of 
gravity and equivalent Brans--Dicke theories. The flowchart 
shows the list of assumptions that are needed to arrive to the 
various versions of $f(R)$ gravity and GR beginning from the 
the general  $f(R)$ action.  It also 
includes the equivalent Brans--Dicke classes. Taken  
from~\cite{Sotiriou:2006hs}.}
\end{figure}
\end{widetext}

\subsection{Why $f(R)$ gravity then?}

Since $f(R)$ gravity in both the metric and the Palatini
formalisms can acquire a Brans--Dicke theory representation, one 
 might
be led to ask two questions: first, why should we consider the
$f(R)$ representation and not just work with the Brans--Dicke one,
and second, why, since we know a lot about Brans--Dicke
theory, should we regard $f(R)$ gravity as unexplored or 
interesting?

The answer to the first question is quite straightforward. There is
 actually no reason to prefer either of the two representations 
 --- at least as far as classical gravity is concerned. There
can be applications where the $f(R)$ representation can be more
convenient and applications where the Brans--Dicke representation is
more convenient. One should probably mention that habit affects
our taste and, therefore, an $f(R)$ representation seems 
more appealing to relativists due to its more apparent geometrical
nature, whereas the Brans--Dicke representation seems more
appealing to particle physicists. This issue can have theoretical
implications. To give an example: if $f(R)$ gravity is considered as a
step towards a more complicated theory, which generalisation would be
more straightforward will depend on the chosen representation [see also \cite{Sotiriou:2007zu} for a discussion]. 

Whether $f(R)$ theories of gravity are unexplored and interesting or
just an already-studied subcase of Brans--Dicke theory, is a more
practical question that certainly deserves a direct answer. It is
indeed true that scalar-tensor theories and, more precisely, 
Brans--Dicke
theory are well-studied theories which have been extensively used in
many applications, including cosmology. However, the specific 
choices
$\omega_0=0,-3/2$ for the Brans--Dicke parameter are quite 
exceptional, as already mentioned in the previous 
section.   It is also
worthwhile pointing out the following: a)  As far as the 
$\omega_0=0$
case is concerned, one can probably speculate that it is the 
apparent
absence of the kinetic term for the scalar in the action which did not
seem appealing and prevented the study of this theory. b) The  
$\omega_0=-3/2$ case leads to a conformally invariant theory in 
the absence of the potential [see \cite{Sotiriou:2006hs} and 
references therein], which constituted the initial form of 
Brans--Dicke theory, and hence it was considered 
non-viable (a 
coupling with non-conformally invariant matter is not feasible). 
However, in the presence of a potential,    the theory no longer 
has this feature. Additionally, most calculations which are done 
for a
general value of $\omega_0$ in the literature actually exclude
$\omega_0=-3/2$, mainly because, merely for simplicity purposes, they are done in such a way that the
combination $2\omega_0+3$ appears in a denominator (see also 
Sec.~\ref{sec:weakfield}). 

In any case,
the conclusion is that the theories in the Brans--Dicke class that
correspond to metric and Palatini $f(R)$ gravity had not yet been
explored before the recent re-introduction of $f(R)$ gravity and, as
will also become clear later, several of their special
characteristics  when compared with more standard Brans--Dicke 
theories
were revealed through studies of $f(R)$ gravity.

\section{Cosmological evolution and constraints}
\label{sec:chapIV}

We now turn our attention to cosmology, which motivated the 
recent surge of interest in $f(R)$ gravity in order to explain 
the current cosmic acceleration without the need for dark 
energy. Before reviewing how $f(R)$ gravity might provide a 
solution to the more recent cosmological riddles, let us stress 
that the following criteria must be satisfied in order for an 
$f(R)$ model to be theoretically consistent and compatible 
with cosmological observations and experiments. The model 
must:

\begin{itemize}

\item have the correct cosmological dynamics;

\item exhibit the correct  behaviour of gravitational 
perturbations;

\item generate cosmological perturbations compatible with the 
cosmological constraints from the cosmic microwave background, 
large scale structure, Big Bang Nucleosynthesis, and 
gravity 
waves.

\end{itemize}

These are independent requirements to be studied
separately, and  they must all be satisfied.


\subsection{Background evolution}

In cosmology, the identification of our universe with a 
Friedmann--Lemaitre--Robertson--Walker (FLRW) spacetime  
is largely based on the high degree of isotropy measured in the 
cosmic microwave background; this identification relies on a 
formal result known as the Ehlers--Geren--Sachs (EGS) theorem 
\cite{Ehlers:1966ad} which is a kinematical 
characterization of FLRW spaces stating that, if a congruence 
of timelike freely falling observers see an isotropic radiation 
field, then (assuming that isotropy holds about every spatial 
point)  the spacetime is spatially homogeneous and isotropic 
and, 
therefore, a FLRW one. This applies to a universe filled with 
any perfect fluid that is geodesic and barotropic 
\cite{Ellis:1983a, Ellis:1983b,Clarkson:1999yj}.
Moreover, an ``almost-EGS theorem'' holds: spacetimes that are 
close to satisfying the EGS conditions are close to FLRW 
universes in an appropriate sense \cite{Stoeger:1994qs}. One 
would expect that the EGS theorem be extended to $f(R)$ gravity; 
indeed, its validity 
for the (metric) theory 
\be
S=\frac{1}{2\kappa} \int d^4x \, \sqrt{-g} 
\left[ R+\alpha R^2+\beta R_{\mu\nu}R^{\mu\nu}\right] +S_M 
\ee
was proved in \cite{Maartens:1994pb, Taylor:1995dy} and 
the generalization to arbitrary metric $f(R)$ gravity was given 
by \cite{Rippl:1995bg}. The validity of the EGS theorem can also 
be seen through the equivalence between $f(R)$ and Brans--Dicke 
theory: the theorem was extended to scalar-tensor theories  in 
\cite{Clarkson:2001qc, Clarkson:2003ts}. Since metric 
and Palatini $f(R)$ gravities are equivalent to $\omega=0$ and 
 $\omega_0=-3/2$ Brans--Dicke theories respectively, it seems that 
 the results of \cite{Clarkson:2001qc, Clarkson:2003ts} can be 
considered as straightforward generalizations of the EGS theorem 
 in both versions of $f(R)$ gravity as well. However, in the 
case of Palatini $f(R)$ gravity there is still some doubt regarding this issue due to complications in averaging \cite{Flanagan:2003rb}.

\subsubsection{Metric $f(R)$ gravity}

Considering the discussion above, it is valid to use the FLRW 
line element 
\be\label{metric}
ds^2=-dt^2+a^2(t)\left[\frac{dr^2}{1-k r^2}+ r^2\left(  
d\theta^2+  \sin^2 \theta d\phi^2 \right)\right]
\ee
as a local
description of spacetime at cosmological scales, where 
$\left( t, r, \theta, \phi \right)$ are comoving coordinates.
We remind the reader that $k=-1,0,1$ according to whether the
universe is hyperspherical, spatially flat, or hyperbolic
 and that $a(t)$ is called the scale factor. Part of the
standard approach, which we follow here  as well, is to use a 
perfect
fluid description for  matter with stress-energy tensor
\be
\label{pf}
T^{\mu\nu}=\left( \rho+P \right) u^\mu u^\nu+ P\, g^{\mu\nu},
\ee
 where $u^\mu$ denotes the four-velocity of an observer comoving with
the fluid and $\rho$ and $P$ are the energy density and  
pressure of the fluid, respectively.

Note that the value of $k$ is an external parameter. As in many other
works in the literature, in what follows we  choose $k=0$, 
{\em
i.e.,~}we focus on a spatially flat universe. This choice in 
made in
order to simplify the equations and should be viewed sceptically. It
is sometimes  claimed in the literature that such a choice is 
favoured
by the data. However, this is not entirely correct. Even though the
data  [{\em e.g.~}\cite{Spergel:2006hy}] indicate that the 
current value of
$\Omega_k$ is very close to zero, it should be stressed that
this does not really reveal the value of $k$ itself. Since
 \be
\Omega_k=-\frac{k}{a^2 H^2},
\ee
 the current value of $\Omega_k$ is sensitive to the current 
value of
$a(t)$, {\em i.e.~}to the amount of expansion the universe has 
undergone  after the Big Bang. A significant amount of expansion 
can easily drive
$\Omega_k$  very close to zero. The success of the inflationary
paradigm is  exactly that it explains the flatness problem --- 
how did the universe become so flat 
--- in a dynamical way, allowing us to avoid  fine-tuning the 
parameter $k$ (the value $k=0$ is statistically  
exceptional).

The above having been said, choosing $k=0$ for simplicity is 
not a dramatic departure from generality when it comes to late 
time cosmology. If it is viewed as an approximation and not as a 
choice of
an initial condition, then one can say that, since $\Omega_k$ as
inferred from observations is very close to zero at current times, the
terms related to $k$ will be subdominant in the Friedmann or
generalised Friedmann equations and, therefore, one could choose 
to
discard them by setting $k=0$, without great loss of accuracy. In any
case, results derived under the assumption that $k=0$ should be
considered preliminary until the influence of  the spatial 
curvature is
precisely determined, since there are indications that even a very
small value of $\Omega_k$ may have an effect on them [see, for
instance \cite{Clarkson:2007bc}].

Returning to our discussion, inserting the flat FLRW metric in 
the field 
equations~(\ref{metf}) and assuming that the stress-energy tensor is that of eq.~(\ref{pf}) yields
\begin{eqnarray}
\label{H_squa}
H^2&=&\frac{\kappa}{3f'}\left[ 
\rho+ \frac{Rf'-f}{2}-3H\dot{R}f''\right] , 
\\
2\dot{H}+3H^2&=&-\frac{\kappa}{f'}\Big[ P + 
(\dot{R})^2 f'''  
+2H\dot{R}f''  \nonumber \\
&& +\ddot{R}f''+\frac{1}{2}(f-Rf')\Big] . 
\label{H_dot} 
\end{eqnarray}
With some hindsight, we assume that $f'>0$ in order to have a 
positive  effective gravitational coupling and $f''>0$ to avoid 
the  Dolgov-Kawasaki instability 
\cite{Dolgov:2003px, Faraoni:2006sy} discussed 
in Sec.~\ref{sec:stabilitysub}.

A significant part of the motivation for $f(R)$ gravity is that 
it can lead to accelerated expansion without the need for 
dark energy (or an inflaton field). An easy way to see this is 
to define an 
effective energy density and
pressure of the geometry as
\bea
\label{density} 
\rho_{eff}\!\!&=& \!\!\frac{Rf'-f}{2f'}-\frac{3H\dot{R}f'' }{f'} ,\\
\label{pressure}
P_{eff}\!\!&=& \!\! \frac{\dot{R}^2 f'''+  
2H\dot{R}f''+\ddot{R}f''+\frac{1}{2} \left( f-Rf' \right)}{f'} ,
\eea
where  $\rho_{eff}$ has to be non-negative 
in a spatially flat FLRW spacetime, as 
follows from the inspection of 
 eq.~(\ref{H_squa}) in the limit $\rho\rightarrow 0$. Then, 
in vacuo
 eqs.~(\ref{H_squa}) and (\ref{H_dot}) can take the form of the 
standard Friedmann equation 
\begin{eqnarray}
H^2&=&\frac{\kappa}{3}\rho_{eff}, 
\\
 \frac{\ddot{a}}{a}& 
=&-\frac{\kappa}{6}\left[\rho_{eff}+3P_{eff}\right] . 
\end{eqnarray}
Hence, in vacuo  the curvature correction can be viewed as an 
effective 
fluid.\footnote{Note the following subtlety though: should  
we have included matter it would enter the Friedmann equations  
with a modified coupling $\kappa/f'$. In general  this effective 
fluid representation is used only for demonstrative  purposes and 
should not be overestimated or misinterpreted.} 

The effective equation of state parameter 
$w_{eff}$ of modified gravity can be expressed as
\begin{equation}\label{eq_state}
w_{eff}  \equiv  \frac{P_{eff}}{\rho_{eff}}= 
\frac{  \dot{R}^2f'''+2H\dot{R}f''+ 
\ddot{R}f''+\frac{1}{2}(f-Rf') }{ \frac{Rf'-f}{2} 
-3H\dot{R}f'' } .
\end{equation}
Since the denominator on the 
right hand side of eq.~(\ref{eq_state}) is strictly positive, 
the sign of $w_{eff}$  is determined by 
its numerator.  In general, for a metric $f(R)$  model to mimic  the de 
Sitter equation of state $w_{eff}=-1$, it must be
\begin{equation}\label{ratio}
\frac{f'''}{f''}=
\frac{ \dot{R}H -\ddot{R}}{( \dot{R})^2 } .
\end{equation}

Let us also give two simple examples that can be found in the 
literature  for demonstrative purposes and without considering 
their viability:
First, one can consider the function $f$ to be of the form
$f(R)\propto R^n$. It is quite straightforward to calculate $w_{eff}$
as a function of $n$ if the scale factor is assumed to be a generic
power law $a(t)=a_0(t/t_0)^\alpha$ (a  general $a(t)$ would lead 
to a time dependent $w_{eff}$) \cite{Capozziello:2003tk}. The 
result is
 \be
w_{eff}=-\frac{6n^2-7n-1}{6n^2-9n+3} 
\ee
for $n\neq 1$, and $\alpha$ is given is terms of $n$ as
\be
\alpha=\frac{-2n^2+3n-1}{n-2}.
\ee
A suitable choice of $n$ can lead to a desired value for 
$w_{eff}$.  For instance, $n=2$  yields $w_{eff}=-1$ and 
$\alpha=\infty$,  as expected, considering that quadratic corrections to the 
Einstein-Hilbert Lagrangian were used in  the 
well known Starobinsky inflation  \cite{Starobinsky:1980te}.

The second example which we will refer to is a model of the form
$f(R)=R-\mu^{2(n+1)}/R^n$, where $\mu$ is a suitably chosen parameter
\cite{Carroll:2003wy}. In this case, and once again if the 
scale factor is assumed to be a generic power law, $w_{eff}$ can again be written as a function of $n$ \cite{Carroll:2003wy}:
 \be
w_{eff}=-1+\frac{2(n+2)}{3(2n+1)(n+1)}.
\ee
 The most typical model within this class is that with $n=1$
\cite{Carroll:2003wy}, in which case $w_{eff}=-2/3$. Note that in this class of
models, a positive $n$ implies the presence of a term inversely
proportional to $R$ in the action, contrary to the situation for the
$R^n$ models.

In terms of the quantity  $\phi (R) \equiv 
f'(R)$ one can  rewrite eq.~(\ref{eq_state}) as
\begin{equation}
w_{eff}= -1+2\, \frac{\left( \ddot{\phi}-H\dot{\phi} 
\right)}{R\phi-f-6H \dot{\phi} }= -1+\frac{\kappa \left( 
\ddot{\phi}-H\dot{\phi}\right) }{3\phi H^2} 
\end{equation}
and
\begin{equation}  \label{questa}
\rho_{eff}+P_{eff}=  \frac{\ddot{\phi}-H\dot{\phi}}{\phi} 
=\frac{\dot{\phi}}{\phi} 
\, \frac{d}{dt} \left[ \ln \left( \frac{\dot{\phi}}{a} \right) 
\right] .
\end{equation}
An exact de  Sitter solution 
corresponds to $\dot{\phi}=f''(R) \dot{R}=0$, or to $\dot{\phi} 
=C a(t)=C a_0 \, \mbox{e}^{H_0 t}$, where $C\neq 0 $ is an 
integration constant. However, the second 
solution for $\phi(t)$ is not  acceptable because it leads to 
 $f''(R)  \dot{R}=C a_0\mbox{e}^{H_0 t}$, which is absurd 
because  the left hand side is  time-independent (for a de 
Sitter solution), while the right 
hand side depends on time.

One could impose energy conditions for the effective  
stress-energy tensor~(\ref{effectivetab}) of $f(R)$ gravity. 
However,  this is not very meaningful from the 
physical  point of view since it is well known that effective 
stress-energy terms originating from the geometry by rewriting 
the field  equations of alternative gravities as effective 
Einstein  equations  do, in general, violate all the energy 
conditions  [{\em e.g.}, \cite{valeriobook}].\footnote{In 
\cite{Santos:2007bs},  the Null Energy Condition and the Strong 
Energy Condition for  metric $f(R)$ gravity have been derived 
by using the Raychaudhuri  equation and imposing that gravity 
be attractive, whereas for  the Weak Energy Condition and the 
Dominant Energy Condition  an effective stress-energy tensor which includes the matter
was used. In  \cite{PerezBergliaffa:2006ni}, a different 
approach was  followed, in which the standard energy 
conditions on matter where used in an attempt to constrain 
$f(R)$ gravity.} Also, the concept of 
gravitational  energy density is, anyway,  ill-defined in GR 
and in all metric 
theories of gravity as a 
consequence of the Equivalence Principle. Moreover, the 
violation of the energy conditions makes it possible to have 
$\dot{H}>0$ and bouncing universes \cite{Carloni:2005ii, 
Novello:2008ra}.

The field equations are clearly of fourth order in $a(t)$. 
When matter is absent 
(a situation of interest in early time inflation or in a very 
late 
universe completely dominated by $f(R)$ corrections), $a(t)$ 
only 
appears in the combination $H\equiv \dot{a}/a$.  Since the 
Hubble parameter 
$H$ is a  cosmological observable, it is convenient to adopt it 
as the (only) dynamical variable; then the field  equations 
(\ref{H_squa}) and (\ref{H_dot})  are of third order in $H$.  
This elimination  of $a$ is not possible when $k\neq  0$, or  
when a fluid with density  $\rho=\rho(a) $  is included in the 
picture.

Regarding the dynamical field content of the theory, the fact 
that quadratic corrections to the Einstein--Hilbert 
action introduce a massive scalar field was noted in 
\cite{Utiyama:1962sn, Stelle:1976gc, Stelle:1978ww, 
Strominger:1984dn, quant1, quant3}; this 
applies to any $f(R)$ gravity 
theory in the metric formalism [see, {\em e.g.},  
\cite{Ferraris:1988zz, Olmo:2006eh, Hindawi:1995cu}].  
The metric tensor contains, in principle, various degrees of 
freedom: spin~2 modes, and vector and scalar modes, 
which can all be massless or massive. In GR we 
find only the massless graviton but, when the action is allowed 
to  include terms that depend on $R$, $R_{\mu\nu}R^{\mu\nu}$, 
$R_{\mu\nu\rho\sigma}R^{\mu\nu\rho\sigma}$, other modes show up. 
In $f(R)$ gravity, a 
massive scalar mode appears, which is evident in the equivalence 
with scalar-tensor theory (see Sec.~\ref{sec:equiv}). As discussed 
in Sec.~\ref{sec:class}, the scalar 
field $\phi=R$ is dynamical in the metric formalism and 
non-dynamical in the Palatini formalism.

\subsubsection{Palatini $f(R)$ gravity}

As already mentioned, some concerns have been expressed on 
whether the homogeneity approximation can justify the use of the 
FLRW metric as a cosmological solution in Palatini $f(R)$ 
gravity \cite{Flanagan:2003rb} [see also \cite{Li:2008bma}]. 
Therefore, even though it is standard practice in the literature 
to assume a FLRW background and a        perfect fluid 
description for matter when studying  cosmology in Palatini 
$f(R)$ gravity  [{\em 
e.g.}~\cite{Vollick:2003aw, 
Meng:2003bk,Meng:2004yf,Meng:2004wg, 
Allemandi:2004ca,Allemandi:2005qs, 
Sotiriou:2005hu, Sotiriou:2005cd,Amarzguioui:2005zq}], and we 
are going to review  this approach here, the reader should 
approach it with some  reasonable skepticism until this issue is 
clarified further.

Under the assumptions that the spacetime is indeed described at 
cosmological scales by the FLRW metric, eq.~(\ref{metric}),  
that the stress-energy tensor of matter is that of 
eq.~(\ref{pf}),  and that $k=0$,  easy manipulations reveal 
that the field eqs.~(\ref{palf12}) and~(\ref{palf22})  yield the 
following modified Friedmann equation [see for instance \cite{Meng:2003bk,Sotiriou:2005hu}]:
\be
\label{pfriedmann}
\left(H+\frac{1}{2}  \frac{\dot{f'}}{f'}\right)^2=\frac {1}{6} 
\frac{\kappa \left( \rho+3 P \right)}{f'}+\frac {1}{6}
\frac{f}{f'},
\ee
 where the overdot denotes differentiation with respect to 
coordinate
time. Note that when $f$ is linear, $f'=1$ and, therefore,
$\dot{f'}=0$. Taking into account eq.~(\ref{paltrace}), one can 
easily
show that in this case eq.~(\ref{pfriedmann}) reduces to the 
standard Friedmann equation.

We will avoid representing the extra terms in 
eq.~(\ref{pfriedmann}) with respect to the standard Friedmann 
equation as a an effective stress energy density and pressure 
since, as it is not that difficult to  see, the former equation 
does  not really carry more dynamics than the latter. Indeed, 
assume 
as usual that the cosmological fluid is composed by pressureless      
dust ($ P_m=0$) and
radiation ($ P_r=\rho_r/3$) and $\rho=\rho_m+\rho_p$ and 
$ P= P_m+ P_r$ where $\rho_m$, $\rho_p$ and $P_m$, $ P_r$ denote 
the energy density and the pressure of dust and radiation, 
respectively. Due to eq.~(\ref{paltrace})  and the fact that for 
radiation $T=0$, it is quite  straightforward to derive an 
algebraic relation between ${\cal R}$ and the energy density of 
the dust. 
Combining this with energy conservation, one obtains 
\cite{Sotiriou:2005hu}
\be
\label{rdot}
\dot{{\cal R}}=-\frac{3H ({\cal R} f'-2f)}{{\cal R} f''-f'}.
\ee
This equation can be used to replace $\dot{{\cal R}}$ in 
eq.~(\ref{pfriedmann}),  yielding
\be
\label{HRform}
H^2=
\frac{1}{6 f'}
\frac{2 \kappa \rho+{\cal R} f'-f}
{\left(1-\frac{3}{2}\frac{f''({\cal R}f'-2f)}{f'({\cal R}f''-f')}\right)^2}
\ee
Considering now that, due to eq.~(\ref{paltrace}), ${\cal R}$ is 
just an algebraic function of $\rho_m$, it is easy to realize 
that  eq.~(\ref{HRform}) is actually just the  usual Friedmann 
equation with a modified source. The functional form of $f$ will determine how the dynamics will be affected by this modification.

It seems, therefore, quite intuitive that by tampering with the 
function $f$ one can affect the cosmological dynamics in a 
prescribed way. Indeed it has been shown that for $f({\cal 
R})={\cal R}-\alpha^2/(3{\cal R}$) one approaches a de Sitter 
expansion  as the density goes to zero \cite{Vollick:2003aw}. In 
order to  match  observations of the expansion history, one 
needs to  choose $\alpha\sim 10^{-67}\,({\rm eV})^2\sim 
10^{-53}\, {\rm m}^{-2}$. Additionally, in regimes for which 
$\kappa \rho\gg\alpha$, eq.~(\ref{HRform}) reduces to high 
precision to the standard Friedmann equation.  The above can 
very easily be verified by replacing this  particular 
choice of $f$ in eq.~(\ref{HRform}). We refer  the reader to 
the literature for more details.

One could, of course,  consider more general functions of ${\cal 
R}$. Of particular  interest would be having positive powers of 
${\cal R}$ higher  than the first power added in the action 
(since one could  think of the Lagrangian as a series 
expansion).  Indeed  this has been considered 
\cite{Meng:2003bk, 
Meng:2004yf,Meng:2004wg,Sotiriou:2005hu,Sotiriou:2005cd}.   
However, it can be shown that such terms do not really lead to 
interesting phenomenology as in metric $f(R)$ gravity: for  
instance they cannot drive inflation as, unlike in the scenario 
proposed by   \cite{Starobinsky:1980te} in the metric formalism, 
here there are  no extra dynamics and inflation cannot end 
gracefully \cite{Meng:2004yf,Sotiriou:2005cd}.  As a matter of 
fact, it is more likely that positive powers  of ${\cal R}$ will 
lead to no interesting cosmological phenomenology unless their 
coefficients are large enough to make the models non-viable 
\cite{Sotiriou:2005cd}.

\subsection{Cosmological eras} 
\label{sec:correctdynamics}

As stated in the Introduction, the recent flurry of 
theoretical activity on $f(R)$ models  derives from the need to 
explain  the present acceleration of  the universe discovered 
with supernovae of type Ia  \cite{Riess:1998cb, Riess:1999th, 
Schmidt:1998ys, Filippenko:1998tv,  Perlmutter:1997zf, 
Tonry:2003zg, Knop:2003iy, Barris:2003dq,   
Riess:2004nr, Astier:2005qq}. We have seen in the previous 
section how $f(R)$ gravity can achieve cosmic acceleration and 
an effective equation of state parameter $w_{eff}\sim -1$; on 
the other hand, it was already known from $R^2$-inflationary 
scenarios of the early universe that this is possible, so we are 
actually witnessing a resurrection of this  theoretical 
possibility in models of the late universe --- this 
parallels the use of scalar fields to drive early 
inflation or late-time acceleration in quintessence models. 
There are also attempts to unify early inflation and late time 
acceleration in modified gravity 
\cite{Nojiri:2008fk,Nojiri:2007cq, Nojiri:2007uq, 
Nojiri:2007as, Nojiri:2008nk, Bamba:2008ja}. However, 
any model attempting to explain the cosmic speed-up at late 
times should not spoil the successes of 
the  standard cosmological model which requires a definite 
sequence of  eras to follow each other, including:

\begin{enumerate}

\item early inflation

\item  a radiation era during which  Big Bang Nucleosynthesis 
occurs;

\item  a matter era;

\item the present accelerated epoch, and 

\item a future era. 

\end{enumerate}

Big Bang Nucleosynthesis is well constrained --- see 
\cite{Lambiase:2006dq, Nakamura:2005qm, Kneller:2004jz, 
Clifton:2005aj, Evans:2007ch, Brookfield:2006mq} for such 
constraints on $f(R)$ models. The matter era must last long 
enough to allow the primordial density perturbations generated 
during inflation to grow and become the structures observed in 
thd universe today. The future era is usually found to be a de 
Sitter attractor solution, or to be truncated at a finite time 
by a Big Rip singularity.

Furthermore, there must be smooth transitions 
between consecutive eras, which may not happen in all $f(R) 
$ models. In particular, the exit from the radiation era has 
been studied and claimed to originate problems for many forms of 
$f(R)$ in the metric formalism, including 
$f=R-\mu^{2(n+1)}/R^n$, $n>0$ \cite{Amendola:2006kh, 
Capozziello:2006dj, Amendola:2006we, 
Nojiri:2006gh, Brookfield:2006mq} [but not in the Palatini 
formalism \cite{Fay:2007gg, Carvalho:2008am}]. However, the 
usual model 
$f(R)=R-\mu^4/R$ with ``bad'' behaviour was studied using 
singular perturbation methods \cite{Evans:2007ch}, 
definitely finding a 
matter era which is also sufficiently long.

Moreover, one 
can always find choices of 
the function $f(R)$  with the  correct cosmological dynamics in 
the following way: one can prescribe the desired form of the 
scale factor $ a(t)$ and 
integrate a differential equation for $f(R)$ that produces the 
desired scale factor \cite{Song:2006ej, 
Nojiri:2006be,  
Fay:2007uy,  Hu:2007pj, Hu:2007nk, 
delaCruzDombriz:2006fj, Capozziello:2005ku, Capozziello:2006dj, 
Nojiri:2006gh, 
Nojiri:2006su,  
Fay:2007gg, Multamaki:2005zs, Faulkner:2006ub}. In general, 
this  
``designer $f(R)$ gravity'' 
produces forms of the  function  $f(R)$ that are rather 
contrived. Moreover, the prescribed evolution of 
the scale factor $a(t)$ does not determine uniquely the form of 
$f(R)$ but, at best, only a class of $f(R)$ models 
\cite{Multamaki:2005zs, Starobinsky:2007hu, 
Sokolowski:2007pk, Sokolowski:2007rd}. 
Therefore, the observational data 
providing information on the history of $a(t)$ are not 
sufficient to reconstruct $f(R)$: one needs additional 
information, which may come from cosmological 
density perturbations. There remains a caveat on being careful 
to terminate the radiation era and allowing a matter era that is 
sufficiently long  for scalar perturbations to grow.

While sometimes it is possible to find exact solutions to the 
cosmological equations, the general behaviour of the solutions 
can only be assessed with a phase space analysis, which 
constitutes a powerful tool in cosmology 
\cite{Wainwright:1997aa, Coley:2003mj}. In a spatially flat 
FLRW universe the dynamical variable is the 
Hubble parameter $H$, and a convenient choice of phase space 
variables in this case is $ \left( H, R \right)$. Then, for any 
form of the function $f(R)$,  the 
phase space is a  two-dimensional curved manifold 
embedded in the three-dimensional  space $\left( H, R, \dot{R} 
\right) $   with de Sitter spaces as fixed points 
\cite{deSouza:2007fq}; the structure of the phase space is 
simplified with respect to that of general scalar-tensor 
cosmology \cite{Faraoni:2005vc}.

Studies of the phase space of $f(R) $ cosmology (not 
limited to 
the spatially flat FLRW case) were common in the pre-1998 
literature on $R^2$-inflation \cite{Starobinsky:1980te, 
Muller:1989rp, Amendola:1992um, Capozziello:1993xn}.  The 
presence or absence of chaos in metric $f(R)$ gravity  was 
studied in \cite{Barrow:1989nb,Barrow:1991hg}. Such studies 
with dynamical system methods have  become widespread with the 
recent surge of interest in $f(R)$ 
gravity to explain the present cosmic acceleration. Of 
course, detailed phase space analyses are only possible for 
specific choices of the function $f(R)$ \cite{Goheer:2007wx, 
Goheer:2007wu, Abdelwahab:2007jp, Carloni:2007br, 
Carloni:2006mr, Carloni:2004kp, Carloni:2007eu, Leach:2007ss, 
Leach:2006br, 
Fay:2007gg, Fay:2007uy, Clifton:2007ih, Clifton:2006jh, 
Nojiri:2003ni, Amendola:2006we, Amendola:2006kh, 
Amendola:2006eh, Amendola:2007nt,  
Li:2007xn, Easson:2004fq, Carroll:2004de, Clifton:2005at, 
Sami:2005zc}.

\subsection{Dynamics of cosmological perturbations} 
\label{sec:correctperts}

Obtaining the correct dynamics of the background cosmological 
model is not sufficient for the theory to be viable: in fact, 
the FLRW metric can be obtained as  a solution of 
the field equations of most gravitation theories, and it is 
practically impossible to discriminate between $f(R)$ gravity 
and dark energy theories (or between different $f(R)$ models) by 
using only the unperturbed FLRW cosmological model, {\em i.e.}, 
by using only probes that are sensitive to the expansion history 
of the universe. By contrast, the growth of cosmological 
perturbations is sensitive to the theory of gravity adopted and 
constitutes a possible avenue to discriminate between dark 
energy and modified gravity. Changing the theory of gravity 
affects the dynamics of cosmological 
perturbations and, among other things,  the imprints that these 
leave in the cosmic microwave background (which currently 
provide the most sensitive cosmological probe) and in galaxy 
surveys \cite{White:2001kt, Shirata:2007qk, Shirata:2005yr, 
Sealfon:2004gz, Stabenau:2006td, Skordis:2005xk, Knox:2006fh, 
Koyama:2005kd, Koivisto:2006ie, Li:2006vi, Song:2006ej, 
Li:2007xn, Zhang:2007nk, Tsujikawa:2007gd}. 
This originated various efforts to constrain 
$f(R)$ gravity with  cosmic microwave background data 
\cite{Starobinsky:2007hu, Li:2007xw, Amendola:2007nt, 
Hu:2007nk, Appleby:2007vb, Tsujikawa:2007xu, Tsujikawa:2007tg, 
Carloni:2007yv,Li:2007xn, Pogosian:2007sw, Wei:2008vw}. 

Most of 
these works are restricted to specific choices of the function 
$f(R)$, but a few general results have also been obtained. The 
growth and evolution of local scalar perturbations, 
which depends on the theory of gravity employed, was studied  
in metric  
$f(R)$ gravity theories which reproduce GR at 
high curvatures  in various papers \cite{Song:2006ej, 
Carroll:2006jn, delaCruzDombriz:2008cp} 
by assuming a scale factor evolution  typical of a $\Lambda$CDM 
model. Vector and tensor modes are  unaffected by $f(R) $ 
corrections. It is found  that 
$f''(R)>0$ is  required for the stability of scalar 
perturbations 
\cite{Song:2006ej}, which matches 
the analysis of Sec.~(\ref{sec:dSstability}) in a locally de Sitter 
background. The corrections to the 
Einstein--Hilbert action produce qualitative differences with 
respect to Einstein gravity: they lower the large angle 
anisotropy of the cosmic microwave background and may help 
explaining the observed low quadrupole; and they produce 
different correlations between the cosmic microwave background 
and galaxy surveys \cite{Song:2006ej}. Further 
studies challenge the viability of $f(R)$ gravity in comparison 
with the $\Lambda$CDM model: in 
\cite{Bean:2006up} it is found that large 
scale density fluctuations are suppressed in comparison to small 
scales by an amount incompatible with the observational data. 
This makes it impossible to fit simultaneously large 
scale data from the cosmic microwave background and small scale 
data from galaxy surveys. Also, a quasi-static approximation 
used in a  previous analysis \cite{Zhang:2007ne} is found to be 
invalid.

In  \cite{delaCruzDombriz:2008cp}, the growth of matter density 
perturbations is studied in the longitudinal gauge using a 
fourth order equation for the density contrast $\delta 
\rho/\rho$, which reduces to a second order one for sub-horizon 
modes. The quasi-static approximation, which does not hold for 
general forms of the function $f(R)$, is however found to  be 
valid for those forms of this function that describe 
successfully the present cosmic acceleration and pass the Solar 
System tests in the weak-field limit.  It is interesting that 
the  relation between the gravitational potentials in the metric 
which are responsible for gravitational lensing, and the matter 
overdensities depends on the theory of gravity; a study of this 
relation in $f(R)$ gravity (as well as in other gravitational 
theories) is contained in \cite{Zhang:2007nk}.

Cosmological density perturbations in the Palatini formalism 
have been studied in \cite{Amarzguioui:2005zq, Lee:2008ek, 
Lee:2007nh, Koivisto:2007sq, Li:2006ag, Koivisto:2006ie, 
Koivisto:2005yc, Uddin:2007gj,  Carroll:2006jn}. Two different  
formalisms developed in \cite{Koivisto:2005yc, Hwang:2001qk} 
and  \cite{Lue:2003ky} were compared for the model 
$f(R)=R-\mu^{2(n+1)}/R^n $ and it was found that the two 
models agree for scenarios that are ``close'' (in parameter 
space) to the standard 
concordance model, but give different results for models that 
differ significantly from the  $\Lambda$CDM model. Although 
this is not something to worry about in practice (all models 
aiming at explaining the observational data are ``close'' to the 
standard concordance model), it signals the need to test the 
validity of perturbation analyses for theories that do differ 
significantly from GR in some aspects.


\section{Other standard viability criteria}
\label{sec:chapV}

In addition to having the correct cosmological dynamics and 
the correct evolution of cosmological perturbations, the 
following criteria must be satisfied in order for 
an $f(R)$ model to be theoretically  consistent and compatible 
with experiment. The model must:

\begin{itemize}

\item  have the correct weak-field limit at both the Newtonian 
and post-Newtonian levels, {\em i.e.}, one that is compatible 
with the available Solar System experiments; 

\item be stable at the classical and semiclassical level 
(the checks performed include the study of a matter 
instability, of gravitational  instabilities for de Sitter 
space,  and of a semiclassical instability with respect to 
black hole nucleation);

\item  not contain ghost fields;

\item admit a well-posed Cauchy problem;

\end{itemize}

These independent requirements are discussed separately in the 
following.

\subsection{Weak-field limit}
\label{sec:weakfield}

It is obvious that a viable theory of gravity must have the 
correct Newtonian and post-Newtonian limits. Indeed, since 
the modified gravity theories of current interest are 
explicitly designed 
to fit the 
cosmological observations, Solar System tests are more stringent 
than the cosmological ones and constitute a real testbed for 
these theories.

\subsubsection{The scalar degree of freedom}

It is clear from the equivalence between $f(R)$ and 
Brans--Dicke gravities  discussed in Sec.~\ref{sec:equiv} that 
the former contains a massive scalar field $\phi$ [see 
eqs.~(\ref{metactionH2}) and (\ref{palactionH2d0})]. While 
in the metric formalism this scalar 
is dynamical and represents a genuine degree of freedom, it is 
non-dynamical in the Palatini case.  Let us, therefore, consider 
the role of the scalar field in the metric forma\-lism as it will 
turn out to be crucial for the weak-field limit.   Using the 
notations of  Sec.~\ref{sec:equivalencemetric}, the action 
is given by eq.~(\ref{metactionH2}) and the corresponding field 
equations by eq.~(\ref{bdmetf1}).

Equation~(\ref{1600}) for $\chi $ has no 
dynamical  content because it only enforces the equality 
$\chi=R$. However, $\chi=R$ is indeed a dynamical 
field that  satisfies the wave equation,
\bea \label{metrictracephi}
3f''(\chi) \Box \chi &+& 3 f'''(\chi) \nabla_{\alpha} \chi \, 
\nabla^{\alpha} \chi  \nn\\
&+&\chi f'(\chi) -2f(\chi)=\kappa \, T .
\eea 
When $f''\neq 0$ a new effective potential $W(\chi)\neq V(\chi)$ 
can be introduced, such that
\be
\frac{dW}{d\chi}=\frac{ \kappa \, T -\chi f'(\chi)+ 
2f(\chi)}{3f''(\chi)} .
\ee
The action can be seen as a Brans--Dicke action with $\omega_0=0$ 
if the field 
$\phi \equiv f'(\chi)=f'(R)$ is used instead of $\chi$ as the 
independent Brans--Dicke field:
\be \label{BD}
S=\frac{1}{2\kappa} \int d^4x \sqrt{-g} \left[ \phi R -V( 
\phi) \right] +S^{(m)} ,
\ee
where $ V\left( \phi \right) $ is given by eq.~(\ref{defV}).

Now one may think  of studying the dynamics and stability  of 
the model by 
looking at the shape and extrema of the effective potential 
$V(\chi)$ but this would be misleading because  the dynamics 
of $\chi$ are  not regulated by $V(\chi) $ (indeed, the wave 
equation~(\ref{metrictracephi}) does not contain $V$), 
but are subject the strong constraint $\chi=R$, and $R$ (or 
$f'(R)$) is ruled by the trace equation~(\ref{metftrace}).

The following  example shows how the use of the potential 
$V(\chi)$ 
can be misleading. As is well known, the effective mass of a 
scalar field (corresponding to the second derivative of   the 
potential evaluated at the minimum) controls the range of  the 
force mediated by this field. Thus,  when studying the  
weak-field limit of the 
theory it is important to know the range of the dynamical scalar 
field $\chi=R$ present in the metric formalism in addition to 
the metric field $g_{\mu\nu}$, as this field can potentially 
violate 
the post-Newtonian constraints obtained from Solar System 
experiments if the scalar field gives observable effects at the relevant scales. One way to avoid Solar System constraints, however, is to have $\chi$ have a 
sufficiently short range (see Sec.~\ref{sec:weakmetricsubsub}  
for more details).  
Consider the example $f(R)=R+aR^2$, 
with $a$ a positive 
constant.  By naively taking the potential, one 
obtains
\be
V(\chi)=a \, \chi^2 \equiv \frac{m_1^2}{2} \, \chi^2 
\ee
with effective mass squared $m_1^2=2a$. Then, the small values 
of $a$ generated by quantum corrections to GR 
imply a small mass $m_1$ and a long range field $\chi$ might be 
detectable at Solar System scales \cite{Chiba:2006jp, 
Olmo:2006eh, Jin:2006if}. However, this 
conclusion is 
incorrect because $m_1$ is not the physical mass of $\chi$. The true 
effective mass is obtained from the trace 
equation~(\ref{metftrace}) 
ruling the evolution of $R$ which, for $f(R)=R+aR^2$, reduces 
to 
\be
\Box R -\frac{R}{6a}=\frac{\kappa \, T}{6a} ,
\ee
and the identification of the mass squared of $\chi=R$ as 
\be
m^2=\frac{1}{6a} 
\ee
is straightforward.\footnote{It was already noted by 
 \cite{Stelle:1978ww} that an $R^2$ correction to the 
Einstein--Hilbert 
Lagrangian generates a Yukawa correction to the Newtonian 
potential --- this has to be kept small at macroscopic scales by 
giving it a short 
range.} A small enough value of $a$ now leads 
to a large value of $m$ and a short range for 
\footnote{The deflection of 
light by the Sun in GR plus  quadratic
corrections was studied by calculating the Feynman amplitudes 
for photon scattering, and it was found that, to 
linearized 
order, this deflection is the same as in GR
\cite{Accioly:1999a}.} $\chi$. The situation is, however, more 
complicated; the chameleon effect due to the dependence of the 
effective mass on the curvature may change the range of the  
scalar \cite{Faulkner:2006ub, Starobinsky:2007hu}.

For a general $f(R)$ model, the effective mass squared of 
$\chi=R$ is obtained in the weak-field limit by considering a 
small, spherically symmetric,  perturbation of de Sitter space 
with constant curvature $R_0$. One finds 
\be
m^2=\frac{1}{3} \left( \frac{f_0'}{f_0''}-R_0 \right) .
\ee
This equation coincides  with eq.~(6) of  \cite{Muller:1989rp}, 
with eq.~(26) of \cite{Olmo:2006eh}, and with eq.~(17) of 
 \cite{Navarro:2006mw}. It also appears 
in a calculation of the propagator for $f(R)$ gravity in a 
locally flat background (eq.~(8) of 
\cite{Nunez:2004ji}). The same expression is  
recovered in a  gauge-invariant stability analysis of de Sitter 
space  \cite{Faraoni:2005vk} reported in 
Sec.~\ref{sec:dSstability} below.

Another possibility is to consider the field $\phi\equiv 
f'(R)$ instead of $\chi =R$, and to define  the 
effective mass of $\phi$ by using the {\em Einstein frame} 
scalar-tensor analog of $f(R)$ gravity instead of its Jordan 
frame cousin already discussed \cite{Chiba:2003ir}. By 
performing the conformal transformation
\be
g_{\mu\nu}\rightarrow \tilde{g}_{\mu\nu}=f'(R) \, g_{\mu\nu}  
\equiv \phi \, g_{\mu\nu}
\ee
and the scalar field redefinition $\phi=f'(R) \rightarrow 
\tilde{\phi} $ with
\be
d\tilde{\phi} =\sqrt{\frac{2\omega_0 +3}{2\kappa} } \, 
\frac{d\phi}{\phi} ,
\ee
a scalar-tensor theory is mapped to the Einstein frame in which 
the ``new'' scalar field $\tilde{\phi}$ couples minimally to the 
Ricci curvature and has canonical kinetic energy, as described 
by the action
\bea
S^{(g)}&=&\int d^4x \, \sqrt{-\tilde{g}} \, \left[ \tilde{R}-\frac{1}{2} 
\, 
\partial^{\alpha}\tilde{\phi}\partial_{\alpha}\tilde{\phi} -U( 
\tilde{\phi} ) \right]+\nn\\
&&\qquad\qquad 
\quad+S_M(e^{-\sqrt{2\kappa/3}\,\tilde{\phi}} 
\tilde{g}_{\mu\nu},\psi) ,
\eea
(note once more the non-minimal coupling of the matter in 
the Einstein frame).
For the $\omega_0=0$ equivalent of metric $f(R)$ gravity we have
\bea
\phi &\equiv & f'(R)=\mbox{e}^{\sqrt{ \frac{2\kappa}{3}} \, 
\tilde{\phi}} ,\\
U( \tilde{\phi} )&=&\frac{R f'(R)-f(R)}{2\kappa \left( f'(R) 
\right)^2} ,
\eea
where $R=R(\tilde{\phi}) $. 
By using $d\tilde{\phi}/d\phi=\sqrt{\frac{3}{2\kappa}} \, 
\frac{f''}{f'} $, the effective mass of $ \tilde{\phi} $ is 
defined by 
\be
\tilde{m}^2_{eff}\equiv \frac{d^2 U}{d\tilde{\phi}^2}=
\frac{1}{3} \left[ \frac{1}{f''} +\frac{\tilde{\phi}}{f'} 
-\frac{4f}{\left( f' \right)^2} \right]
\ee
[this equation appears in the footnote on p.~2 of 
\cite{Chiba:2003ir}]. By assuming a de Sitter background with 
constant curvature $R_0=12H_0^2=f_0/(6f_0')$, this turns into
\be
\tilde{m}^2_{eff} = \frac{1}{3f_0'} \left( 
\frac{f_0'}{f_0''} -R_0 \right) = \frac{ m^2_{eff}}{f_0'}  .
\ee
In the Einstein frame, it is not the mass $\tilde{m}$ of a 
particle or a field that is measurable, but rather the ratio $ 
\tilde{m}/\tilde{m}_u$  between $ \tilde{m}$ and the Einstein 
frame 
unit of  mass $\tilde{m}_u$, which is varying, scaling as 
$\tilde{m}_u=\left[ f'(R) \right]^{-1/2} m_u =\phi^{-1/2}\, 
m_u$, where $m_u$ is 
the constant unit of mass in the Jordan frame 
\cite{Dicke:1961gz, Faraoni:1998qx, Faraoni:2006fx}. Therefore, 
\be
\frac{\tilde{m}^2_{eff}}{\tilde{m}^2_u} = 
\frac{m^2_{eff} }{ m^2_u} .
\ee  
In practice, $ \phi\equiv f'(R)$ is dimensionless and its value 
must be of  order  unity  in order to obtain the 
gravitational coupling strength measured in the Solar System; as 
a result, the Einstein frame metric $\tilde{g}_{\mu\nu}$ and the 
Jordan frame metric $g_{\mu\nu}$ are almost equal, and the same 
applies to $\tilde{m}_u, m_u$ and to $\tilde{m}_{eff}, 
m_{eff}$, respectively. 
Then, the only relevant difference between Einstein and Jordan 
frames is the scalar field redefinition $\phi\rightarrow 
\tilde{\phi}$.

\subsubsection{Weak-field limit in the metric formalism}

\label{sec:weakmetricsubsub}

Having discussed the field content of the theory, we are now 
ready to discuss the weak-field limit. Having the correct 
weak-field limit at the Newtonian and post-Newtonian levels is 
essential for theoretical viability.

From the beginning, works on the weak-field (Newtonian and 
post-Newtonian) limit of $f(R)$ gravity led to  
opposite results appearing in the  literature 
\cite{Accioly:1999a, Rajaraman:2003st, Soussa:2003re, 
Dick:2003dw, Easson:2004fq,  
Navarro:2005da, Navarro:2005gh, Cembranos:2005fi, Shao:2005wt, 
 Olmo:2005zr, Olmo:2005jd,  
Capozziello:2005bu,  Clifton:2005aj,Clifton:2006kc, 
Barrow:2005dn, 
Capozziello:2006jj, Multamaki:2006zb, Zhang:2007ne,
Hu:2007nk, Baghram:2007df, Capozziello:2007id, 
Capozziello:2007eu, Multamaki:2007jk, Iorio:2007ee, 
Capozziello:2007ms, Ruggiero:2006qv}. Moreover, a 
certain lack of rigour in checking the convergence of series 
used in the expansion around a de Sitter background often left 
doubts even on results that, {\em a posteriori}, turned out to 
be correct \cite{Sotiriou:2005xe}.

By using the equivalence between $f(R)$ and  
scalar-tensor gravity, Chiba originally suggested that all $f(R) 
$ theories are 
ruled out \cite{Chiba:2003ir}. This claim was based on the fact that 
metric $f(R)$ 
gravity is equivalent to an $\omega_0=0$
Brans--Dicke theory, while the observational 
constraint is $\left| \omega_0 \right|>40000 $ 
\cite{Bertotti:2003rm}. This is not quite the case and the 
weak-field limit is more subtle than it appears, as the 
discussion of the previous section might have already revealed: 
The value of the parametrized post-Newtonian (PPN) parameter $\gamma$, on which the 
observational  bounds are directly applicable, is practically 
independent of the mass of the   scalar only when the latter is 
small \cite{Wagoner:1970vr}. In  this case, the constraints on 
$\gamma$ can indeed be turned into constraints on $\omega_0$. 
However, if the mass of this scalar is large, it dominates over 
$\omega_0$  in the expression of $\gamma$ and drives its value 
to unity. The physical explanation of this fact, as mentioned 
previously, is that the scalar becomes short-ranged and, 
therefore, has no effect at Solar System scales.  Additionally, 
there is even the possibility that the effective mass of  the 
scalar field itself is actually scale-dependent. In this 
case, the scalar  may acquire a large 
effective mass at terrestrial and Solar System scales, 
shielding it from experiments performed there while 
being effectively light at cosmological scales. This is the 
{\em chameleon mechanism}, well-known in quintessence models 
\cite{Khoury:2003aq, Khoury:2003rn}. 

Given the above, it is worth examining these issues in more 
detail. Even though early doubts about the validity of the 
dynamical 
equivalence with scalar-tensor theory in the Newtonian limit 
\cite{Faraoni:2006hx, Kainulainen:2007bt} have now been 
dissipated  \cite{Faraoni:2007yn},  a 
direct approach  which does not resort to the 
scalar-tensor equivalence is preferable as the former  could in 
principle hide things 
\cite{Olmo:2005jd}. This was given  
in the metric formalism, first in 
the special  case \cite{Erickcek:2006vf}  $f(R)=R-\mu^4/R$  
[which  is already ruled out by the Ricci 
scalar instability \cite{Dolgov:2003px, Faraoni:2006sy}]
and  in the case $f(R)=R^n$ using light 
deflection  and other Solar System experiments\footnote{The 
perihelion precession in modified gravity is studied in 
\cite{Schmidt:2008qi, Iorio:2007rk, Iorio:2007ee, 
Iorio:2007zz, Baghram:2007df}.}  
\cite{Clifton:2005aj,Clifton:2006kc, 
Barrow:2005dn, Zakharov:2006uq}. Only later 
was the case of a general function $f(R)$ discussed   
\cite{Chiba:2006jp, Olmo:2006eh, Jin:2006if}.  Chiba's result  
based on the 
scalar-tensor equivalence eventually turns out to 
be valid subject to certain assumptions which are not always 
satisfied \cite{Chiba:2006jp, Olmo:2006eh, Jin:2006if}  --- see 
below. This  method, however, does not apply to the Palatini 
version of  $f(R)$ gravity.

In what follows we adhere to, but streamline, the discussion of 
\cite{Chiba:2006jp} with minor modifications, 
in order to compute the PPN parameter $\gamma$ for metric $f(R)$ 
gravity [see also \cite{Olmo:2006eh}]. We consider a spherically 
symmetric, static, non-compact  
body embedded in a background de Sitter universe; the latter  
can exist in an adiabatic  approximation in which 
the evolution of the universe is very 
slow in comparison with local dynamics. The 
condition for the existence 
of a de Sitter space with $R_{\mu\nu}=R_0 g_{\mu\nu}/4$ and 
constant 
curvature $R_0=12H_0^2$ is
\begin{equation}
f_0'R_0 - 2f_0= 0 , \;\;\;\;\;\;\;\;\; H_0=\sqrt{ 
\frac{f_0}{6f_0'} } .
\end{equation}
The  line element is
\begin{eqnarray} \label{weakfieldlineelement}
 ds^2 & = & -\left[ 1+2\Psi(r) -H_0^2r^2 \right] dt^2 \nonumber 
\\
& &+ \left[  1+2\Phi(r) +H_0^2r^2 \right] dr^2 
 +r^2 d\Omega^2
\end{eqnarray}
in Schwarzschild coordinates, 
where the post-Newtonian potentials $\Psi(r)$ and $\Phi(r)$ are 
treated as small perturbations.\footnote{Isotropic 
coordinates are usually employed in the study of the weak-field 
limit of spherically symmetric metrics; however, the difference 
is irrelevant to first order in $\Psi$ 
and $\Phi$ \cite{Olmo:2006eh}.} The goal is to compute 
the PPN parameter $\gamma=-\Psi/\Phi$ by solving the 
equations satisfied by these potentials. A linearized 
analysis is performed assuming 
\begin{equation}
\left| \Psi(r) \right|, \left| \Phi(r) \right| \ll 1 , 
\;\;\;\;\;r<<H_0^{-1} ,
\end{equation}
and
\begin{equation} 
R(r)=R_0+R_1(r) ,
\end{equation}
where the deviation $R_1(r)$ of the Ricci curvature from 
the constant $R_0$ is also a  small 
perturbation.\footnote{The solution  derived  for the 
spherically symmetric metric  is only valid 
when $mr<<1$, where $m$ is the effective 
mass of the  scalar.  If this 
assumption is not made, then (for example, according to 
\cite{Jin:2006if}), it would seem that quantum 
corrections in 
$f(R)=R+aR^2$ with $a\simeq 10^{-24} \, \mbox{GeV}^{-2}$ are 
ruled out by Solar System 
constraints, which is not the case because these corrections are 
equivalent to a {\em massive} scalar field with short range that 
is  not constrained by the available data.}

Three  assumptions  are made:

\noindent {\em Assumption~1:} $f(R)$ is analytical at $ 
R_0$.

\noindent {\em Assumption~2:}~~$mr<<1$, where $m$ is the 
effective mass of 
the scalar  degree of freedom of the theory. In other 
words, this scalar field (the Ricci 
curvature, which is an extra dynamical quantity in the metric 
formalism) must have a  range longer than 
the  size of the Solar System---if it is much shorter 
than, say, 0.2~mm \cite{Hoyle:2000cv}, the presence of this 
scalar 
is effectively hidden  from Solar System and terrestrial 
experiments. In this case, 
this field could not have cosmological effects at late times, 
but could only be important in the very early universe at high 
curvatures, {\em e.g.}, in Starobinsky-like inflation 
\cite{Starobinsky:1980te}.

\noindent {\em Assumption~3:} the pressure $P\simeq 0$ for the 
energy-momentum of the local  star-like object. The trace of 
the corresponding energy-momentum tensor reduces  to $T_1 \simeq
-\rho$.

By expanding $f(R)$ and $f'(R)$ around $R_0$, the trace 
equation~(\ref{trace})  reduces to  
\begin{equation}\label{reducedtrace}
3f_0'' \, \Box R_1+\left( f_0''R_0-f_0' \right) R_1=\kappa \, 
T_1 ,
\end{equation}
where $T= T_1$, since $T$ is zero in the background. For a 
static, spherically symmetric 
body, $R_1=R_1(r)$ and $\Box R_1=\nabla^2 
R_1=\frac{1}{r^2}\frac{d}{dr}\left( r^2 \frac{d R_1}{dr}  
\right)$.
The reduced trace equation (\ref{reducedtrace}) then becomes
\begin{equation} \label{eqforR1}
\nabla^2R_1 -m^2 R_1=-\frac{\kappa\, \rho}{3f_0''} ,
\end{equation}
where 
\begin{equation} 
m^2 = \frac{f_0'-f_0'' R_0}{3f_0''} .
\end{equation}
By using $R_0=12H_0^2=2f_0/f_0'$, this reduces to 
\begin{equation}\label{effectivemsquared}
m^2=\frac{ \left(f_0'\right)^2 -2f_0 f_0''}{3f_0' f_0''}   .
\end{equation}
This equation is found in 
various other treatments of perturbations of de Sitter space 
\cite{Olmo:2006eh, Navarro:2006mw,  
Nunez:2004ji, Faraoni:2005vk}. 

Assumption~2 that the scalar $R_1$ is light, which 
enables the $f(R)$ theory to produce significant cosmological 
effects at late times, also  allows one to  
neglect\footnote{Although  \cite{Chiba:2006jp} provide Green 
functions in both cases $m^2>0$ and  $m^2<0$, the latter 
corresponds to a spacetime instability and is unphysical.   This 
is is irrelevant in the end because only the case 
$m^2\rightarrow 0$ is necessary and used in the calculation 
\cite{Faraoni:2007ke}.} 
the term $m^2 R_1$ in eq.~(\ref{eqforR1}). 
The  Green function of the equation 
$\nabla^2 R_1=-\frac{ \kappa \, \rho}{3f_0''} $ is then 
$G(r)=-\frac{1}{4\pi r} $ and the solution is 
$ R_1 \simeq \int d^3 \vec{x}' \, \frac{ -\kappa \, \rho 
(r')}{3f_0''} G(r-r') $, which yields 
\begin{equation} \label{R1}
R_1\simeq \frac{\kappa \, M}{12\pi f_0'' r} 
\;\;\;\;\;\;\;\;\;\;\;\;\;\;\;\; \left( 
mr<<1 \right) . 
\end{equation}
Now, the condition $m^2 r^2 <<1$ yields 
\begin{equation}
\frac{1}{3}\left| \frac{f_0'}{f_0''}-R_0 \right| r^2 <<1
\end{equation}
and, using $H_0r<<1$, 
\begin{equation}
\left| \frac{f_0'}{f_0''} \right| r^2 <<1 .
\end{equation}

Let us use now the full field equations (\ref{metf}); 
by expanding $f(R)$ and $ f'(R)$ and using $f_0=6H_0^2 f_0'$ we 
get 
\begin{eqnarray}
\delta^{\alpha}_{\beta} f_0'' \Box R_1  \!\!&+&\!\! f_0' \left( 
R^{\alpha}_{\beta} -3H_0^2 
\delta^{\alpha}_{\beta} 
\right)-\frac{f_0'}{2}\, 
R_1 \delta^{\alpha}_{\beta}
\nonumber \\\!\!&-&\!\! f_0'' \nabla^{\alpha} 
\nabla_{\beta} R_1 +f_0''R_1 
R^{\alpha}_{\beta} =\kappa \, T^{\alpha}_{\beta} .
\end{eqnarray}
By using again the assumption $H_0r<<1$, the d'Alembertian 
$\Box $ becomes $ \nabla^2_{\eta}$ and, for $\left( \mu ,\nu 
\right)=\left(0,0 \right)$, 
\begin{equation}
f_0'\left( R^0_0-3H_0^2 \right)-\frac{f_0'}{2} R_1 
+f_0''R_1R^0_0+f_0''\nabla^2 R_1=-\kappa\, \rho .
\end{equation}
By computing $R^0_0=3H_0^2-\nabla^2\Psi(r) $ and 
dropping terms $f_0''H_0^2R_1 << f_0'\nabla^2\Psi$, {\em etc.}, 
we obtain 
\begin{equation}
f_0' \nabla^2 \Psi(r) +\frac{f_0'}{2}\, R_1 -f_0'' \nabla^2 
R_1=\kappa \, \rho .
\end{equation}
Recalling that $\nabla^2 R_1\simeq -\, \frac{\kappa \, 
\rho}{3f_0''} $ 
for $mr<<1$, one obtains 
\begin{equation} \label{25ofCSE}
f_0' \nabla^2 \Psi(r)=\frac{2\kappa\, \rho}{3} -\frac{f_0'}{2}\, 
R_1 .
\end{equation}
Eq.~(\ref{25ofCSE}) can  be integrated from $r=0$ to $r>r_0$ 
(where $r_0$ is the radius of the star-like object) to obtain,  
using Gauss' law,
\begin{equation}
\frac{d\Psi}{dr}=\frac{\kappa }{6\pi f_0'}\, 
\frac{\kappa M }{48\pi f_0 '' \, r^2}-\frac{C_1}{r^2} ,
\end{equation}
 where 
$ M(r)=4\pi \int_0^{r_0}  dr'\, (r')^2 \rho( r')$. The 
integration constant $C_1$ must be set to zero to guarantee 
regularity of the Newtonian potential at $r=0$.  The 
potential  $\Psi(r)$ then becomes 
\begin{equation}
\Psi(r)= - \, \frac{\kappa\, M}{6\pi f_0' \, r} 
 -\, \frac{\kappa \, M}{48\pi f_0''}\, r .
\end{equation}
The second term on the right hand side  is negligible; in 
fact,
\begin{equation}
\left| \frac{  \frac{\kappa\, M r}{48\pi f_0''} 
}{ \frac{-\kappa\, M}{6\pi f_0' r}} \right|=
\left| \frac{ f_0'}{8f_0''} \right| r^2<<1 ,
\end{equation} 
and
\begin{equation} \label{Psi}
\Psi(r)\simeq  - \frac{\kappa\, M}{6\pi f_0' \, r}  .
\end{equation}

Let us now find the second potential $\Phi(r)$ appearing in the 
line element (\ref{weakfieldlineelement}). By using the field 
equations (\ref{metf}) with $\left( a,b 
\right)=\left(1,1 \right)$,
\begin{eqnarray}
 f_0'\left( R^1_1-3H_0^2 \right) \!\!&-&\!\!\frac{f_0'}{2}\, R_1 
-f_0''\nabla^1\nabla_1R_1 \nonumber \\
\!\!&+&\!\!f_0''R_1 R^1_1
+f_0''\Box R_1  =\kappa \, T^1_1 
\end{eqnarray}
with $  T^1_1 \simeq 0$ outside the star, and
\begin{eqnarray}
&& R^1_1  \simeq  3H_0^2 -\frac{d^2 
\Psi}{dr^2}+\frac{2}{r}\frac{d\Phi}{dr} , \\
&& g^{11} \nabla_1 \nabla_1 R_1  \simeq  \frac{d^2 
R_1}{dr^2} ,
\end{eqnarray}
and neglecting higher order terms, one obtains 
(eq.~(22) of \cite{Chiba:2006jp})
\begin{equation}\label{22ofCSE}
f_0'\left( -\frac{d^2\Psi}{dr^2}+\frac{2}{r}\frac{d\Phi}{dr} 
\right) -\frac{f_0'R_1}{2}+\frac{2f_0''}{r} \frac{dR_1}{dr} 
\simeq 0 .
\end{equation}
Now, using eq.~(\ref{R1}) for $R_1$, one concludes that the 
third 
term in eq.~(\ref{22ofCSE}) is negligible in comparison with 
the fourth term. In fact,
\begin{equation}
\left| \frac{  
\frac{f_0'R_1}{2} }{ 
\frac{2f_0''}{r}\frac{dR_1}{dr} } \right|\simeq \left| 
\frac{f_0'}{f_0''} \right|r^2 <<1 .
\end{equation}
Then, 
using again the expression (\ref{R1}) for $dR_1/dr$ and 
eq.~(\ref{Psi}) for $\Psi(r)$, one obtains
\begin{equation}
\frac{d\Phi}{dr}=- \frac{\kappa\, M}{12\pi f_0' \, r}  ,
\end{equation}
which is immediately integrated to
\begin{equation} \label{Phi}
\Phi(r)=  \frac{\kappa\, M}{12\pi f_0' \, r}  .
\end{equation}
The post-Newtonian metric (\ref{weakfieldlineelement})  
therefore gives  the PPN parameter $\gamma$ as 
\begin{equation}
\gamma =-\frac{\Phi(r)}{\Psi(r)}=\frac{1}{2} .
\end{equation}
This is a gross violation of the experimental bound $\left| 
\gamma -1 \right|< 2.3\cdot 10^{-5} $ 
\cite{Bertotti:2003rm} and agrees with the calculation of the 
PPN parameter $\gamma=\frac{\omega_0+1}{\omega_0+2}$ found by using 
the equivalence of metric $f(R)$ gravity with an $\omega_0=0$ 
Brans--Dicke theory \cite{Chiba:2003ir}.

The results of \cite{Chiba:2006jp} have been reproduced
by \cite{Olmo:2006eh}, who works in isotropic 
coordinates with a  slightly different approach. 
\cite{Kainulainen:2007bt} have 
obtained spherically symmetric interior solutions matched to the 
exterior solutions of metric $f(R)$ gravity and have confirmed 
the result $\gamma=1/2$.

{\em Limits of validity of the previous analysis:} One can 
contemplate various circumstances  in which the 
assumptions above are not satisfied and the previous analysis 
breaks  down. It is important to ascertain whether these 
are  physically relevant situations. There are three main 
cases to consider.

{\em The case of non-analytic $f(R)$:} While \cite{Chiba:2006jp} 
consider  
functions $f(R)$ that are analytic at the background 
value  $R_0$ of the Ricci curvature, the 
situation in which this function is 
not analytical has been contemplated briefly in 
\cite{Jin:2006if}. Assuming that 
$f(R)$ has an isolated singularity at $R=R_s$, it can be 
expressed as  the sum of a Laurent series,
\begin{equation}
f(R)=\sum_{n=0}^{+\infty} \, a_n \left( 
R-R_s  \right)^n  .
\end{equation}
\cite{Jin:2006if} note that it must be $R\neq 
R_s$ in the dynamics of the 
universe because a constant curvature space with $R=R_s$ can  
{\em not} be a solution of the field equations. Therefore, one 
can approximate the solution adiabatically with a de Sitter 
space with constant curvature $R_0\neq R_s$. The function $f(R)$ 
is analytical here  and the previous discussion applies. This is 
not possible if $f(R)$ has an essential singularity, for  
example, if  $f(R)=R-\mu^2 \sin\left( 
\frac{\mu^2}{R-\lambda} \right) $ \cite{Jin:2006if}. There is, 
of course, no reason other than Occam's razor to exclude this 
possibility.

{\em Short range scalar field:} If the assumption 
$mr<<1$ is not satisfied, the scalar  is massive. If its range 
is sufficiently 
short, it is effectively hidden from experiments probing 
deviations from Newton's law and from other Newtonian and 
post-Newtonian  experiments in the solar neighbourhood. This is 
the case of 
quadratic quantum corrections to Einstein's gravity, {\em e.g.}, 
$f(R)=R + \alpha R^2$. If the effective mass is $m\geq 
10^{-3}$~eV 
(corresponding to a fifth force range less than $\sim 0.2$~mm, the shortest scale currently accessible to weak-field experiments), 
this correction is undetectable and yet it can still have large 
effects in the early inflationary universe 
\cite{Starobinsky:1980te}. 
However, it can not work as a model for late time 
acceleration.

{\em Chameleon behaviour:} The {\em chameleon effect} ß
\cite{Khoury:2003aq, Khoury:2003rn}, 
originally 
discovered in 
scalar field models of dark energy, consists of 
the effective mass $m$ of the scalar degree of freedom 
being a function of the curvature (or, better, of the  energy 
density of the local environment), so that $m$ can be large at 
Solar System and terrestrial curvatures and densities, 
and small at 
cosmological curvatures and densities --- effectively, it is 
short-ranged in the Solar System and it becomes long-ranged at 
cosmological densities thus causing the acceleration of the 
universe. The chameleon effect can be applied to metric $f(R)$ 
gravity 
\cite{Cembranos:2005fi,Navarro:2006mw, Faulkner:2006ub, Starobinsky:2007hu}, with 
the 
result that theories of the kind 
\cite{Carroll:2003wy, Amendola:2006kh, Amendola:2006eh, 
Amendola:2006we}
\begin{equation} \label{chameleontheories}
f(R)=R-(1-n) \mu^2 \left( \frac{R}{\mu^2} \right)^n
\end{equation}
are compatible with the observations in the region of the 
parameter space $ 0 < n \leq 0.25 $ with $\mu$  
sufficiently small \cite{Faulkner:2006ub}. Precisely, using the 
Cassini bound on the PPN parameter $\gamma$ 
\cite{Bertotti:2003rm}, the constraint 
\begin{equation}
\frac{\mu}{H_0}\leq \sqrt{3} \left[ 
\frac{2}{n(1-n)}\right]^{\frac{1}{2(1-n)}} 
10^{\frac{-6-5n}{2(1-n)}}
\end{equation}
is obtained \cite{Faulkner:2006ub}.  
Fifth force experiments give the bounds 
\begin{equation}
\frac{\mu}{H_0}\leq  \sqrt{1-n} \, \left[  
\frac{2}{n(1-n)}\right]^{\frac{1}{2(1-n)}}
10^{\frac{-2-12n}{1-n} } .
\end{equation}
Preferred values seem to be  $m\simeq 10^{-50}$~eV$\sim 
10^{-17}H_0$ \cite{Faulkner:2006ub}. Note that $n>0$, which 
guarantees $f''>0$, is required for Ricci scalar stability 
($n=0$ reduces the model to GR with a 
cosmological constant, but avoiding the latter  was exactly the 
reason why dark  energy and modified gravity  were 
introduced in the first 
place).

These models work to explain the current cosmic 
acceleration because,  for small curvatures $R$, the 
correction in $R^n$ with $n<1$ is 
larger than the Einstein--Hilbert term $R$ and comes to dominate 
the dynamics. On the negative side, these theories 
are observationally indistinguishable from a cosmological 
constant and they have been dubbed ``vanilla 
$f(R)$ gravity'' \cite{Faulkner:2006ub, Amendola:2006kh, 
Amendola:2006eh, Amendola:2006we, Amendola:2007nt}. 
However, they still have  the advantage of 
avoiding a 
fine-tuning problem in $\Lambda$ at the price of a much smaller 
fine-tuning of the parameter $\mu$. As for all modified gravity 
and dark energy models, they 
do not address the cosmological constant problem.

The weak-field limit of metric $f(R)$ theories which admit a 
{\em global} Minkowski solution around  which to linearize, was 
studied by \cite{Clifton:2008jq}.  These theories (including, 
{\em e.g.}, analytic functions  $f(R)=\sum_{n=1}^{+\infty} a_n 
R^n$) are not motivated by late time cosmology and the Minkowski 
global solution, although present, may not be stable 
\cite{Clifton:2005aj}, which in practice detracts from the 
usefulness of 
this analysis. Several new post-Newtonian potentials are found 
to appear in addition to the two usual ones \cite{Clifton:2008jq}.


\subsubsection{Weak-field limit in the Palatini formalism}
\label{sec:weakPalatini}

Early works  on the weak-field limit of Palatini $f(R)$ gravity 
often  led to contradictory results and to several technical 
problems as well \cite{Meng:2003sx, Barraco:2000dx, 
Dominguez:2004ds, 
 Olmo:2005jd, Olmo:2005zr, 
Olmo:2006eh, Allemandi:2005tg, 
Sotiriou:2005xe, Allemandi:2006bm, 
Kainulainen:2006wz, Bustelo:2006ms, Ruggiero:2006qv, 
Ruggiero:2007jr} which seem to have been clarified by now.

First of all, there seems to have been some confusion in the 
literature about the  fact that Palatini $f(R)$ gravity reduces  
to GR with a cosmological constant in vacuum and the 
consequences that this can have on the weak-field limit and 
Solar System tests. It is, of course, true (see  
Sec.~\ref{sec:palatinifield}) that in vacuo Palatini $f(R)$ gravity 
will have the same solutions of GR plus a cosmological constant 
and, therefore, the Schwarzschild-(anti-)de Sitter solution will 
be the unique  vacuum spherically symmetric solution (see also 
Sec.~\ref{sec:astro} for a discussion of the Jebsen-Birkhoff
theorem).  This was interpreted in 
\cite{Allemandi:2006bm,Ruggiero:2006qv} as an indication that 
the only parameter that can be constrained is  the
effective cosmological constant and, therefore, models that are
cosmologically interesting (for which this parameter is very 
small) trivially satisfy Solar System tests. However, even if 
one sets aside the fact that a weak gravity regime is possible 
inside matter as well, such claims cannot be correct: they 
would completely defeat the purpose of performing a 
parametrized 
post-Newtonian expansion for any theory for which one can 
establish uniqueness of a spherically symmetric solution, as in 
this case we would be able to judge Solar System viability just 
by considering this vacuum solution (which would be much 
simpler). 

Indeed, the  existence of a spherically symmetric vacuum 
solution, irrespective of its uniqueness, does not suffice to 
guarantee a good Newtonian limit. For instance, 
the Schwarzschild-de Sitter solution has two free parameters; 
one of them can be associated with the effective cosmological 
constant in a straightforward manner (using the asymptotics).
However, it is not clear how the second parameter, which in 
GR is identified with the mass of the object in the 
Newtonian regime, is related to the internal structure of the 
object in Palatini $f(R)$ gravity. The {\em assumption} that it 
represents the mass defined in the usual way is not, of course, 
sufficient. One would have to actually match the exterior 
solution to a solution describing the interior of the Sun within 
the realm of the theory in order to express the undetermined 
parameter in the exterior solution in terms of known physical  
quantities, such as Newton's constant and the Newtonian mass. 
The essence of the derivation of the Newtonian  limit of the  
theory consists also  in deriving such an  explicit relation for 
this quantity and showing that it agrees 
with the Newtonian expression. The parametrized post-Newtonian 
expansion is nothing but an alternative way to do that without 
having to solve the full field equations. Therefore, it is 
clear that more information than the form of the vacuum 
solution is needed in order to check whether the theory can 
satisfy the Solar System constraints.

However, some early attempts towards a Newtonian and 
post-Newtonian expansion  were also flawed. In  
\cite{Meng:2003sx} and \cite{Barraco:2000dx} for instance, a  
series expansion around a de Sitter 
background was performed in order to derive the Newtonian limit. 
Writing 
 \be
{\cal R}={\cal R}_0+{\cal R}_1,
\ee
where ${\cal R}_0$ is the Ricci curvature of the background  
and ${\cal R}_1$ is a correction,  one is tempted to 
expand in powers of ${\cal R}_1/{\cal R}_0$ regarding the 
latter as  a small quantity.  Since one  needs the quantities  
$f({\cal R}_0+{\cal R}_1)$  and  $f'({\cal R}_0+{\cal R}_1)$, 
the usual
approach is to Taylor-expand around ${\cal R}={\cal R}_0$ and 
keep only the leading order  terms in ${\cal R}_1$. However, it 
has been shown in \cite{Sotiriou:2005xe} that this can not be 
done for most cosmologically interesting models 
because ${\cal R}_1/{\cal R}_0$ is not small. 

Take as an example  the model   
\cite{Vollick:2003aw} 
 \be
\label{cdtt2}
f({\cal R})={\cal R}-\frac{\epsilon_2^2}{{\cal R}},
\ee
and $\epsilon_2\sim 10^{-67}(\textrm{eV})^2\sim 
10^{-53}\textrm{m}^{-2}$.
Expanding as
\be
f({\cal R})=f({\cal R}_0)+f'({\cal R}_0) {\cal 
R}_1+\frac{1}{2}f''({\cal R}_0) {\cal R}_1^2+\ldots
\ee
and using eq.~(\ref{cdtt2}) yields
\be
\label{expans}
f({\cal R})=f({\cal R}_0)+\left(1+\frac{\epsilon_2^2}{{\cal 
R}_0^2}\right) {\cal R}_1-\frac{1}{2}\frac{2 
\epsilon_2^2}{{\cal R}_0^3} {\cal R}_1^2+\ldots ,
\ee
 where now ${\cal R}_0=\epsilon_2$. It is then easy to see that 
the second term on the right hand side is 
of the order of ${\cal R}_1$, whereas the third term is of the 
order of ${\cal R}_1^2/\epsilon_2$. Therefore, in order to 
truncate before the third term, one needs ${\cal R}_1\gg {\cal 
R}_1^2/\epsilon_2$ or
 \be
\label{cond}
\epsilon_2\gg {\cal R}_1. 
\ee
This is not a stringent constraint:  ${\cal R}_0\sim \epsilon_2$ 
and so this is the usual condition for linearization.

Let us return now to the trace equation~(\ref{paltrace}). For 
the model under consideration, 
 \be
\label{rt}
{\cal R}=\frac{1}{2}\left(-\kappa\,T\pm\sqrt{ \kappa^2\, 
T^2+12 \epsilon_2^2} \, \right).
\ee

According to eq.~(\ref{rt}), the value of ${\cal R}$, and
consequently, ${\cal R}_1$, is algebraically related to $T$ and,  
whether or not the condition~(\ref{cond}) is
satisfied or not critically depends on the value of the 
energy density. To demonstrate this, pick the mean density of  
the Solar System, $\rho\sim 10^{-11} \textrm{gr/cm}^3$, which 
satisfies the weak-field limit criteria. 
For this value, $\left|{\epsilon_2/\kappa\,T}\right|\sim 
10^{-21}$, where $T\sim -\rho$. The ``physical'' branch of the 
solution~(\ref{rt}) is the one with positive sign 
because, given that $T<0$, it ensures 
that matter leads to a standard positive curvature in strong
gravity. Then, 
 \be
\label{r}
{\cal R}\sim -\kappa\,T-\frac{3\epsilon_2^2}{\kappa\, T}
\ee
 and ${\cal R}_1\sim -\kappa\,T\sim \kappa\,\rho$.  Thus, 
$\epsilon_2/{\cal R}_1\sim 10^{-21}$ and it is evident that
the required condition does not hold for some typical densities 
related to the Newtonian limit.

The situation does not improve even with the
``unphysical'' branch of eq.~(\ref{rt}) with a negative 
sign. In fact, in this case, ${\cal R}_1\sim
\epsilon_2 [3 \epsilon_2/(\kappa\,T)+\sqrt{3}]$ and  the 
correction to the background curvature is of the order 
$\epsilon_2$ and not much smaller than
that, as it would be required in order to truncate 
the expansion~(\ref{expans}). In \cite{Barraco:2000dx}, this fact 
was overlooked and only linear terms in
${\cal R}_1$ were kept in the expansion of $f({\cal R})$ and 
$f'({\cal R})$ around ${\cal R}_0$. In \cite{Meng:2003sx}, even 
though  this fact is noticed in the final stages of the analysis 
and is actually used, the authors do not take it into account
properly from the outset, keeping again only linear terms
[see, {\em e.g.}, eq.~(11) of \cite{Meng:2003sx}].

However, the algebraic dependence of ${\cal R}$ on the density  
does not only signal a problem for the approaches just 
mentioned. It actually implies that the outcome  of 
the post-Newtonian expansion itself depends on the density, as 
shown in \cite{Sotiriou:2005xe,  Olmo:2005jd, Olmo:2005zr}. 
Consider  for instance, along the lines of 
\cite{Sotiriou:2005xe,Sotiriou:2005cd}, the conformal metric
 \be
h_{\mu\nu}=f'({\cal R})g_{\mu\nu}
\ee
that was introduced in Sec.~\ref{sec:palatinifield} [{\em 
cf.}~eq.~(\ref{hgconf})].  In terms of this metric, the field 
equations can be written in the form
 \be
\label{eingen}
{\cal R}_{\mu\nu}-\frac{1}{2}{\cal R} 
h_{\mu\nu}+(f'-1)\left({\cal R}_{\mu\nu}-\frac{{\cal 
R}}{2f'}h_{\mu\nu}\right)= \kappa \,T_{\mu\nu},
\ee
and ${\cal R}_{\mu\nu}$ is the Ricci tensor of the metric 
$h_{\mu\nu}$. It is evident that, if $f'=1$, then $h_{\mu\nu}$ 
and $g_{\mu\nu}$  
coincide and eq.~(\ref{eingen})  yields Einstein's 
equation. However, since  ${\cal R}$ and consequently
$f'({\cal R})$ are  functions of the energy density, due to 
eq.~(\ref{paltrace}), the deviation of
$f'$ from unity will  always depend on the energy density and 
the functional form of  $f$. Therefore, one can definitely find 
some function $f$ which,  for some range of energy densities, 
will give $f'=1$ to high  precision. However, for the same 
function $f$, there will be large deviations from $f'=1$ at a 
different density range. This dependence of the 
weak-field limit on the energy density is a  novel 
characteristic  of Palatini $f(R)$ gravity. 

This dependence can be made explicit if the problem is 
approached via the equivalent Brans--Dicke theory 
\cite{Olmo:2005jd,  Olmo:2005zr}. Note that the usual bounds 
coming from Solar System experiments do  not
apply in the $\omega_0=-3/2$ case, which is equivalent to Palatini
$f(R)$ gravity. This is because the standard treatment of the
post-Newtonian expansion of Brans--Dicke theory, which one uses to
arrive to such bounds, is critically based on the assumption 
that
$\omega_0\neq -3/2$ and the term $(2\omega_0+3)$ frequently 
appears as a denominator. Making this assumption is not 
necessary, of course, in
order to derive a post-Newtonian expansion, but is a convenient 
choice, which allows for this  otherwise general treatment. 
Therefore, a different approach, such as the one followed in 
\cite{Olmo:2005jd}, was
indeed  required for the $\omega_0=-3/2$ case. Following the 
standard
assumptions of a post-Newtonian expansion around a background
specified by a cosmological solution \cite{willbook}, the 
following relations were derived for the post-Newtonian limit:
 \bea
\label{olmo11}
- \frac{1}{2}\nabla^2\left[h^1_{00}-\Omega(T)\right]& 
=&\frac{\kappa \,\rho-V(\phi)}{2\phi}, \\
\label{olmo21}
-\frac{1}{2}\nabla^2\left[h^1_{ij}+\delta_{ij} 
\Omega(T)\right]&=&\left[\frac{ 
\kappa \,\rho+V(\phi)}{2\phi}\right],
\eea
where $V$ is the potential of the scalar field $\phi$ and
$\Omega(T)\equiv \log[\phi/\phi_0]$. The subscript $0$ in 
$\phi_0$, and in any other quantity in the rest of this 
subsection, denotes that it is evaluated
at $T=0$.

The solutions of eqs.~(\ref{olmo11}) and~(\ref{olmo21}) are
\bea
\label{olmo1s}
h^{(1)}_{00} \left( t,\vec{x} \right)\!\!\!&=&\!\!\! \frac 
{2 G_{\rm eff} M_\odot}{r}+\frac{V_0}{6\phi_0}r^2+\Omega(T),\\
h^{(1)}_{ij} \left( t,\vec{x} \right)\!\!\!&=&\!\!\!\left[  \frac{ 2\gamma 
G_{\rm 
eff} M_\odot}{r}-  
\frac{V_0}{6\phi_0}r^2-\Omega(T)\right]\delta_{ij}, \label{olmo2s}
\eea
where $M_\odot \equiv \phi_0 \int d^3 \vec{x}' 
\rho \left(t, \vec{x}'\right)/\phi$. The
effective Newton constant $G_{\rm eff}$ and the post-Newtonian
parameter $\gamma$ are defined as
 \bea
G_{\rm eff}& 
\equiv &\frac{G}{\phi_0}\left(1+\frac{M_V}{M_\odot}\right),\\
&&\nonumber \\
\gamma& \equiv &\frac{M_\odot-M_V}{M_\odot+M_V},
\eea
where $M_V\equiv \kappa^{-1} \phi_0 \int d^3 
\vec{x}'\left[V_0/\phi_0-V(\phi)/\phi\right]$. 

As stated in different words in \cite{Olmo:2005jd}, if  the
Newtonian mass is defined as $M_N\equiv \int d^3 \vec{x}' 
\rho \left(t,\vec{x}'\right)$, the requirement
that a theory has a good Newtonian limit is that $G_{\rm 
eff} M_{\odot}$ equals $G M_N$, where $N$ denotes Newtonian, and
$\gamma\simeq 1$ to very high precision.  Additionally, the 
second term on the right hand side of both eqs.~(\ref{olmo1s})  
and~(\ref{olmo2s}) should be negligible, since it plays the role 
of a cosmological constant term. $\Omega(T)$ should also be 
small and have a negligible dependence on $T$. 

Even though it is not impossible, as mentioned before, to 
prescribe $f$ such that all of the above are satisfied for some 
range of densities within matter \cite{Sotiriou:2005xe}, this 
does not seem possible over the wide range of densities 
relevant for the Solar System tests. As a matter of fact, 
$\Omega$ is nothing but an algebraic  function of $T$ and, 
therefore, of the density (since $\phi$  is an algebraic 
function of ${\cal R}$). The presence of  the $\Omega(T)$ term 
in eqs.~(\ref{olmo1s}) and~(\ref{olmo2s}) signals an algebraic 
dependence of the post-Newtonian metric on the density.  This 
direct dependence of the metric on the matter field is  not only 
surprising but also seriously problematic. Besides the fact that 
it is evident that the theory cannot have  the proper Newtonian 
limit for all densities (the range of densities  for which it 
will fail depends on the functional form of $f$),  consider 
the following: What happens to the post-Newtonian metric if a 
very weak point source (approximated by a delta function) is 
taken into account as a perturbation? And will the 
post-Newtonian metric be  continuous when going from the 
interior of a source to the exterior, as it should? 

We will refrain from further analysis of these issues  here, 
since  evidence coming from considerations different than the 
post-Newtonian limit,  which we will review shortly, will be of 
significant help. We will,  therefore, return to this discussion 
in Sec.~\ref{sec:surfsing}.

\subsection{Stability issues}
\label{sec:stabilitysub}

In principle, several kinds of instabilities need to be 
considered  to make sure that $f(R)$ 
gravity is a viable alternative to GR  \cite{Sokolowski:2007pk, 
Sokolowski:2007zz,  Sokolowski:2007rd, Calcagni:2006ye, 
DeFelice:2006pg, Wang:2005bi, 
Chiba:2005nz}.

The Dolgov-Kawasaki \cite{Dolgov:2003px} instability in 
the 
matter sector, specific to metric $f(R)$ gravity, imposes 
restrictions on the functional form of $f$, and  is 
discussed below.  More generally, it is believed that a stable 
ground state, the 
existence of  which is necessary in a gravitational theory, 
should be highly 
symmetric, such as the de Sitter, or Minkowski, or perhaps the 
Einstein static space. Instabilities 
of de Sitter space in the gravity sector have been 
found in \cite{Faraoni:2005ie,
Faraoni:2004dn, Faraoni:2004bb, Faraoni:2005vk, Dolgov:2005se, 
Barrow:2005qv} [see also 
\cite{Barrow:1983rx, Muller:1989rp} for pre-1998 discussions], 
while stability in first loop quantization of $f(R)$ gravity 
and with respect to black hole nucleation was 
studied in \cite{Cognola:2005sg, Paul:2005wb, 
Paul:2005bk, Cognola:2005de, Cognola:2007vq}.  The linear 
stability of de Sitter space with respect to 
homogeneous perturbations in generalized theories of the form
$ f\left( R, R_{\mu\nu}R^{\mu\nu}, 
R_{\mu\nu\alpha\beta}R^{\mu\nu\alpha\beta} \right)$ was studied 
in \cite{Cognola:2008wy}. The stability of the Einstein static 
space in metric $f(R)$ gravity with respect to homogeneous 
perturbations was studied in \cite{Boehmer:2007tr}, while 
stability of this space with respect to inhomogeneous isotropic 
perturbations 
was established, with a gauge-invariant formalism and under 
certain conditions, in \cite{Goswami:2008fs}.


\subsubsection{Ricci stability in the metric formalism}
\label{sec:stabilitymetricsubsub}

In the metric formalism, Dolgov and Kawasaki 
discovered  an  instability in the prototype  model 
$f(R)=R-\mu^4/R$ (now 
called ``Dolgov-Kawasaki'', or ``Ricci scalar'' or ``matter'' 
instability), which   manifests itself on an extremely short 
time scale  and is sufficient to rule out  this model 
\cite{Dolgov:2003px}.  Their  result was confirmed 
in \cite{Nojiri:2003ft, Nojiri:2003ni}, in 
which it was also shown that adding to this specific $f(R)$ an 
$R^2 $ term removes this instability. The instability was 
rediscovered in \cite{Baghram:2007df} for a
specific form of the function $f(R)$.  The 
analysis of this instability is generalized to 
arbitrary $f(R)$ theories in the metric formalism 
in the following way \cite{Faraoni:2006sy}.

We  parametrize the deviations from Einstein gravity  as
\begin{equation}\label{1ter}
f(R)=R+ \epsilon \, \varphi(R) ,
\end{equation}
where $\epsilon $ is a small parameter with the dimensions of a 
mass squared and  $\varphi$ is arranged to be 
dimensionless (in the example $f=R-\mu^4/R$, one has  
$\epsilon=\mu^2$, $\varphi=-\mu^2/R$, and $\mu\simeq H_0\approx 
10^{-33}$~eV).

By using the trace equation~(\ref{metftrace})
\begin{equation} \label{trace}
3\Box f'(R)+ f'(R) R -2f(R)=\kappa T ,
\end{equation}
and evaluating $\Box f'$,  
\begin{equation}\label{300}
\Box R +\frac{\varphi '''}{\varphi ''} \, \nabla^{\alpha} R \, 
\nabla_{\alpha} 
R+\frac{\left( \epsilon \varphi 
'-1\right) }{3\epsilon \, \varphi ''}\, R=\frac{\kappa \, T 
}{3\epsilon  \, \varphi''}\, +\, \frac{2\varphi}{3\varphi ''} .
\end{equation}
We assume that  $\varphi '' \neq 0$:  if $\varphi ''=0$ on an 
interval then the theory 
reduces to GR. Isolated zeros of $\varphi ''$, 
at  which the theory is  ``instantaneously GR'',  
are  in principle possible but will not be considered here.

Consider a small region of  spacetime in the weak-field regime 
and approximate {\em locally} the metric and the curvature by
\begin{equation}\label{4}
g_{\mu\nu}=\eta_{\mu\nu}+h_{\mu\nu} , 
\;\;\;\;\;\;\;\;\;\;\;\;\; 
R=-\kappa\, T +R_1 ,
\end{equation}
where $\eta_{\mu\nu} $ is the Minkowski metric and $\left| 
R_1/\kappa\, T\right| \ll 1$. This 
inequality excludes the case of conformally 
invariant matter with $T=0$, a situation considered later. 
Equation~(\ref{4}) yields, to first order in $R_1$,
\begin{eqnarray}
&& \ddot{R}_1 -\nabla^2 R_1 -\frac{2\kappa\, \varphi 
'''}{\varphi ''}\, \dot{T}\dot{R}_1+\, 
\frac{2\kappa \, \varphi '''}{\varphi ''}\, \vec{\nabla}T \cdot 
\vec{\nabla}R_1  \nn\\
&& +\frac{1}{3\varphi ''} \left( \frac{1}{\epsilon}-\varphi' 
\right) R_1=
\kappa \, \ddot{T}-\kappa \nabla^2 T -\, \frac{\left(\kappa\, 
T\varphi '+ 2 \varphi 
\right)}{3\varphi ''} , \nn\\\label{5}
\end{eqnarray}
where $\vec{\nabla}$ and $\nabla^2$ are the gradient and 
Laplacian  in Euclidean three-dimensional space, 
respectively, and an overdot denotes differentiation with respect 
to time. The function $\varphi$ and its derivatives are now 
evaluated at $R=-\kappa\, T$. The 
coefficient of $R_1$ in the fifth 
term on the  left hand side is the square of an effective mass 
and is dominated by the term  $\left( 
3\epsilon \,  \varphi '' \right)^{-1}$ due to the 
extremely small  value of  $\epsilon$ needed for these theories 
to reproduce the correct 
cosmological dynamics. Then, the scalar mode $R_1$ of the $f(R)$ 
theory  is stable if $\varphi '' =f''>0$, and unstable   if this 
effective mass is negative, {\em i.e.}, if  $\varphi''=f''<0$. 
The 
time scale for this instability  to manifest is estimated to be 
of the order of the inverse effective mass $\sim 10^{-26}$~s in 
the 
example $\epsilon \varphi(R)=-\mu^4/R$ 
\cite{Dolgov:2003px}. The small value  of $\varphi ''$ gives a 
large  effective mass and is responsible for the  small time 
scale over which the instability develops.

Let us  consider now matter with vanishing trace $T$ of the 
stress-energy tensor. In this case eq.~(\ref{5}) becomes
\begin{eqnarray}
&& \ddot{R}_1+\frac{\varphi '''}{\varphi ''} \, \dot{R_1}^2 
-\nabla^2 R_1  -\frac{\varphi '''}{ \varphi ''} \, 
\left(\vec{\nabla} R_1 \right)^2 \nonumber \\
&& \nonumber \\
&& +\frac{1}{3\varphi 
''}\left( \frac{1}{\epsilon}-\varphi '\right) R_1  
=\frac{2\varphi}{3\varphi ''} .\label{6}
\end{eqnarray}
Again, the effective mass term is $\sim \left( 3\epsilon 
\varphi '' \right)^{-1}$, which  has the sign of $f''$ and the 
previous stability criterion is recovered. The stability  
condition $f''(R) \geq 0$  is useful to veto $f(R)$  gravity 
 models.\footnote{Refs.~\cite{Nojiri:2004dw, 
Multamaki:2005zs} hinted towards the stability 
criterion, but did not fully derive it because  
a decomposition in orders of $\epsilon^{-1}$ was not performed.}

When $f''<0$, the instability of these theories can 
be interpreted, following  eq.~(\ref{5}), as an instability in 
the gravity sector. Equivalently, since it appears inside matter when $R$ starts deviating from $T$ [see eq.~(\ref{4})], 
it can be seen as a matter instability [this is the 
interpretation taken in  \cite{Dolgov:2003px}]. Whether the 
instability arises in the gravity or matter sector seems 
to be  a matter of interpretation.

The instability of stars made of any type of matter in theories 
with 
$f''<0$ and sufficiently small is confirmed, with a different 
approach (a generalized variational principle) in 
\cite{Seifert:2007fr}, in which the time scale for instability 
found by Dolgov and Kawasaki in the $1/R $ model is also 
recovered. The stability condition $f''\geq 0$ is recovered in 
studies of cosmological perturbations \cite{Sawicki:2007tf}.

The stability condition $f''(R)\geq 0$, expressing the 
fact  that the scalar degree of freedom is not 
a ghost, can be given a simple 
physical interpretation \cite{Faraoni:2007yn}. Assume that the 
effective  gravitational coupling $G_{eff}(R) \equiv G/f'(R)$ is 
positive; 
then, 
if $G_{eff}$ increases with the curvature, {\em i.e.},
\begin{equation}
\frac{dG_{eff}}{dR}=\frac{-f''(R) G}{\left( f'(R) \right)^2}>0 
,
\end{equation}
at large curvature the effect of gravity becomes stronger, 
and since $R$ itself generates larger and larger curvature via 
eq.~(\ref{trace}), the effect of which becomes stronger and 
stronger because of an increased $G_{eff}(R)$, a positive 
feedback mechanism acts to destabilize the theory. There is no 
stable ground state if a small curvature grows and grows 
without limit and the system  runs away. If instead the 
effective gravitational coupling {\em decreases} when $R$ 
increases, which is achieved when $f''(R) >0$, a negative 
feedback mechanism operates which compensates for the increase 
in  $R$ and there is no running away of the solutions. These 
considerations have to be inverted if $f'<0$, which can only 
happen if 
the effective energy density $\rho_{eff}$ also becomes 
negative. This is not a physically meaningful 
situation because the effective gravitational coupling 
becomes negative and the tensor field and the scalar field of 
metric  $f(R)$ gravity become ghosts 
\cite{Nunez:2004ji}.

GR, with $f''(R)=0$ and 
$G_{eff}=$~constant, is 
the  borderline case between the two behaviours corresponding 
to stability ($f''>0$) and instability ($f''<0$), respectively.  

Remarkably, besides the Dolgov-Kawasaki instability which 
manifest itself in the linearized version of equation 
(\ref{trace}), there are also recent claims that $R$ can be 
driven to infinity due to strong non-linear effects related to 
the same equation \cite{Tsujikawa:2007xu, 
Frolov:2008uf,Appleby:2008tv}.  More specifically, in 
\cite{Tsujikawa:2007xu} an oscillating mode is found as a 
solution to the perturbed version of eq.~(\ref{trace}) . This 
mode appears to dominate over the matter-induced mode as one 
goes back into the past and, therefore, it can violate the 
stability conditions. In \cite{Frolov:2008uf}, 
eq.~(\ref{trace}) was studied, with the use of a convenient 
variable redefinition but without resorting to any  
perturbative approach. It was found that there exists a 
singularity at a finite field value and energy  level. The 
strongly non-linear character of the equation allows $R$ 
to easily reach the singularity in the presence of matter. 
As noticed in \cite{Appleby:2008tv}, since when it comes 
to  cosmology the singularity lies in the past, it can in 
principle be avoided by choosing appropriate initial 
conditions and evolving forward in time. This, of course, 
might result in a hidden fine-tuning issue. 

All three studies mentioned consider models in which 
$f(R)$ includes, besides the linear term, only terms which 
become important at low curvatures.  It is the form of the 
 effective potential governing the motion of $R$, which depends 
on the functional form of $f(R)$,  that determines how 
easy it is to drive $R$ to infinity \cite{Frolov:2008uf}.  
Therefore,  it seems interesting to study how  the presence of 
terms which become important at large curvatures, such as 
positive powers of $R$, could affect these results. Finally, 
it would be interesting to see in detail how these findings 
manifest themselves in the case of compact objects, and whether 
there  is any relation between this issue and the 
Dolgov-Kawasaki instability.

\subsubsection{Gauge-invariant stability of de Sitter space 
in the metric formalism}
\label{sec:dSstability}

One can consider the generalized gravity action
\begin{equation} \label{dSonly2}
S=\int d^4x \, \sqrt{-g} \left[ \frac{f \left( \phi, R 
 \right)}{2}\, -\frac{\omega( \phi)}{2}\, 
\nabla^{\alpha}\phi\nabla_{\alpha}\phi 
-V(\phi) \right] ,
\end{equation}
incorporating both scalar-tensor gravity (if $f\left( 
\phi, R \right)=\psi(\phi) R $) and modified gravity (if 
the scalar  field $\phi$ is absent and $f_{RR} \neq 0$). 
In a spatially flat FLRW 
universe the vacuum field equations assume the form
\begin{eqnarray} 
&& H^2  =  \frac{1}{3 f'} \left( \frac{\omega}{2} \, 
\dot{\phi}^2  +\frac{Rf'}{2} -\frac{f}{2} +V  -3H\dot{f'} 
\right) , \label{dSonly4} \\
&& \dot{H} =   - \, \frac{1}{2f'}  \left( \omega \dot{\phi}^2 
+ \ddot{F}  -H\dot{f'} \right)  , \label{dSonly5} \\
&&  
\ddot{\phi } +3 H \dot{\phi} +\frac{1}{2\omega} \left(
\frac{d\omega}{d\phi} \,  \dot{\phi}^2 - \frac{\partial 
f}{\partial \phi} +2\, \frac{dV}{d\phi}  
\right) =0 ,\label{dSonly6}
\end{eqnarray}
where $ f' \equiv \partial f/\partial \phi$, $ F \equiv \partial 
f/\partial R$,   
and an overdot denotes differentiation with respect to $t$. We 
choose $\left( H, \phi \right) $ as 
dynamical variables; then, the stationary points of the 
dynamical  system (\ref{dSonly4})-(\ref{dSonly6}) are de Sitter 
spaces with 
constant  scalar field $\left( H_0, \phi_0 \right) $. The 
conditions for these de Sitter solutions to exist are 
\begin{eqnarray} \label{dSonly7}
&& 6H_0^2 \, f'_0 - f_0+2V_0 =0 , \\
&& \left. \frac{df}{d\phi}\right|_0- 
\left.2\frac{dV}{d\phi}\right|_0 =0 , \label{dSonly8} 
\end{eqnarray}
where $ 
f'_0 \equiv f' \left( \phi_0, R_0 \right) $, $ f_0\equiv f
\left( \phi_0, R_0 \right) $, $ V_0 \equiv  V  \left( \phi_0 
\right) $, and $ R_0 = 12 H_0^2 $. The phase space is a curved 
two-dimensional surface embedded in a three-dimensional space 
\cite{deSouza:2007fq}.

Inhomogeneous perturbations of de Sitter space have 
been studied using the covariant and gauge-invariant 
formalism of  \cite{Bardeen:1980kt, 
Ellis:1989jt, Ellis:1989ju, Ellis:1990gi} in 
a 
version provided  by \cite{Hwang:1990re, Hwang:1990jh, 
Hwang:1996bc, Hwang:1997uc, Hwang:1996xh} for generalized 
gravity. The metric  perturbations are defined by 
\begin{eqnarray} 
&& g_{00} =  -a^2 \left( 1+2AY \right) , \;\;\;
g_{0i} =  -a^2 \, B \, Y_i  ,  \\
&& g_{ij}  = a^2 \left[ h_{ij}\left(  1+2H_L  Y \right) +2H_T \, 
Y_{ij}  \right] .\label{dSonly11}
\end{eqnarray}
Here the $Y$ are scalar spherical harmonics, $h_{ij} $  is 
the three-dimensional metric of the FLRW 
background,  $ \hat{\nabla}_i 
$ is the covariant derivative of  $h_{ij}$, and  
$k$ is the eigenvalue of $
\hat{\nabla}_i\hat{\nabla}^i \, Y =-k^2 \, Y $. The $Y_i$ and 
$Y_{ij}$ are vector and tensor 
harmonics satisfying
\begin{equation} \label{dSonly13}
Y_i= -\frac{1}{k} \, \hat{\nabla}_i Y , \;\;\;\;\;\;
Y_{ij}= \frac{1}{k^2} \, \hat{\nabla}_i\hat{\nabla}_j Y 
+\frac{1}{3} \, Y \, h_{ij} ,
\end{equation}
respectively. The Bardeen gauge-invariant potentials 
\begin{eqnarray} 
&& \Phi_H = H_L +\frac{H_T}{3} +\frac{ \dot{a} }{k} \left( 
B-\frac{a}{k} \, \dot{H}_T \right) , \label{dSonly15} \\
&& \Phi_A = A  +\frac{ \dot{a} }{k} \left( B-\frac{a}{k} \, 
\dot{H}_T \right)\nn\\
&& \qquad\quad\quad\!+ \frac{a}{k} \left[ \dot{B} -\frac{1}{k} \left( a \dot{H}_T 
\right)\dot{}  \right] ,  \label{dSonly16}
\end{eqnarray}
the Ellis-Bruni variable 
\begin{equation} \label{dSonly17}
 \Delta \phi = \delta \phi  +\frac{a}{k} \, \dot{\phi}  \left( 
B-\frac{a}{k} \, \dot{H}_T 
\right) ,
\end{equation}
and analogous gauge-invariant variables $ \Delta f$, $\Delta f' 
$,  and $\Delta R$ satisfy first order equations 
given in \cite{Hwang:1990re, Hwang:1990jh, 
Hwang:1996bc, Hwang:1997uc, Hwang:1996xh}, which simplify  
significantly in the de Sitter 
background  $\left(H_0,\phi_0 \right)$  
\cite{Faraoni:2004bb, Faraoni:2005ie, Faraoni:2005vk}.

To first order and in the absence of ordinary 
matter, vector perturbations do not appear   \cite{Hwang:1990re, 
Hwang:1990jh, 
Hwang:1996bc, Hwang:1997uc, Hwang:1996xh}, and de  Sitter space 
is 
always stable with respect to first order tensor 
perturbations. Focusing on 
scalar  perturbations, modified gravity corresponds to  $\phi 
\equiv 1$ and $f=f(R)$ with $f''(R)\neq 0 $ in~(\ref{dSonly2}). 
The gauge-invariant perturbations 
$\Phi_H$ (from which one easily obtains $ \Phi_A$ and $\Delta 
R$)  satisfy  
\begin{equation} \label{dSonly20}
\ddot{\Phi}_H+3H_0 \dot{\Phi}_H+\left( 
\frac{k^2}{a^2}-4H_0^2 +\frac{f'_0}{3f_0'' }\right) \Phi_H=0 ,
\end{equation}
\cite{Faraoni:2004bb, 
Faraoni:2005ie, Faraoni:2005vk},
where  the term $k^2/a^2$ can be dropped at late times and 
for long wavelength modes. Linear stability ensues if the 
coefficient of $\Phi_H$  is 
non-negative, {\em i.e.} (using eq.~(\ref{dSonly7})), 
if\footnote{The generalization of the 
condition~(\ref{dSonly200}) to $D$ spacetime dimensions, derived 
in \cite{Rador:2007gq} for homogeneous perturbations, is
\be
\frac{ (D-2)\left( f_0'\right)^2- D f_0 f_0''}{f_0' f_0''} \geq 
0 .
\ee }
\begin{equation}\label{dSonly200}
\frac{ (f'_0)^2-2f_0 f''_0 }{f'_0 f''_0}\geq 0 .
\end{equation}
The only term containing the comoving wave vector $k$  in 
eq.~(\ref{dSonly20})  becomes negligible at late times and/or 
for  zero-momentum modes and thus the spatial dependence 
effectively 
disappears. In fact, eq.~(\ref{dSonly200}) coincides with the 
stability condition that can 
be obtained by a straightforward homogeneous perturbation 
analysis of eqs.~(\ref{dSonly4}) and (\ref{dSonly5}). As a 
result, in the  stability analysis of de Sitter space in 
modified gravity,  
inhomogeneous perturbations can be ignored and the study can 
be restricted to the simpler homogeneous 
perturbations, which are free of the notorious gauge-dependence 
problems.  This result, which  could not be reached {\em a 
priori} but relies on the inhomogeneous 
perturbation analysis,  holds only for de 
Sitter spaces and not for different attractor 
({\em e.g.}, power-law) solutions that  may be present in the 
phase space. The stability condition~(\ref{dSonly200}) is 
equivalent to  the condition that the scalar field potential in 
the  Einstein frame of the equivalent Brans--Dicke theory has a 
minimum at the configuration identified by the de Sitter space 
of curvature $R_0$ \cite{Sokolowski:2007rd}.

As an example, let us consider the prototype model 
\begin{equation}   
f(R)=R-\, \frac{\mu^4}{R} .
\end{equation}
The background de Sitter space has $R_0=12H_0^2=\sqrt{3}\, 
\mu^2$ and the stability condition 
(\ref{dSonly200}) is never satisfied: this de Sitter solution is 
always unstable. An improvement is obtained by adding a  
quadratic correction to this model: 
\begin{equation} \label{Dsonlylast}
f(R)=R-\, \frac{\mu^4}{R} +aR^2 .
\end{equation}
Then, the condition for the existence of a de Sitter solution 
is again
$R_0=\sqrt{3}\, \mu^2$, while the  stability 
condition  (\ref{dSonly200}) is satisfied if $
a> \frac{1}{ 3\sqrt{3} \, \mu^2} $, in  agreement  
with \cite{Nojiri:2003ft, Nojiri:2003ni} who use independent 
methods.

Different definitions of stability lead to different, albeit 
close, stability criteria for de Sitter space [see 
\cite{Cognola:2005de, Cognola:2007vq} for the  semiclassical 
stability of 
modified 
gravity, \cite{Bertolami:1987wj} for scalar-tensor gravity, and 
\cite{Seifert:2007fr} for a variational approach applicable to 
various alternative gravities].

\subsubsection{Ricci stability in the Palatini formalism}
\label{sec:stabilityPalatinisubsub}

For Palatini $f(R)$ gravity the field equations~(\ref{palf12}) 
and~(\ref{palf22}) are of second 
order and the trace equation~(\ref{paltrace}) is  
\begin{equation} \label{Palatinitrace}
f' ( {\cal R} ) {\cal R} -2 f ( {\cal R} ) =\kappa \, T ,
\end{equation}
where ${\cal R}$ is the Ricci scalar of the 
non-metric connection 
$\Gamma^{\mu}_{\nu\sigma} $ (and not that of the metric 
connection $\left\{ {}^{\mu}_{\nu\sigma} \right\}$ of 
$g_{\mu\nu}$). Contrary to the metric case, 
eq.~(\ref{Palatinitrace}) is not  an evolution equation for 
$\cal{R}$; it is not even a 
differential equation, but rather an {\em algebraic} equation in 
$\cal{R}$ once the function $f \left( \cal{R} \right)$ is 
specified. 
This is also the case in GR, in which the 
Einstein field equations are of second order and taking their  
trace  yields $R=-\kappa \,  T$. Accordingly, the scalar field 
$\phi$ of the 
equivalent $\omega_0=-3/2$  Brans--Dicke theory is not dynamical. 
Therefore, the  Dolgov-Kawasaki instability can not occur in 
Palatini $f(R)$  gravity \cite{Sotiriou:2006sf}.

\subsubsection{Ghost fields}
\label{sec:ghosts}

Ghosts (massive states of negative norm that cause 
apparent lack of unitarity) appear easily in higher order 
gravities. A viable theory should be 
ghost-free: the  presence of ghosts in $f(R, 
R_{\mu\nu}R^{\mu\nu}, 
R_{\mu\nu\alpha\beta}R^{\mu\nu\alpha\beta}) $ gravity has been 
studied in \cite{quant3, 
quant1,  
Strominger:1984dn, Utiyama:1962sn, Codello:2006in, 
Stelle:1976gc,  DeFelice:2007zq, 
Stelle:1978ww, DeFelice:2007zz}. 
Due to the  Gauss--Bonnet identity, if the initial action is linear in $R_{\mu\nu\alpha\beta}R^{\mu\nu\alpha\beta}$, one can reduce the theory 
under  consideration to\footnote{Furthermore, 
$R_{\mu\nu}R^{\mu\nu}$ can be expressed in terms of $R^2$ in a 
FLRW background \cite{Wands:1993uu}.} 
$f(R, R_{\mu\nu}R^{\mu\nu})$  which, in general, 
contains a massive  spin~2 ghost field in addition to the usual 
massless graviton and the massive scalar. $f(R)$ theories  have 
no ghosts \cite{quant3, 
quant1,  
Strominger:1984dn, Utiyama:1962sn,  Stelle:1976gc, 
Stelle:1978ww, 
Ferraris:1988zz}, and the stability condition $f''(R) \geq 0$ of 
\cite{Dolgov:2003px, Faraoni:2006sy} essentially amounts to 
guarantee that the scalaron is not a ghost. Theories of the kind 
$f(R, 
R_{\mu\nu}R^{\mu\nu}, R_{\mu\nu\rho\sigma}R^{\mu\nu\rho\sigma}) 
$ in general are plagued 
by ghosts [this is the case, for example, of conformal 
gravity, as noticed long before the 1998 discovery of the  
cosmic acceleration \cite{Riegert:1984hf}], but models 
with only $f \left( R, R^2-4R_{\mu\nu}R^{\mu\nu} + 
R_{\mu\nu\rho\sigma}R^{\mu\nu\rho\sigma} 
\right) $ terms in the action have been claimed to be ghost-free  
\cite{Navarro:2005da,Comelli:2005tn}.

\subsection{The Cauchy problem}
\label{sec:Cauchy}

A physical theory must have predictive power and, to this 
extent, a  well-posed  initial value problem is a required 
feature. GR satisfies this requirement for most 
reasonable forms of matter \cite{wald}. The 
well-posedness of 
the Cauchy problem for $f(R)$ gravity is an open issue. Using 
harmonic coordinates, Noakes  showed that 
theories with action
\begin{equation} 
S = \frac{1}{2\kappa} \int d^4x \, \sqrt{-g}\left( R+\alpha 
R_{\mu\nu}R^{\mu\nu}+\beta R^2\right) +S_M
\end{equation}
in the metric formalism have  a well posed initial value problem 
in vacuo \cite{Noakes:1983xd}. By using the dynamical 
equivalence with the  
scalar-tensor theory~(\ref{metactionH2}) when 
$f''(R)\neq 0$,  the well-posedness of the  Cauchy 
problem can be reduced to the  analogous problem  for 
Brans--Dicke gravity with $\omega_0=0$ (metric formalism) or 
$\omega_0=-3/2$ (Palatini formalism). The fact that the  
initial value  problem  is well-posed was demonstrated for 
particular scalar-tensor theories 
in \cite{Cocke:1968a, Noakes:1983xd} and a general analysis has  
recently been presented in \cite{Salgado:2005hx, 
Salgado:2008xh}. This work, 
however, does not cover the $\omega_0=0, -3/2 $ cases.

 A system of $3+1$ equations of motion is 
said to  be {\em well-formulated} if it can be rewritten as a 
system of equations which are of only first order  in 
both time and space derivatives. When this set can be put in 
the full  first order form 
\begin{equation}
\partial_t \vec{u} + M^i \nabla_i \vec{u}=\vec{S}\left( 
\vec{u}\right) ,
\end{equation}
where $\vec{u}$ collectively denotes the fundamental variables 
$h_{ij}, K_{ij}$, {\em etc.} introduced below, $M^i$ is 
called the {\em characteristic matrix} of the system, and 
$\vec{S}\left(  \vec{u} \right)$ describes source terms and 
contains only the fundamental variables but not their 
derivatives. The initial 
value formulation is {\em well-posed} if the system of partial 
differential equations is {\em symmetric hyperbolic} ({\em  
i.e.}, the matrices $M^i$ are symmetric)  and {\em strongly 
hyperbolic} if $ s_iM^i$ has a real set of 
eigenvalues and a complete 
set of eigenvectors for  any 1-form $s_i$, and obeys some 
boundedness conditions [see \cite{Solin:2006aa}]. 

The Cauchy problem for metric $f(R)$ gravity is 
well-formulated   and is well-posed in vacuo and with matter, 
as shown below.  For  Palatini 
$f(R)$ gravity,  instead, the  Cauchy problem is unlikely to be  well-formulated or well-posed unless the trace of the matter energy-momentum tensor is constant, due to the presence of 
higher derivatives of  the 
matter  fields in the field equations and to the impossibility 
of eliminating them (see below). 

A systematic covariant 
approach to scalar-tensor 
theories of the form
\begin{equation}
S = \int d^{4}x\sqrt{-g}\left[ \frac{\psi(\phi) R}{2\kappa}  - 
\frac{1}{2}\partial^{\alpha}\phi\partial_{\alpha}\phi  - 
W(\phi)\right] + S_M 
\end{equation}
is due to \cite{Salgado:2005hx}, who 
showed that the Cauchy  problem of these theories is well-posed 
in the 
absence of matter and  well-formulated 
otherwise.  With the exception of $\omega_0=-3/2$,  as we 
will see later, most 
of Salgado's 
results can be extended to the more 
general action  
\begin{equation}
 S \!=\! \int \! d^{4}x\sqrt{-g}\left[ \frac{\psi(\phi) R}{2\kappa} - 
\frac{\omega
(\phi)}{2}\partial^{\alpha}\phi\partial_{\alpha}\phi - W(\phi)\right] \!+\!  S_M 
,
\end{equation}
which contains the additional coupling function $\omega(\phi)$  
(which is different  from the Brans--Dicke parameter 
$\omega_0$)
\cite{LanahanTremblay:2007sg}. 

The field equations, after setting $\kappa = 1$ for this section, are 
\begin{eqnarray}
 &&G_{\mu\nu} = \frac{1}{\psi}\Big[ \psi^{\prime\prime} \left( 
\nabla_{\mu}\phi\nabla_{\nu}\phi  - 
g_{\mu\nu}\nabla^{\alpha}\phi\nabla_{\alpha}\phi \right) \nn\\
 &&\qquad  \quad+ \psi^\prime \left( \nabla_{\mu}\nabla_{\nu} \phi - 
g_{\mu\nu}\square\phi \right)\Big] \nonumber \\
 && \qquad\quad+ \frac{1}{\psi}\Big[ \omega \left( 
\nabla_{\mu}\phi\nabla_{\nu}\phi - \frac{1}{
2} g_{\mu\nu}\nabla^{\alpha}\phi\nabla_{\alpha}\phi \right) \nn\\&&\qquad\quad - 
W(\phi)g_{\mu\nu} + 
T_{\mu\nu}^{(m)}
\Big] ,  \label{CP24}\\
&& \omega\square\phi + \frac{\psi^\prime}{2}  R -  
W^\prime (\phi) +
\frac{\omega^\prime}{2}\nabla^{\alpha} 
\phi\nabla_{\alpha}\phi = 0 ,
\end{eqnarray}
where a  prime denotes differentiation with respect to $\phi$. 
Eq.~(\ref{CP24}) can be cast in the form of  the effective 
Einstein equation $
G_{\mu\nu} = T_{\mu\nu}^{(eff)} $, 
with the effective stress-energy tensor \cite{Salgado:2005hx}
\begin{equation}
T_{\mu\nu}^{(eff)}  =  \frac{1}{\psi(\phi)} \left( 
T_{\mu\nu}^{\left( \psi 
\right)} + T_{\mu\nu}^{\left(  \phi  
\right)} + T_{\mu\nu}^{(m)}\right) ,
\end{equation}
and
\begin{eqnarray}
&&T_{\mu\nu}^{\left( \psi \right)} = \psi^{\prime\prime}(\phi) \left( 
\nabla_{\mu} \phi \nabla_{\nu} \phi -  
g_{\mu\nu}\nabla^{\beta}\phi \nabla_{\beta}\phi \right) 
\nonumber \\
&&\qquad\quad+  \psi^\prime  (\phi)\left( \nabla_{\mu}\nabla_{\nu} \phi - 
g_{\mu\nu}\square\phi \right) ,\\
&&T_{\mu\nu}^{(\phi)} = \omega(\phi) \left( 
\nabla_{\mu}\phi\nabla_{\nu}\phi - 
\frac{1}{2} g_{\mu\nu}\nabla^{\beta}\phi\nabla_{\beta}\phi 
\right) \nn\\ &&\qquad\quad- 
W(\phi)g_{\mu\nu}.
\eea
The trace of  the effective Einstein equations  yields 
\begin{eqnarray}
\label{boxphi}
&& \square\phi = 
\left\{ \psi\left[ 
\omega+ \frac{ 3\left( \psi^\prime \right)^{2} }{2\psi} \right] 
\right\}^{-1}
\left\{ \frac{\psi^\prime T^{(m)}}{2} -2\psi^\prime 
W(\phi)  \right. \nonumber \\
&& \left.  + \psi W^\prime (\phi) + \left[  -\frac{\omega^\prime 
\psi}{2} -  
\frac{\psi^\prime}{2} \left( \omega + 
3\psi^{\prime\prime} \right) \right] \nabla^{c}\phi\nabla_{c}\phi 
 \right\} . \nonumber \\
&&
\end{eqnarray}
The $3+1$ Arnowitt--Deser--Misner (ADM) formulation of the theory proceeds by introducing 
 lapse, shift,  extrinsic curvature, and gradients of $\phi$ 
\cite{wald, Reula:1998ty,  Salgado:2005hx}. Assume that a 
time 
function $t$ 
exists such that the spacetime ($M$,$g_{\mu\nu}$) admits a  
foliation with  hypersurfaces $\Sigma_{t}$ of constant $t$ with 
unit timelike 
normal $n^{a}$.  The 
3-metric and projection operator on $\Sigma_t$ are $h_{\mu\nu} = 
g_{\mu\nu}+n_{\mu}n_{\nu}$ and  ${h^{\alpha}}_{\beta}$, 
respectively. Moreover, 
\begin{eqnarray}
&& n^{\mu}n_{\mu} = -1 , \qquad h_{\alpha\beta}n^{\beta} = 
h_{\alpha\beta} \;, \nn\\
&& n^{\alpha} = 0 , \qquad {h_{\alpha}}^{\beta} 
h_{\beta\gamma} = h_{\alpha\gamma} . 
\label{nolabel}
\end{eqnarray}
The metric is then
\begin{equation}
ds^{2} = -\left( N^{2} - N^{i}N_{i} \right) dt^{2} - 
2N_{i}dtdx^{i} + h
_{ij}dx^{i}dx^{j}
\end{equation}
$\left( i,j = 1,2,3 \right)$, where $N > 0$, $n_{\alpha} = - N 
\nabla_{\alpha} t
$, $
N^{\alpha} = -{h^{\alpha}}_{\beta}t^{\beta} $
is the shift vector, while  $t^{\alpha}$ obeys 
$t^{\alpha}\nabla_{\alpha}t = 1$ 
and $
t^{\alpha} = -N^{\alpha}+Nn^{\alpha} $ 
so that $N = -n_{\alpha}t^{\alpha}$ and $N^{\alpha}n_{\alpha} = 
0$. The extrinsic 
curvature 
 of $\Sigma_{t}$ is
\begin{equation}
K_{\alpha\beta} = -{h_{\alpha}}^{\gamma} 
{h_{\beta}}^{\delta}\nabla_{\gamma}n_{\delta} 
\end{equation}
and the 3D covariant derivative of $h_{\alpha\beta}$ on 
$\Sigma_{t}$ is 
defined by 
\begin{equation}
 D_{i} ^{\left(3\right)}{T^{\alpha_{1}\ldots}}_{\beta_{1}\ldots} 
= { 
h^{\alpha_{1}}  }_{\gamma_{1}}  \ldots 
{h^{\delta_{1}}}_{\beta_{1}}\ldots {h^{\mu}}_{i}\nabla_{\mu} ^{ 
\left(3\right)} {T^{\gamma_{1}\ldots}}_{\delta_{1}\ldots}
\end{equation}
for any 3-tensor $^{(3)} {T^{\mu_{1} \ldots} }_{\nu_{1}\ldots}$ 
, 
with $D_i h_{\mu\nu}= 0$.  The 
 spatial gradient of the scalar $\phi$ is $
Q_{\mu} \equiv  D_{\mu} \phi $ (where $D_{\mu}$ denotes the 
covariant derivative of $h_{\mu\nu}$), 
while its momentum is $
\Pi = {\cal L}_n\phi = n^{\nu}\nabla_{\nu}\phi $ 
and 
\bea
\!\!K_{ij} \!\!\!&=&\!\!\! -\nabla_i n_j \!= - \frac{1}{2N}\! 
\left(\frac{\partial h_{ij} }{\partial t}  + 
D_i N_j 
+ D_j N_i \right) ,\\
&&\qquad\Pi = \frac{1}{N}\left(\partial_t\phi+N^{\gamma} 
Q_{\gamma}\right) ,\\
&& \partial_t Q_i + 
N^l\partial_l Q_i +Q_l \partial_iN^l=D_i\left(N\Pi\right) .
\eea
The effective stress-energy tensor 
$T_{\alpha\beta}^{(eff)}$ is decomposed as  
\begin{equation}
T_{\alpha\beta}^{(eff)} = \frac{1}{\psi}\left( 
S_{\alpha\beta}+J_{\alpha} n_{\beta}+J_{\beta} 
n_{\alpha} + E n_{\alpha} n_{\beta} 
\right) ,
\end{equation}
where
\bea
 \label{CP41}
\!\!\!\!\!\!S_{\alpha\beta} \!\! &\equiv& \!\! {h_{\alpha}}^{\gamma} 
{h_{\beta}}^{\delta} T_{\gamma\delta}^{(eff)} = 
\frac{1}{\psi}\left( S_{\alpha\beta}^{(\psi)}+S_{\alpha\beta}^{(\phi)} 
+S_{\alpha\beta}^{(m)}\right) ,\\
\label{CP42}
\!\! J_{\alpha} \!\! &\equiv &\!\!  -{h_{\alpha}}^{\gamma} 
T_{\gamma\delta}^{(eff)} n^{\delta}= \frac{1}{\psi} 
\left(J_{\alpha}^{(\psi)}+J_{\alpha}^{(
\phi)}+J_{\alpha}^{(m)}\right) ,
\\
\label{CP43}
E \!\! &\equiv & \!\! n^{\alpha} n^{\beta} T_{\alpha\beta}^{(eff)} = 
\frac{1}{\psi} \left(E^{(\psi)}+E^{(\phi)}+E^{(m)}\right) .
\eea 
Its trace is  $ T^{(eff)}  = 
S-E $, 
where
$ S \equiv { S^{\mu} }_{\mu}  $. The Gauss--Codazzi equations 
then 
yield  the Hamiltonian constraint  \cite{wald, Salgado:2005hx}
\begin{equation}
^{(3)}R + K^2 - K_{ij}K^{ij} = 2E ,
\end{equation}
the vector  constraint 
\begin{equation}
\label{CP45}
D_l {K^l}_i - D_i K = J_i ,
\end{equation}
and the dynamical equations
\begin{eqnarray}
\label{CP46}
&& \partial_t {K^i}_j + N^l \partial_l {K^i}_j + {K^i}_l 
\partial_j N^l -  {K^l}_j 
\partial_l N^i + D^i D_j N \nonumber \\
&&- ^{(3)}{R^i}_j N - NK{K^i}_j = \frac{N}{2} \left[ 
\left(S-E\right) \delta^i_j -2S^i_j 
\right] ,
\end{eqnarray}
where $K \equiv {K^i}_i$. The trace of eq.~(\ref{CP46}) yields
\begin{equation}
\partial_t K + N^l \partial_l K + ^{(3)}\Delta N - 
NK_{ij}K^{ij} = \frac{N}{2} \left( S + E \right) ,
\end{equation}
where $^{(3)}\Delta \equiv D^i D_i$.  
Our purpose is to eventually eliminate all second 
 derivatives. $\Box \phi$ which is present in 
 eqs.~(\ref{CP41})-(\ref{CP43}) can actually be eliminated using 
eq.~(\ref{boxphi}), provided that $\omega\neq 
-3(\psi')^2/(2\psi)$.

To be more precise, a direct 
calculation yields the $\psi$- and $\phi$-quantities 
of eqs.~(\ref{CP41})-(\ref{CP43})  
\bea
E^{(\psi)} & =& \psi^\prime\left( D^{\mu}Q_{\mu} + K\Pi \right) + 
\psi^{\prime\prime}Q^2 ,\\
J_{\alpha}^{(\psi)} &=& -\psi^\prime\left( K_{\alpha}^{\gamma} 
Q_{\gamma} +D_{\alpha}\Pi \right) - 
\psi^{\prime\prime
}\Pi Q_{\alpha} ,\\
 S_{\alpha\beta}^{(\psi)} &=& \psi^\prime\left( D_{\alpha} Q_{\beta} + 
\Pi K_{\alpha\beta} - 
h_{\alpha\beta}\square\phi 
\right)   \nonumber \\
&&  - \psi^{\prime\prime} \left[ 
h_{\alpha\beta}\left( Q^2 - \Pi^2 \right) - Q_{\alpha} Q_{\beta} 
\right] ,
\end{eqnarray}
where $Q^2 \equiv Q^{\nu} Q_{\nu}$, while 
\begin{equation}
S^{(\psi)} = \psi^\prime \left( D_{\nu} Q^{\nu} + K\Pi - 3\square\phi 
\right) 
+  \psi^{\prime\prime}\left( 3\Pi^2 - 2Q^2 \right) ,
\end{equation}
and 
\bea
E^{(\phi)} \!\! &=& \!\! \frac{\omega}{2} \left( \Pi^2 + Q^2 \right) + 
W(\phi) ,\\
 J_{\mu}^{(\phi)} \!\!&=&\!\! -\omega\Pi Q_{\mu} ,\\
S_{\alpha\beta}^{(\phi)} \!\!&=&\!\! \omega Q_{\alpha} Q_{\beta}  - 
h_{\alpha\beta}\left[ \frac{\omega}{2} 
\left( 
Q^2 - \Pi^2 \right) + W(\phi) \right] ,\\
S^{(\phi)} \!\! & = &\!\! \frac{\omega}{2} \left( 3\Pi^2 - Q^2 \right) - 
3W(\phi) .
\eea
The Hamiltonian and the vector constraints become
\bea
\label{CP64}
&& ^{(3)}R + K^2 - K_{ij}K^{ij} - \frac{2}{\psi} \left[ 
\psi^\prime\left( D_{\mu} Q^{\mu} + 
 K\Pi \right) + \frac{\omega}{2} \Pi^2 \right. \nonumber \\
&& \qquad\left.  + \frac{Q^2}{2} \left( \omega + 2\psi^{
\prime\prime} \right) \right] 
= \frac{2}{\psi} \left( E^{(m)}+W(\phi) \right) ,\\
\label{CP65} 
&& D_l {K^l}_i - D_i K + \frac{1}{\psi} \big[ \psi^\prime \left( 
{K_i}^c 
Q_c + D_i\Pi 
\right) \nonumber \\
&&\qquad\qquad+ \left( \omega + \psi^{\prime\prime} \right) \Pi Q_i 
\big] = 
\frac{J_i^{(m)}}{\psi} , 
\end{eqnarray}
respectively, and the dynamical equation~(\ref{CP46}) is 
\begin{eqnarray}
&& \partial_t {K^i}_j + N^l \partial_l {K^i}_j + {K^i}_l 
\partial_j N^l - {K_j}^l  \partial_l N^i 
\nonumber \\
&& + D^i D_j N - ^{(3)}{R^i}_j N - NK{K^i}_j \nonumber 
\\
&& + \frac{N}{2\psi} \left[ \psi^{\prime\prime}\left( Q^2 - \Pi^2 
\right) + 2W(\phi) +  \psi^\prime\square\phi  \right] 
\delta^i_j  \nonumber \\
&&  + \frac{N\psi^\prime}{\psi}\left( D^i 
 _j  + \Pi {K^i}_j \right) \frac{N}{\psi} \left( \omega + 
\psi^{\prime\prime} \right) Q^i Q_j \nonumber \\
&& = \frac{N}{2\psi} \left[\left( S^{(m)} - E^{(m)} \right) 
\delta^i_j 
- 2{S^{(m) \,\, i}}_j \right] ,
\end{eqnarray}
with trace
\begin{eqnarray}
\label{CP67}
&& \partial_t K + N^l \partial_l K + ^{(3)}\Delta N - 
NK_{ij}K^{ij} \nn\\&&- \frac{N\psi^\prime}{\psi} \left( D^{\mu} Q_{\mu} + \Pi 
K 
\right) \nonumber \\ 
&& + \frac{N}{2\psi} \left[ \psi^{\prime\prime}Q^2 - \left( 2\omega + 
3\psi^{\prime\prime}  
\right)\Pi^2 \right] \nonumber \\
&& = \frac{N}{2\psi} \left( -2W(\phi)  
-3\psi^\prime\square\phi + 
S^{(m)} + E^{(m)} \right) 
\end{eqnarray}
where \cite{Salgado:2005hx}
\begin{eqnarray}
\label{CP68}
&& {\cal L}_n\Pi - \Pi K - Q^{\mu} D_{\mu} \left( \ln N \right) 
- D_{\mu}Q^{\mu}  = -\square\phi  \nonumber \\
&& =  -\frac{1}{ \psi\left[ \omega + \frac{3(\psi^\prime)^2}{2\psi} 
\right] } \Bigg\{  \frac{\psi^\prime T^{(m)} }{2} -2\psi^\prime 
W(\phi) + \psi W^\prime (\phi) \nonumber \\
&&\quad+\left[
\frac{-\omega^\prime \psi}{2} - \left( \omega + 3\psi^{\prime\prime} 
\right)
\frac{\psi^\prime}{2} \right] \nabla^{\mu} \phi\nabla_{\mu} \phi 
\Bigg\}  
\end{eqnarray}
In vacuo, the  initial data $(h_{ij}, K_{ij}, \phi, Q_i, \Pi)$ 
on an initial  hypersurface $\Sigma_0$ obey~(\ref{CP64}),  
(\ref{CP65}), and $
Q_i = D_i\phi $, $
D_i Q_j = D_j Q_i $. 
In the presence  of matter, the variables $E^{(m)}$, 
$J_a^{(m)}$, $S_{ab}^{(m)}$ must  also 
be assigned on the  initial hypersurface $\Sigma_0$. 
Fixing a gauge corresponds to  specifying the lapse and the 
shift vector. The system (\ref{CP64})-(\ref{CP67}) contains only  
first-order  derivatives 
in both space  and time once the d'Alembertian $\square\phi$ is 
written in terms of 
$\phi, \nabla^{\mu} \phi\nabla_{\mu} \phi$, $\psi$, and its 
derivatives by 
means of   eqs.~(\ref{boxphi}) or (\ref{CP68}).  As 
mentioned  earlier, this can be done whenever $\omega\neq 
-3(\psi')^2/(2\psi)$.  As already pointed out in 
\cite{Salgado:2005hx}  for the specific case $\omega=1$, and can 
be now generalized for  any $\omega\neq -3(\psi')^2/(2\psi)$, 
the reduction to 
a first-order system  shows that the Cauchy problem is 
well-posed in vacuo and 
well-formulated 
 in the presence of reasonable matter. 

Let us now consider the  results specific to $f(R)$ gravity. 
Recall that Brans--Dicke  theory, which is of interest for us 
due to its equivalence with $f(R)$ gravity,  corresponds to 
$\omega(\phi)=\omega_0/\phi$, with 
$\psi(\phi)=\phi$, and $W\rightarrow 2V$.  
 This yields the constraints
\begin{eqnarray}
&& ^{(3)}R + K^2 - K_{ij}K^{ij} - \frac{2}{\phi}\left[ D^{\mu} 
Q_{\mu}
+ K\Pi \right. \nonumber \\
&& \quad\left. +   \frac{\omega_0}{2\phi}\left( \Pi^2 + Q^2 \right) 
\right] = \frac{2}{\phi}\left[ E^{(m)} + V(\phi) \right] , 
\label{CP71}\\
&& D_l {K^l}_i - D_i K \nonumber \\
&& \quad+ \frac{1}{\phi}\left( {K_i}^l Q_l + D_i \Pi + 
\frac{\omega_0}{ \phi}\Pi 
Q_i \right)= \frac{J_i^{(m)}}{\phi} ,
\end{eqnarray}
and the dynamical equations
\begin{eqnarray}
&& \partial_t {K^i}_j + N^l\partial_l {K^i}_j + 
{K^i}_l\partial_j N^l - 
{K_j}^l\partial_l  N^i 
+ D^i D_j N \nonumber \\
&& \quad- ^{(3)}{R^i}_j N - NK{K^i}_j + \frac{N}{2\phi}\delta^i_j 
\left( 2V(\phi) +  \square \phi \right)  \nonumber \\
&& \quad+ \frac{N}{\phi} \left( D^i Q_j + \Pi {K^i}_j  \right) 
+ \frac{N\omega_0}{\phi^2}Q^i Q_j \nonumber 
\\
&&\qquad = \frac{N}{2\phi}\left[ 
\left( S^{(m) } - E^{(m)} \right) \delta^i_j - 2 {S^{(m) \,\, 
i}}_j \right] ,\\
&& \partial_t K + N^l\partial_l K + ^{(3)}\Delta N - 
NK_{ij}K^{ij}\nonumber \\
&& \quad - \frac{N}{\phi}\left( D^{\nu}Q_{\nu} + 
\Pi K \right) - \frac{\omega_{0}N}{\phi^2}\Pi^2 
\nonumber \\
&&  \qquad
 = \frac{N}{2\phi}\left[ -2V(\phi) - 3\square\phi + S^{(m)} + 
E^{(m)} 
\right] ,\label{CP74}
\end{eqnarray}
with
\begin{equation}
\label{CP75}
\left( \omega_0 + \frac{3}{2} \right)\square\phi = 
\frac{T^{(m)}}{2} - 
2V(\phi) + \phi V^{\prime}(\phi) + \frac{\omega_{0}}{\phi} 
\left( \Pi^{2}
 - Q^{2} \right) . 
\end{equation}
The condition  $\omega\neq -3(\psi')^2/(2\psi)$, which needs to 
be satisfied in  order to for one to be able to use 
eq.~(\ref{boxphi})  in order to eliminate $\Box\phi$ can be 
written in the Brans--Dicke theory  notation as $\omega_0\neq 
-3/2$. One could of course had guessed  that by looking at 
eq.~(\ref{CP75}).
Therefore, metric $f
(R)$ gravity, which is equivalent to $\omega_0 = 0$  Brans--Dicke 
gravity,  has 
a well-formulated Cauchy problem in general and is  well-posed 
in vacuo. Further work by \cite{Salgado:2008xh} 
established 
the well-posedness of the Cauchy problem for scalar-tensor 
gravity with $\omega=1$ in the presence of matter; this can be 
translated into the well-posedness of metric $f(R)$ gravity with 
matter along the lines established above.

How about Palatini $f(R)$ gravity, which,  corresponding to 
$\omega_0=-3/2$, is exactly the case that the  constraint 
$\omega_0\neq -3/2$ excludes?
 Actually, for this value of the Brans--Dicke parameter, 
eq.~(\ref{bdf2}), and  consequently at eq.~(\ref{CP75}), include 
no derivatives of $\phi$.  Therefore, one can actually solve 
algebraically for $\phi$. [The same  could be done using 
eq.~(\ref{boxphi}) in the more general  case where $\omega$ is a 
function of $\phi$, when $\omega= -3(\psi')^2/(2\psi)$.]  
We will not consider cases for which  eq.~(\ref{bdf2})  has no 
roots or when it is identically satisfied in vacuo. These cases 
lead to theories for which, in the Palatini $f(R)$ formulation,  
 eq.~(\ref{paltracev}) has no roots or when it is identically 
satisfied  in vacuo respectively. As already mentioned in 
Sec.~\ref{sec:palatinifield},  the first case leads to 
inconsistent field equations, and  the second to a conformally 
invariant theory \cite{Ferraris:1992dx},  see also 
\cite{Sotiriou:2006hs} for a discussion. 

 Now, in vacuo one can easily show that the solutions of 
 eq.~(\ref{bdf2}) or (\ref{CP75}) will be of the form 
$\phi={\rm 
  constant}$. Therefore, all derivatives of $\phi$ vanish and one 
 conclude that $\omega_0=-3/2$ Brans--Dicke theory or Palatini 
 $f(R)$ gravity have a well-formulated and well-posed Cauchy 
 problem.\footnote{This has been missed in 
\cite{LanahanTremblay:2007sg},  where it is claimed that 
the Cauchy problem is not well-posed because  the constraint 
$\omega_0\neq -3/2$ does not allow  for the use of 
eq.~(\ref{CP75}) in order to eliminate  $\Box\phi$. Note also 
that in the absence of a potential (there  is no corresponding 
Palatini $f(R)$ gravity)  $\omega_0=-3/2 $ Brans--Dicke 
theory does not have a well-posed Cauchy problem, as noticed 
in \cite{Noakes:1983xd}, because this theory is  conformally 
invariant and $\phi$ is  indeterminate.} This could  
have been 
expected, as noticed in \cite{Olmo:2008nf}, considering  that 
Palatini $f(R)$ gravity reduces to GR with a cosmological  
constant in vacuo.

In the presence of matter, things are more complicated. The 
solutions of eq.~(\ref{bdf2})  or (\ref{CP75}) will give $\phi$ 
as a function of $T$, the trace of the stress-energy tensor. This can still be used to replace $\phi$
in all equations but it will lead to terms such as $\Box T$. 
Therefore, for the Cauchy problem to be  well-formulated  in the 
 presence of matter, one does not only have to assume that the 
 matter is ``reasonable", in the sense that the matter fields 
 satisfy a quasilinear, diagonal, second order hyperbolic system 
 of equations [see \cite{wald}], but also to require that the 
 matter field equations are such that allow to express all 
derivatives  of $T$ present in eq.~(\ref{CP71}) to (\ref{CP74}) 
for $\omega_0=-3/2$  in terms of only first derivatives of the 
matter fields. It seems  highly implausible that this 
requirement can be fulfilled for generic  matter fields. This 
seems to imply that $\omega_0=-3/2$ Brans--Dicke theory  and 
Palatini $f(R)$ gravity are unlikely to have a well-formulated  
Cauchy problem in the presence of matter fields. However, more  
precise conclusions can only be drawn if specific matter fields  
are considered on a case by case basis. The complication 
arising  
from the appearance of derivatives of $T$, and consequently 
higher    derivatives of the matter fields in the equations, and 
seem to be  critical on whether the Cauchy problem can be 
well-formulated  in the presence of matter will be better 
understood in Sec.~\ref{sec:surfsing}.


\section{Confrontation with particle physics and astrophysics}

\subsection{Metric $f(R)$ gravity as dark matter}

Although most recent motivation for $f(R)$ gravity originates 
from the need to find alternatives to the mysterious dark energy 
at cosmological scales, several authors adopt the same 
perspective at galactic and cluster scales, using metric $f(R)$ 
gravity 
as a substitute for dark matter 
\cite{Zhao:2008ta, Jhingan:2008ym, Nojiri:2008nk, 
Capozziello:2005tf, 
Capozziello:2006ph, Capozziello:2007yb, 
Capozziello:2006uv, 
Capozziello:2004km, Capozziello:2004us,
 Martins:2007uf, 
 Saffari:2007xc, Iorio:2007rk, 
Iorio:2007zz}.
Given the equivalence between $f(R)$ and scalar-tensor gravity, 
these efforts resemble previous attempts to model dark matter 
using scalar fields \cite{RodriguezMeza:2007pd, 
RodriguezMeza:2005yd, RodriguezMeza:2004xb, 
CervantesCota:2007ys, CervantesCota:2007uc,  
Bernal:2005hw, Bernal:2006ci, Bernal:2006cj, Bernal:2006it,
Matos:2007zza, Matos:2001ps, 
Matos:2000jx, Matos:2000ss,  Alcubierre:2002et, 
Alcubierre:2001ea}.

Most works concentrate on models of the form $f(R)=R^n$. A 
theory of this form with $n=1-\alpha/2$ was studied in 
\cite{Saffari:2007xc, Mendoza:2006hs} by using 
spherically 
symmetric solutions to approximate galaxies. The fit to galaxy 
samples yields 
\be
\alpha=(3.07\pm 0.18) \times 10^{-7} \left( 
\frac{M}{10^{10}M_{\odot}}\right)^{0.494} ,
\ee
 where $M$ is the mass 
appearing in the spherically symmetric metric (the mass of the 
galaxy). Notice that having $\alpha$ depending on the mass of each individual galaxy straightforwardly implies that one cannot fit the data for all galactic masses with the same choice of $f(R)$. This make the whole approach highly implausible.

\cite{Capozziello:2007yb,  Capozziello:2006ph, 
Capozziello:2006uv, Capozziello:2004km, Capozziello:2004us} 
computed weak-field limit corrections to the Newtonian 
galactic potential and the resulting rotation curves; when 
matched to galaxy samples,  a best fit yields $n\simeq 1.7$. 
\cite{Martins:2007uf} performed a $\chi^2$ fit using two broader 
samples, finding $n\simeq 2.2$ [see also \cite{Boehmer:2007kx}  
for a variation of this approach focusing on  the constant 
velocity  tails of the rotation curves].
All these values of the parameter $n$ are in violent contrast 
with the bounds  obtained by \cite{Clifton:2005aj,Clifton:2006kc, 
Barrow:2005dn, Zakharov:2006uq} and have been shown to 
violate also the current constraints on the precession of 
perihelia of several Solar System planets \cite{Iorio:2007zz, 
Iorio:2007rk}.  In addition, the consideration of vacuum metrics 
 used in these works in order to model the gravitational field 
of galaxies is highly questionable.

The potential obtained in the weak-field limit of $f(R)$ 
gravity can affect other aspects of galactic dynamics as well: 
the scattering probability of an intruder star and the 
relaxation time of a stellar system were studied by  
\cite{Hadjimichef:1996ax}, originally motivated by  quadratic 
corrections to the  Einstein--Hilbert action.

\subsection{Palatini $f(R)$ gravity and  the conflict with the 
Standard Model}
\label{sec:conflict}

One very important and unexpected shortcoming of Palatini $f(R)$ 
gravity is that it appears to be in conflict with the Standard 
Model of particle physics, in the sense that it introduces 
non-perturbative corrections to the matter action (or the field 
equations) and strong couplings between gravity and matter in 
the local frame and at low energies. The reason why we call this 
shortcoming unexpected is that, judging by the form of the 
action~(\ref{palaction}), Palatini $f(R)$ gravity is, as we 
mentioned, a 
metric theory of gravity in the sense that  matter is only 
coupled minimally to the metric. Therefore, the stress energy 
tensor is divergence-free with respect to the metric covariant 
derivative, the metric postulates \cite{willbook} are fulfilled, 
the theory apparently satisfies the Einstein Equivalence 
principle, and the matter action should trivially reduce to that 
of Special Relativity locally.

Let us see how this conflict comes about. This issue was first 
pointed out in~\cite{Flanagan:2003rb} using Dirac particles for 
the matter action as an example, and later on studied again in 
\cite{Iglesias:2007nv} by assuming that the matter action is 
that of the Higgs field [see also \cite{Olmo:2008ye}]. Both 
calculations use the equivalent Brans--Dicke theory and are 
performed in the Einstein frame.  
Although the use of the Einstein frame has been criticized 
\cite{Vollick:2003ic},\footnote{Note that in the case of 
\cite{Flanagan:2003rb} in which fermions are used as the matter 
fields, one could decide to couple the independent connection 
to them by allowing it to enter the matter action and define the 
covariant derivative (which would be equivalent to assuming that 
the spin connection is as independent variable in a tetrad 
formalism), as noted in~\cite{Vollick:2004ws}. 
Although the results of~\cite{Flanagan:2003rb} 
would cease to hold in this case, this can not be considered a 
problem: clearly in this case we would be talking about a 
different theory, namely metric-affine $f(R)$ gravity 
\cite{Sotiriou:2006qn}.}, this frame is equivalent to the Jordan 
frame and both are perfectly suitable for performing 
calculations~\cite{Flanagan:2004bz} [see also the relevant 
discussion in Sec.~\ref{sec:equiv} and 
\cite{Faraoni:2006fx,Sotiriou:2007zu}].

Nevertheless, since test particles are supposed to follow 
geodesics of the Jordan frame metric, it is this metric which 
becomes approximately flat in the laboratory reference 
frame.  Therefore, when the calculations are performed in the 
Einstein frame they are less transparent since the actual 
effects could be confused with frame effects and vice-versa. 
Consequently, for simplicity and clarity, we present the 
calculation in the Jordan frame, as it appears in 
\cite{Barausse:2007ys}. We begin from the 
action~(\ref{palactionH2d0}), which is the Jordan frame 
equivalent of  Palatini $f(R)$ gravity, and we take matter 
to 
be represented by  a scalar field $H$ (\textit{e.g.,} the Higgs 
boson), the action 
of which reads 
\be\label{higgs} 
S_{M}= \frac{1}{2\hbar}\int d^4 
x \sqrt{-g}\left(g^{\mu\nu}\partial_\mu H \partial_\nu H 
-\frac{m_{\rm H}^2}{\hbar^2} H^2\right) 
\ee 
(in units in which  
$G=c=1$). As an example, we choose $f(R)={\cal R}-\mu^4/{\cal R}$ 
\cite{Carroll:2003wy,Vollick:2003aw}. For this choice of $f$, 
the  potential is  $V(\phi)=2\mu^2 (\phi-1)^{1/2}$. To 
go to 
the local frame, we want to expand the action to second order 
around vacuum. The vacuum of the action~(\ref{palactionH2d0}) 
with~(\ref{higgs}) as a matter action   is $H=0$, $\phi=4/3$ 
[using eq.~(\ref{bdf2})] and $g_{\mu\nu}\simeq 
\eta_{\mu\nu}$ ($\mu^2\sim \Lambda$ acts as an effective 
cosmological constant, so its contribution in the local frame 
can be safely neglected).

However, when one tries to use a perturbative expansion for 
$\phi$, things stop being straightforward: $\phi$ is 
algebraically related to the matter fields as is obvious from 
eq.~(\ref{bdf2}). Therefore, one gets $\delta\phi\sim 
T/\mu^2\sim m_{\rm H}^2 \delta\! H^2/(\hbar^3 \mu^2)$ at 
energies lower than the Higgs mass ($m_{\rm H} \sim 100-1000$ 
GeV). Replacing this expression in the action 
(\ref{palactionH2d0}) perturbed to second order, one immediately 
obtains that the effective action for the Higgs scalar is
\begin{eqnarray}\label{higgs2} 
S_{M}^{\rm effective}&\simeq &\int 
d^4 x 
\sqrt{-g}\frac{1}{2\hbar}\left(g^{\mu\nu}\partial_\mu\delta\! H 
\partial_\nu \delta\! H -\frac{m_{\rm H}^2}{\hbar^2} \delta\! 
H^2\right)\nonumber\\&&\times \left[1+ \frac{m_{\rm H}^2 \delta\! 
H^2}{\mu^2\hbar^3}+\frac{m_{\rm H}^2 (\partial \delta\! 
H)^2}{\mu^4\hbar^3}\right] 
\end{eqnarray}
at energies $E\ll m_{\rm H}$. Taking into account the fact that 
$\mu^2\sim\Lambda\sim H_0^{2}$ where $H_0^{-1}=4000$ Mpc 
is the Hubble radius and $\delta \! H\sim m_{\rm H}$ because 
$E\ll m_{\rm H}$, it is not difficult to estimate the 
order of magnitude of the corrections: at an energy $E=10^{-3}$ 
eV (corresponding to the length scale $L=\hbar/E=2\times 10^{-4}$ 
m), the first correction is of the order $m_{\rm H}^2 \delta\! 
H^2/(\mu^2\hbar^3)\sim (H_0^{-1}/\lambda_{\rm H})^2(m_{\rm 
H}/M_{\rm P})^2\gg1$, where $\lambda_{\rm H}=\hbar/m_{\rm 
H}\sim2\times 10^{-19}-2\times 10^{-18}$ m is the Compton 
wavelength 
of the Higgs and $M_{\rm P}=(\hbar 
c^5/G)^{1/2}=1.2\times 10^{19}$ GeV is the Planck mass.
The second correction is of the order $m_{\rm H}^2 (\partial 
\delta\! H)^2/(\mu^4\hbar^3)\sim (H_0^{-1}/\lambda_{\rm X
H})^2(m_{\rm H}/M_{\rm P})^2 (H_0^{-1}/L)^2\gg 1$. Clearly, 
having such non-perturbative corrections to the local frame 
matter action is unacceptable.

An alternative way to see the same problem would be to replace 
$\delta\phi\sim m_{\rm H}^2 \delta\! H^2/(\hbar^3 \mu^2)$ 
directly in (\ref{palactionH2d0}). Then the coupling of matter 
to gravity is described by the interaction Lagrangian \bea {\cal 
L}_{\rm int}&\sim & \frac{m_{\rm H}^2 \delta\! H^2}{\hbar^3} 
\left(\delta g+\frac{\partial^2 \delta 
g}{\mu^2}\right)\nn\\&\sim & \frac{m_{\rm H}^2 \delta\! 
H^2}{\hbar^3} \delta g\left[1+ 
\left(\frac{H_0^{-1}}{L}\right)^2\right]\,. \eea This clearly 
exhibits the fact that gravity becomes non-perturbative at microscopic 
scales.

It is obvious that the algebraic dependence of $\phi$ on the 
matter fields stands at the root of this problem. We have still 
not given any explanation for the ``paradox'' of seeing such a 
behavior in a theory which apparently satisfies the metric 
postulates both in the $f(R)$ and the Brans--Dicke 
representation. However, this will become clear in 
Sec.~\ref{sec:surfsing}.

\subsection{Exact solutions and relevant constraints}

\subsubsection{Vacuum and non-vacuum exact solutions}
\label{sec:astro}

Let us now turn our attention to exact solutions starting from metric $f(R)$ gravity. We have already mentioned in Sec.~\ref{sec:actionmetric} 
that, as it can be seen easily from the form of the field 
equations~(\ref{metf}), the maximally symmetric vacuum solution 
will be either Minkowski spacetime, if $R=0$ is a root of
eq.~(\ref{metftr}), or de Sitter and anti-de Sitter spacetime, 
depending on the sign of the root of the same equation. Things 
are slightly more complicated for vacuum solutions with less 
symmetry:  by using eq.~(\ref{metf}) it is easy to verify that 
any vacuum solution ($R_{\mu\nu}=\Lambda g_{\mu\nu}$, 
$T_{\mu\nu}=0$) of Einstein's theory with a (possibly vanishing) cosmological 
constant, including black hole solutions, is a solution of 
metric $f(R)$ gravity (except for pathological cases for which 
eq.~(\ref{metftr}) has no roots). However, the converse is not 
true.

For example, when spherical symmetry is imposed, the 
Schwarzschild metric is a solution of metric $f(R)$ gravity if 
$R=0$ in vacuum. If $R$ is constant in vacuo, then 
Schwarzschild-(anti-)de Sitter spacetime is a solution. As we 
have already mentioned though, the Jebsen-Birkhoff theorem 
\cite{wald, weinberg} does not hold in metric $f(R)$ gravity 
[unless, of course one wishes to impose further conditions, such 
as that $R$ is constant \cite{Capozziello:2007id}]. Therefore, 
other solutions can exists as well. An interesting finding is 
that the cosmic no-hair theorem valid in GR and in pure $f(R)$ 
gravity is not valid, in general, in theories of the form 
\be
S=\frac{1}{2\kappa} \int d^4 x \sqrt{-g} \, \left( R +\alpha 
R^2+\beta R_{\mu\nu}R^{\mu\nu} -2\Lambda \right)\, ,
\ee
 for which 
exact anisotropic solutions that continue to inflate 
anisotropically have been found \cite{Barrow:2005qv, 
Barrow:2006xb} [see also \cite{Maeda:1987xf, Kluske:1995vj}]. 
However, isotropization during inflation 
occurs in mixed $f\left( \phi, R \right)$ models 
\cite{Maeda:1988rb}

In addition to the exact cosmological solutions explored for the 
purpose of explaining the current cosmic acceleration [see, 
{\em e.g.}, \cite{Abdalla:2004sw, Clifton:2006jh, 
Clifton:2007ih, Clifton:2006kc, Clifton:2005at, Barrow:2005dn}  
and \cite{Modak:2005xz, Capozziello:2008ch, 
Capozziello:2007wc, Vakili:2008ea} for 
an approach based on Noether 
symmetries; see \cite{Carloni:2005ii} for bouncing solutions 
and the conditions that they satisfy], exact 
spherically symmetric solutions of metric $f(R)$ gravity 
have been explored in the literature, with most recent 
studies being motivated by the need to understand the 
weak-field limit of cosmologically-motivated  theories. 

Regarding non-vacuum solutions, the most common matter source is a perfect fluid. Fluid dynamics in metric $f(R)$ gravity was studied in \cite{Maartens:1994pb,Taylor:1995dy,Rippl:1995bg,MohseniSadjadi:2007qk}.
Spherically symmetric solutions were found in
 \cite{Whitt:1985ki, Mignemi:1991wa, Clifton:2006ug, 
Clifton:2006jh, Bustelo:2006ms, Bronnikov:2005zr,
Bronnikov:2006kh, Bronnikov:2005kr,  
Capozziello:2007id,Multamaki:2007jk,Multamaki:2006zb, 
Multamaki:2006ym}. We regret not being able to present these 
solutions extensively here due to space limitations and refer 
the reader to the literature for more details.

Stability issues for spherically symmetric 
solutions were discussed in \cite{Seifert:2007fr}. In 
the  theory 
\be \label{Whitt}
S=\int d^4 x \, \frac{\sqrt{-g}}{\kappa} \left[ R-\alpha R^2 
-\beta R_{\mu\nu}R^{\mu\nu} \right] +\epsilon \chi ,
\ee
where $\alpha, \beta$, and $\epsilon $ are constants and $\chi$ 
is the Gauss--Bonnet invariant, the Schwarzschild metric is a 
solution and the  stability 
of Schwarzschild black holes was studied in \cite{Whitt:1985ki}.  
Surprisingly, it was found that the massive ghost graviton 
(``poltergeist'') present in this theory stabilizes small mass 
black holes against quantum instabilities [see also 
\cite{Myers:1988ze, Myers:1989kt}]. In the case 
$\beta=\epsilon=0$, which reduces the theory~(\ref{Whitt}) to a 
quadratic $f(R)$ gravity, the stability criterion 
found in \cite{Whitt:1985ki} reduces to $\alpha<0$, which 
corresponds again to $f''(R)> 0$. For $\alpha=0$ we recover 
GR, in which black holes are stable classically 
(but not quantum-mechanically, due to Hawking radiation and 
their negative specific heat, a feature that persists in $f(R)$ 
gravity), so the classical stability condition for Schwarzschild 
black holes can be enunciated as $f''(R) \geq 0$.

Let us now turn our attention to Palatini $f(R)$ gravity. In 
this case things are simpler in vacuo: as we saw in 
Sec.~\ref{sec:palatinifield}, the theory reduces in this case (or 
more 
precisely even for matter fields with $T=$const., where $T$ is 
the trace of the stress energy tensor) to GR with a cosmological 
constant, which might as well be zero for some models 
\cite{Barraco:1998eq, Ferraris:1994af, Borowiec:1996kg, 
Ferraris:1992dx}. Therefore, it is quite straightforward that 
Palatini $f(R)$ gravity will have the same vacuum solutions as 
GR with a cosmological constant. Also, the Jebsen-Birkhoff 
theorem \cite{wald, weinberg} is valid in the Palatini formalism 
\cite{Kainulainen:2006wz, Barausse:2007pn, Barausse:2007ys, 
Barausse:2008nm}.

Cosmological solutions in quadratic gravity were obtained in 
\cite{Shahid-Saless:1990aaa, Shahid-Saless:1991aaa}. Spherically 
symmetric  interior solutions in the Palatini formalism can be 
found by 
using the generalization of the Tolman-Oppenheimer-Volkoff 
equation valid for these theories, which was found in 
\cite{Bustelo:2006ms, Kainulainen:2006wz,Barraco:2000dx}. 
Indeed, such solutions have been found and matched with the 
unique exterior (anti-)de Sitter solution \cite{Barraco:2000dx, 
Barraco:1998hc, Bustelo:2006ms, Kainulainen:2006wz, 
Kainulainen:2007bt}. Nevertheless, a matching between exterior 
and interior which can lead to a sensible solution throughout 
spacetime is not always feasible and this seems to have serious 
consequences for the viability of $f(R)$ gravity 
\cite{Barausse:2007pn, Barausse:2007ys, Barausse:2008nm}. This 
is discussed extensively in the next section.

Let us close this section with some remarks on black hole 
solutions. As is clear from the above discussion, all black hole 
solutions of GR (with a cosmological constant) will also be 
solutions of $f(R)$ in both the metric and the Palatini 
formalism [see also \cite{Psaltis:2007cw,Barausse:2008xv}]. 
However, in the Palatini formalism they will constitute the 
complete set of black hole solutions of the theory, whereas in 
the metric formalism other black hole solutions can exist in 
principle, as the Jebsen-Birkhoff theorem does  not hold. For a 
discussion on black hole entropy in $f(R)$ gravity  see 
\cite{Jacobson:1993vj,Jacobson:1995uq, 
Vollick:2007fh}.\footnote{See 
also \cite{Eling:2006aw} for a derivation of the field  
equations of metric $f(R)$ gravity based on thermodynamical  
arguments applied to local Rindler horizons.}

\subsubsection{Surface singularities and the incompleteness of 
Palatini $f(R)$ gravity} \label{sec:surfsing}

In Secs.~\ref{sec:weakPalatini}, \ref{sec:Cauchy} and~\ref{sec:conflict}, we 
already spotted three serious shortcomings of Palatini $f(R)$ 
gravity, namely the algebraic dependence of the post-Newtonian 
metric on the density, the complications with the initial 
value problem 
in the presence of matter, and a conflict with particle physics. In 
this section we will study static spherically symmetric interior 
solutions and their matching to the unique exterior with the 
same symmetries, the Schwarzschild-de Sitter solution, along 
the lines of \cite{Barausse:2007pn, Barausse:2007ys, 
Barausse:2008nm}.  As we will see, the three problems mentioned 
earlier are actually very much related and stem from a very 
specific characteristic of Palatini $f(R)$ gravity, which the 
discussion of this section will help us pin down.

A common way of arriving to a full description of a spacetime 
which includes matter is to solve separately the field equations inside  
and outside the sources, and then match the interior 
and exterior solutions using appropriate junction conditions 
[called Israel junction condition in GR \cite{israeljunct}]. This is 
what 
we are going to attempt here. We already know the exterior 
solution so, for the moment, let us focus on the interior. Since 
we assume that the metric is static and spherically symmetric, 
we 
can write it in the form 
\begin{equation} 
\label{metric2} 
ds^2 
\equiv -e^{A(r)}{\rm d}t^2 + e^{B(r)}{\rm d}r^2 + r^2{\rm 
d}\Omega^2 \,. 
\end{equation} 
We can then replace this metric in 
the field equations of Palatini $f(R)$ gravity, preferably in 
eq.~(\ref{eq:field}) which is the simplest of all the possible 
reformulations. Assuming also a perfect fluid description for 
matter with $T_{\mu\nu}=\left( \rho+P \right)u_\mu 
u_{\nu}+Pg_{\mu\nu}$, 
where $\rho$ is the energy density, $P$ is the pressure, and 
$u^\mu$ is the fluid 4-velocity, and representing $d/dr$ with a 
prime,\footnote{In this section we modify our standard notation 
for economy 
and a prime  denotes differentiation with 
respect to the radial coordinate instead of differentiation with 
respect to the 
argument of the function.} one arrives at the 
equations
\bea\label{eq:Ap}
A' & = & \frac{-1}{1 + \gamma} \left(
	\frac{1 - e^B}{r} - \frac{e^B}{F}8\pi GrP
			+ \frac{\alpha}{r}
		\right), \\
\label{eq:Bp}	B' & = &  \frac{1}{1 + \gamma} \left(
			\frac{1 - e^B}{r} + \frac{e^B}{F}8\pi Gr\rho
			+ \frac{\alpha + \beta}{r}
		\right),\\
\alpha  &\equiv&  r^2 \left[
			\frac{3}{4}\left(\frac{F'}{F} \right)^2  
+ \frac{2F'}{rF}
			+ \frac{e^B}{2} \left( {\cal R}  - 
\frac{f}{F} \right)
		\right], \\
	\beta  &\equiv&  r^2 \left[
	 		\frac{F''}{F} - 
\frac{3}{2}\left(\frac{F'}{F}\right)^2
		\right],\qquad
\gamma \equiv \frac{rF'}{2F} ,\label{eq:abc}
\eea
where $F\equiv \partial f/\partial R$. 
To determine an interior solution we need a generalization of 
the Tolman-Oppenheimer-Volkoff (TOV) hydrostatic equilibrium 
equation. This has been derived for Palatini $f(R)$ gravity in 
\cite{Bustelo:2006ms, Kainulainen:2006wz,Barraco:2000dx}:  
Defining $m_{\rm tot}(r)\equiv r(1 - e^{-B})/2$ and using 
Euler's equation
\begin{equation}\label{eq:euler} 
P'=-\frac{A'}{2}\left( P+\rho \right), 
\end{equation} 
one can use eqs.~(\ref{eq:Ap}) and (\ref{eq:Bp}) 
to arrive to the generalised TOV equations: 
\bea \label{eq:OVB} 
\!\!\!\!\!\!\!\!\!\!\! P'\!\!& =&\!\!
 -\frac{1}{1 + \gamma} \frac{\left(\rho + P \right)}{r(r - 
2m_{\rm tot})}\times\\
&&\qquad\times\left( m_{\rm tot} + \frac{4\pi r^3 P}{F}  
          - \frac{\alpha}{2} (r - 2m_{\rm tot} ) \right),
\nn\\
\label{eq:mass}
\!\!\!\!\!\!\!\!\!\!\!m_{\rm tot}' \!\!&=& \!\!\frac{1}{1 + 
\gamma} \bigg(\frac{4\pi r^2\rho}{F} + 
\frac{\alpha\!+\!\beta}{2}
- \frac{m_{\rm tot}}{r}(\alpha\! +\! \beta\! - \!\gamma) \bigg).
\eea

We now have three differential equations, namely (\ref{eq:euler}), 
(\ref{eq:OVB}) and (\ref{eq:mass}), and four unknown functions, 
namely $A$, $m_{\rm tot}$ (or $B$), $P$, and $\rho$. The missing 
piece is the information about the microphysics of the matter 
configuration under investigation.  In the case of a perfect 
fluid, this is effectively given by an equation of state (EOS). A 
one-parameter EOS relates the pressure directly to the energy 
density, {\em i.e.,~}$ P=P(\rho)$. A simple form of such an EOS 
is  a polytropic equation of state  $ P=k\rho_0^\Gamma$, 
where $\rho_0$ is the rest-mass density and $k$ and $\Gamma$ are 
constants. This is the case that we will consider here. Note 
that the rest-mass density can be expressed in terms of the 
energy density $\rho$ and the internal energy $U$ as 
$\rho_0=\rho-U$. Assuming an adiabatic transformation and using 
the first law of thermodynamics, one can express the internal 
energy in terms of the pressure, {\em i.e.}, $U = P/(\Gamma-1)$. 
Therefore, the polytropic EOS can be rewritten as 
\begin{equation} 
\rho=\left(\frac{P}{k}\right)^{1/\Gamma}+\frac{P}{\Gamma-1}, 
\end{equation} 
giving a direct link between $ P $ and $\rho$.

Without specifying the interior solution, we can already  
examine the appropriate matching conditions needed. One needs 
continuity of the metric and  of $A'$ on the surface 
of the matter configuration ($A$ is given by a second order 
differential equation). Since we know that the exterior 
solution is unique and it is the Schwarzschild-de Sitter 
solution with a cosmological constant equal to ${\cal R}_0/4$, 
where ${\cal R}_0$ is the vacuum value of ${\cal R}$ (see 
Sec.~\ref{sec:palatinifield}), we can directly write for the 
exterior
 \be
\exp(-B(r))=\ell\exp(A(r))=1-2m/r-{\cal R}_0 r^2/12,
\ee
 where  $\ell$ and $m$ are integration constants to be fixed by
requiring  continuity of the metric coefficients across the 
surface, which is implicitly defined by $r =
r_{\rm out}$ where $P = \rho = 0$. 
Using the  definition of $m_{\rm tot}(r)$ this gives, in the 
exterior,
 \begin{equation} \label{eq:m_ext} 
m_{\rm tot}(r) = m+\frac{r^3}{24}{\cal R}_0 . 
\end{equation} 
On the other hand, based on the exterior solution, one gets on 
the surface
\be
\label{eq:requested_A_prime}
A'(r_{\rm out})=\frac{2 \left(r_{\rm out}^3 {\cal R}_0-12 m\right)}{r_{\rm out} 
\left({\cal R}_0 r_{\rm out}^3-12 r_{\rm out}+24 m\right)}\:.
 \ee

Assuming that, approaching the surface from the interior, $A$ 
and 
$m_{\rm  tot}$ indeed take the correct values required for the 
matching,  it can be shown that continuity of $A'$ across the 
surface requires $F'(r_{\rm out})=
0$ for $ r\to r_{\rm out}^-$ \cite{Barausse:2007pn}. 
Additionally,  if this is the case then \cite{Barausse:2007pn}
\begin{equation}
\label{eq:mprime} 
 m_{\rm tot}'(r _{\rm out})=\frac{2 F_0 {\cal R}_0 r_{\rm 
out}^2+\left(r_{\rm 
out}^3 {\cal R}_0 -8 m_{\rm tot}\right) {\cal C}'}{16 F_0},
 \end{equation}
where
\be
{\cal{C}}=\frac{dF}{dP} \left( P+\rho \right)= 
\frac{dF}{d\rho}\frac{d\rho}{dP} \left( P +\rho \right) .
\label{eq:calC}
\ee

Let us now examine the behavior of $m'_{\rm tot}$ at the surface 
for different values of the polytropic index $\Gamma$. For 
$1<\Gamma<3/2$, ${\cal C}'=d{\cal C}/dP \,P'\propto d{\cal 
C}/dP\,\left( P +\rho \right)\to 0$ at the surface so that 
the expression~(\ref{eq:mprime}) is finite and it gives 
continuity of $m_{\rm 
tot}'$ across the surface [\textit{cf.} eq.~(\ref{eq:m_ext})]. 
However, for $3/2<\Gamma<2$, ${\cal C}'\to\infty$ as the surface 
is approached, provided that $dF/d{\cal R}({\cal R}_0)\neq0$ and 
$d{\cal R}/dT(T_0)\neq0$ (note that these conditions are 
satisfied by generic forms of $f({\cal R})$, {\em 
i.e.,~}whenever 
an ${\cal R}^2$ term or a term inversely proportional to ${\cal 
R}$ is present).  Therefore, even though $m_{\rm tot}$ remains 
finite (as can be shown using the fact that $ P'=0$ at the 
surface), the divergence of $m_{\rm tot}'$ drives to infinity 
the Riemann tensor of the metric, $R_{\mu\nu\sigma\lambda}$, and 
curvature invariants, such as $R$ or 
$R^{\mu\nu\sigma\lambda}R_{\mu\nu\sigma\lambda}$, as can easily 
be checked.\footnote{This fact seems to have been missed in 
\cite{Barraco:2000dx}.} Clearly, such a singular behaviour is 
bound to give rise to unphysical phenomena, such as infinite 
tidal forces at the surface (\textit{cf.} the geodesic deviation 
equation) which would destroy anything present there. We are, 
therefore, forced to conclude that no physically relevant 
solution exists for any polytropic EOS with $3/2<\Gamma<2$.

The following points about the result just presented should be 
stressed: 
\begin{itemize} 
\item The sufficient condition for the 
singularity to occur is that a polytropic EOS with 
$3/2<\Gamma<2$ should adequately describe just the outer layer 
of the 
matter configuration (and not necessarily the whole 
configuration). 
\item In practice, there is no dependence of the 
result on the functional form of $f({\cal R})$ [a few 
unrealistic exceptions can be found in \cite{Barausse:2007pn}]
so 
what is revealed is a generic aspect of Palatini $f(R)$ gravity 
as a class of theories. 
\item The singularities discussed are 
not coordinate, but true singularities, as can be easily 
verified by checking that curvature invariants diverge. 
\item The only 
assumptions made regard the EOS and the symmetries. Thus, the 
result applies to all regimes ranging from Newtonian to 
strong gravity. 
\end{itemize}

Let us now interpret these results. Obviously, one could object 
to the use of the polytropic EOS. Even though it is 
extensively used  for simple stellar models, it is 
not a very realistic description for stellar configurations. 
However, one does not necessarily have to refer to stars in order 
to check whether the issue discussed here leaves an observable 
signature. Consider two very well known matter configurations 
which are exactly describes by a polytropic EOS: a monoatomic 
isentropic gas and a degenerate non-relativistic electron gas. 
For both of those cases $\Gamma=5/3$, which is well within the 
range for which the singularities appear. Additionally, both of 
these configuration can be very well described even with 
Newtonian gravity. Yet, Palatini $f(R)$ gravity fails to provide 
a reasonable description. Therefore, one could think of such 
matter 
configurations as gedanken experiments which reveal that 
Palatini 
$f(R)$ gravity is at best incomplete \cite{Barausse:2007pn, 
Barausse:2007ys, Barausse:2008nm}.

On the other hand, the use of the polytropic EOS requires a 
perfect fluid approximation for the description of matter. 
One may, therefore, wish to question whether the length scale 
on which the tidal forces become important is larger than the 
length scale for which the fluid approximation breaks down 
\cite{Kainulainen:2007bt}. However, quantitative results for  
tidal forces have been given in~\cite{Barausse:2007ys}, and it 
has been shown that the length scales at which the tidal forces become relevant are 
indeed larger than it would be required for the fluid 
approximation to break down. The observable consequences on 
stellar configurations have also been discussed there. To this,  
one could also add that a 
theory which  requires a full description of the microscopic 
structure of the system in order to provide a macroscopic 
description of the dynamics is not very appealing anyway.

In any case, it should be stated that the problem discussed is 
not specific to the polytropic EOS. The use of the latter only 
simplifies the calculation and allows an analytic approach. The 
root of the problem actually lies with the differential 
structure of Palatini $f(R)$ gravity.

Consider the field equations in the form~(\ref{eq:field}): it is 
not difficult to notice that these 
are second order partial differential equations in the metric. 
However, since $f$ is a function of ${\cal R}$, and ${\cal R}$ 
is 
an algebraic function of $T$ due to eq.~(\ref{paltrace}), the right 
hand side of eq.~(\ref{eq:field}) includes second derivatives 
of $T$. Now, $T$, being the trace of the stress energy tensor, 
will include up to first order derivatives of the matter fields 
(assuming that the matter action has to lead to second order 
field equations when varied with respect to the matter fields). 
Consequently, eq.~(\ref{eq:field})  can be up to third order in 
the matter fields!

In GR and most of its alternatives, the field equations are 
only of 
first order in the matter fields. This guarantees that gravity 
is a cumulative effect: the metric is generated by an integral 
over the 
matter sources and, therefore, any discontinuities (or even 
singularities) in the latter and their derivatives, which are 
allowed, will not become discontinuities or singularities of the 
metric, which are not allowed [see \cite{Barausse:2007ys} for a 
detailed discussion]. This characteristic is not present in 
Palatini $f(R)$ gravity and creates an algebraic dependence of 
the metric on the matter fields.

The polytropic description not only does not cause this problem 
but, as a matter of fact, it makes it less acute than it is, 
simply because in the fluid approximation the stress-energy 
tensor does not include derivatives of the matter fields and effectively ``smoothes out'' the matter distribution. 
Actually, the fact that the metric is extremely sensitive to 
rapid changes of the matter field has been exhibited also in the 
interior of stars described by realistic tabulated EOSs in 
\cite{Barausse:2007pn}.

One should not be puzzled by the fact that this awkward 
differential structure of Palatini $f(R)$ gravity is not 
manifest in the $f({\cal R})$ formulation of the theory [and the 
field eqs.~(\ref{palf12}) and (\ref{palf22})]. We have already 
mentioned that the independent connection is actually an 
auxiliary field and the presence of auxiliary fields can always 
be misleading when it comes to the dynamics. In fact, it just 
takes a closer look to realize that the Palatini $f({\cal R})$ 
action does not contain any derivatives of the metric and is 
 of only first order in the derivatives of the connection. Now,  
given 
that the connection turns out to be an auxiliary field and can 
be algebraically related to derivatives of the matter and of the 
metric, it no longer comes as a surprise that the outcome is a 
theory with higher differential order in the matter than the 
metric.

By now, the fact that the post-Newtonian metric turns out to be 
algebraically dependent on the density, as discussed in  
Sec.~\ref{sec:weakPalatini}, should no longer sound surprising: it is 
merely a manifestation of the problem discussed here in the weak 
field regime. The fact that  it is unlikely that the Cauchy 
problem will be well-formulated in the presence of matter  
 also originates from the same feature of Palatini $f(R)$ 
gravity,  as already mentioned. Similarly, the fact that a 
theory which manifestly 
satisfies the metric postulates and, therefore, is expected to 
satisfy the Equivalence Principle, actually exhibits  
unexpected phenomenology in local non-gravitational experiments 
and conflicts with the Standard Model, as shown in 
Sec.~\ref{sec:conflict},  also ceases to be a puzzle: the 
algebraic 
dependence of the connection on the derivatives of matter fields 
(as the former is an auxiliary field) makes the matter enter the 
gravitational action through the ``back door". This introduces 
strong couplings between matter and gravity and 
self-interactions of the matter fields which manifest 
themselves in 
the local frame. Alternatively, if one completely 
eliminates the connection (or the scalar field in the equivalent 
Brans--Dicke representation) at the level of the action, or 
attempts to write down an action which leads to the field 
eqs.~(\ref{eq:field}) directly through metric variation, then 
this action would have to include higher order derivatives of 
the matter field and self-interactions in the matter sector. In 
this sense, the $f({\cal R})$ representation is simply 
misleading 
[see also \cite{Sotiriou:2007zu} for a general discussion of 
representation issues in gravitational theories].

\subsection{Gravitational waves in $f(R)$ gravity}

By now it is clear that the metric tensor of $f(R)$ gravity 
contains, in addition to the usual massless spin 2 graviton, a 
massive scalar that shows up in gravitational waves in the 
metric version of these theories (in the Palatini version, this 
scalar is not dynamical and does not propagate).  A scalar 
gravitational wave mode   is 
familiar from scalar-tensor gravity \cite{willbook}, to which 
$f(R)$ gravity is 
equivalent.  Because this scalar is massive, it propagates at a 
speed lower than the speed of light and of massless tensor modes 
and is, in principle, detectable in the arrival times of 
signals from an exploding supernova when gravitational wave 
detectors are sufficiently sensitive [this possibility has 
been pointed out as a discriminator between 
Tensor-Vector-Scalar theories and GR  
\cite{Kahya:2007zy}]. This massive scalar mode is longitudinal 
 and is of dipole nature to lowest 
order \cite{willbook, Corda:2007hi}. The 
study of its generation, propagation, and detection falls within 
the purview of 
scalar-tensor gravity \cite{willbook}. The propagation of 
gravitational waves in the specific model $f(R)=R^n$ was studied 
in \cite{Mendoza:2006hs} where the massive scalar mode is, 
however, missed.

The generation of  gravitational waves specifically in $f(R)$ 
gravity has not  received much attention in the literature. Even 
though the fact  that the black hole solutions of GR will also 
be solutions of metric $f(R)$ gravity (without the  converse 
being true) implies that determining the geometry around  a 
black hole is unlikely to provide evidence for such  
modifications of gravity \cite{Psaltis:2007cw}, 
solutions describing perturbed black holes  do behave 
differently and could, therefore, leave a detectable  imprint on 
gravitational wave radiation \cite{Barausse:2008xv}. Note the 
analogy to the fact that cosmological FLRW solutions
are shared by most gravitational theories,  but cosmological 
perturbations reveal more about the  underlying theory of 
gravity than  the exact solutions 
themselves.   Additionally, gravitational radiation from binary 
systems would  probably  be  more revealing than that coming 
from perturbed black holes when it comes to modified gravity.

The detection of gravitational waves generated in the theories 
$f(R)=1/R$ [already ruled out by Solar System data 
\cite{Clifton:2005aj, Clifton:2006kc, Barrow:2005dn}]  and 
$f(R)=R+aR^2$ were studied in \cite{Corda:2007tz, Corda:2007hi} 
and \cite{Corda:2007nr}, respectively.

The study of  cosmological gravitational 
waves in $f(R)$ gravity is perhaps more promising than that of 
astrophysically generated waves. The stochastic gravitational 
wave background produced in the early universe was analyzed in 
\cite{Capozziello:2007zz, Capozziello:2007vd}.  The authors of 
this last reference consider the model $f(R)=R^{1+\delta}$ 
and 
derive an evolution equation for the metric 
perturbations $h_{ij}=h(t) \, \mbox{e}^{i k_l k^l} e_{ij}$ in a 
background FLRW universe with 
scale factor $a(t)=a_0 \left( \frac{t}{t_0} \right)^n $:
\be
\ddot{h}+\frac{\left( 3n-2\delta\right)}{t} \, \dot{h} +k^2 a_0 
\left( \frac{t}{t_0} \right)^2n \, h=0 .
\ee
This can be solved in terms of Bessel functions; plots of these 
wave amplitudes are reported in \cite{Capozziello:2007vd} for 
various values of the parameter $\delta$, but 
the limit $0\leq \delta < 7.2\cdot 10^{-19}$ obtained by 
\cite{Clifton:2005aj, Clifton:2006kc, Barrow:2005dn} leaves 
little hope of detecting $f(R)$ effects in the gravitational 
wave background.

\cite{Ananda:2007xh} give a covariant and gauge-invariant 
description of gravitational waves in a perturbed FLRW universe 
filled with a barotropic perfect fluid in the toy model $f(R)=R^n$. The 
perturbation equations are solved  (again, in terms of Bessel 
functions of the first and second kind) in the approximation of 
scales much larger or much smaller than the Hubble radius 
$H^{-1}$, finding a high sensitivity of the tensor modes 
evolution to the value of the parameter $n$. In particular, a 
tensor mode 
is found that grows during the radiation-dominated era, with 
potential implications for detectability in advanced space 
interferometers. This study, and others 
of this kind expected to appear in future literature, are in the 
spirit of discriminating between dark energy and dark gravity, 
or even between different $f(R)$ theories (if this class is 
taken seriously), when gravitational wave observations will be 
available: as already remarked, this is not possible by 
considering only unperturbed FLRW solutions.

\section{Summary and Conclusions}
\subsection{Summary}

While we have presented $f(R)$ gravity as a class of 
toy theories, various authors elevate modified gravity, in one 
or the other of its  incarnations corresponding to specific choices of the function 
$f(R)$, to  the role  of a fully realistic model to be compared 
in detail with  cosmological observations, and to be 
distinguished from other models.  A large fraction of the works 
in the literature is actually devoted to 
specific models corresponding to definite choices of the 
function $f(R)$, and to specific parametrizations. 

Besides the power law and power series of $R$ models which we have already mentioned extensively, some other typical examples are functions which contain terms such as $\ln (\lambda R)$ \cite{Nojiri:2003ni,PerezBergliaffa:2006ni}  or ${\rm e}^{\lambda R}$ \cite{Carloni:2005ii,Song:2006ej,Bean:2006up,Abdelwahab:2007jp} or are more involved functions of $R$, such as $ f(R)=R - a\left( R-\Lambda_1 \right)^{-m} +b 
\left( R 
-\Lambda_2 \right)^n $ with $n,m,a,b >0$ \cite{Nojiri:2003ft} . 
Some models have actually been tailored to pass all or most of 
the known constraints, such as  the one proposed in 
\cite{Starobinsky:2007hu} where  $f(R)=R+\lambda 
R_0[(1+R^2/R_0^2)^{-n}-1]$ with  $n,\lambda> 0$ and $R_0$ being 
of the  order of $H_0^2$.
Here we have tried 
to avoid considering specific  models and we have attempted to 
collect  general, 
model-independent results, with  the viewpoint that these 
theories are to be seen more as toy theories than definitive and 
realistic models.

We are now ready to summarize the main results on $f(R)$ 
gravity. On the theoretical side, we have explored all three versions of $f(R)$ gravity: metric, Palatini and metric-affine. Several issues concerning dynamics, degrees of freedom, matter couplings, {\em etc.} have been extensively discussed. The dynamical equivalence between both metric/Palatini $f(R)$ gravity and Brans--Dicke theory has been, and continues to be, 
a very useful tool to study these theories given some 
knowledge of the aspects of interest in scalar-tensor gravity. At the same time, the study of $f(R)$ gravity itself has provided new insight in the two previously unexplored cases of Brans--Dicke theory with $\omega_0=0$ and $\omega_0=-3/2$. We have also considered most of the applications of $f(R)$ gravity to both cosmology and astrophysics. Finally, we have explored a large number of possible ways to constrain $f(R)$ theories and check their viability. In fact,
many avatars of $f(R)$ have been shown to be 
subject to potentially fatal troubles, such as a grossly  
incorrect post-Newtonian limit, short time scale instabilities, the absence of a matter era, conflict with particle physics or astrophysics, {\em etc.}

To avoid repetition, we will not attempt to summarize here all of the theoretical issues, the applications or the constraints discussed. This, besides being redundant, would not be very helpful to the reader, as, in most cases, the insight gained cannot be summarized in a sentence or two. Specifically, some of the constraints that have been derived in the literature are not model or parametrization independent (and the usefulness of some parametrizations is questionable).  This does not allow for them to be expressed in a straightforward manner through simple mathematical equations applicable directly to a general function $f(R)$. Particular examples of such constraints are those coming from cosmology (background evolution, perturbations, {\em etc.}).  

However, we have encountered cases in which clear-cut viability criteria are indeed easy to derive. We would, therefore, like to make a specific mention of those. A brief list of quick-and-easy-to-use results is:
\begin{itemize}

\item In metric $f(R)$ gravity, the Dolgov-Kawasaki instability 
is avoided if and only if $f''(R) \geq 0$. The stability condition of de 
Sitter space is expressed by eq.~(\ref{dSonly200}).

\item Metric $f(R)$ gravity might pass the weak-field limit test 
and at the same time constitute an alternative to dark energy only if the chameleon mechanism is effective---this restricts 
the possible forms of the function $f(R)$ in a way that can not 
be specified by a simple formula. 

\item Palatini $f(R)$ gravity suffered multiple deaths, due to 
the differential structure of its field equations. These conclusions 
are essentially model-independent. (However, this theory could 
potentially be fixed by adding extra terms quadratic in the 
Ricci and/or Riemann tensors, which would raise the order of the 
equations.)

\item Metric-affine gravity as an extension of the Palatini 
formalism is not sufficiently developed yet. At the moment of 
writing, it is not clear whether it suffers or not of the same 
problems that afflict the Palatini formalism.
\end{itemize}

Of course, as already mentioned, the situation is often more involved and cannot be 
summarized with a quick recipe. We invite the reader to consult 
the previous sections and especially the references that 
they contain.

\subsection{Extensions  and new perspectives on $f(R)$ gravity}

We have treated $f(R)$ gravity here as a toy theory and, as  stated in the Introduction, one of its merits is its 
relative 
simplicity. However, we have seen a number of viability issues 
related to such theories. One obvious way to address this issue 
is to generalize the action even further in order to avoid these 
problems, at the cost of increased complexity. Several extensions 
of $f(R)$ gravity exist. Analyzing them in detail goes beyond 
the scope of this review, but let us make a brief mention of the 
most straightforward of them.

We have already discussed the possibility of having higher order 
curvature invariants, such as $R_{\mu\nu}R^{\mu\nu}$, in the 
action. In fact, from a dimensional analysis perspective, the 
terms $R^2$ and $R_{\mu\nu}R^{\mu\nu}$ should appear at the same 
order. However, theories of this sort seem to be burdened with 
what is called the Ostrogradski instability 
\cite{Woodard:2006nt}. Ostrogradski's theorem states  that 
there is a linear instability in the Hamiltonians 
associated with Lagrangians which depend upon higher than first 
order derivatives in such a way that the dependence cannot be 
eliminated by partial integration \cite{ostro}. $f(R)$ gravity 
seems to be the only case that manages to avoid this theorem 
\cite{Woodard:2006nt} and it obviously does not seem very 
appealing to extend it in a way that will spoil 
this.\footnote{However, one could consider adding a function of 
the 
Gauss--Bonnet invariant ${\cal G}=R^2-4 
R_{\mu\nu}R^{\mu\nu}-R_{\alpha\beta\mu\nu}R^{\alpha\beta\mu\nu}$ 
\cite{Nojiri:2005jg,Cognola:2006eg}.}

The alert reader has probably noticed that the above holds true 
only for 
metric $f(R)$ gravity. In Palatini $f(R)$ gravity (and 
metric-affine $f(R)$ gravity), as it was mentioned earlier, one 
could add more dynamics to the action without having to worry about 
 making it second order in the fields. Recall that, in 
practice, the 
independent connection is an auxiliary field. For 
instance, the term ${\cal R}_{\mu\nu}{\cal R}^{\mu\nu}$ still 
contains only first derivatives of the connection. In fact, 
since we have traced the root of some of the most crucial 
viability issues of Palatini $f(R)$ gravity to the lack of 
dynamics in the gravity sector, such generalizations could 
actually help by promoting the connection from the role of an 
auxiliary 
field to that  of a truly dynamical field 
\cite{Barausse:2007ys}. Such 
generalizations have been considered in \cite{Li:2007xw}.

Another extension of metric $f(R)$ gravity that appeared 
recently is that in which the action includes also an  
explicit coupling 
between $R$ and the matter fields. In \cite{Bertolami:2007gv, Boehmer:2007fh, Bertolami:2007vu} 
the following action was considered:
\be \label{BBHLaction} 
S=\int d^4x\, \sqrt{-g} 
\left\{ \frac{f_1(R)}{2} +\left[ 1+\lambda f_2(R) \right] L_m 
\right\} , 
\ee 
where $L_m$ is the matter Lagrangian and $f_{1,2} 
$ are ({\em a priori} arbitrary) functions of the Ricci 
curvature $R$.
Since the matter is not minimally coupled to $R$, such theories 
will not lead to energy conservation and will generically 
exhibit a violation of the Equivalence Principle (which could 
potentially be controlled by the parameter $\lambda$).

The motivation for considering such an action spelled out in 
\cite{Bertolami:2007gv} was that the non-conservation of energy 
could lead to extra forces, which in turn might give rise to 
phenomenology similar to MOdified Newtonian Dynamics (MOND) \cite{Milgrom:1983ca} on 
galactic scales. Other variants  of this action have also been 
considered elsewhere:
 in \cite{Nojiri:2004bi},  as  an alternative to dark energy by setting $f_1(R)=R$ 
and keeping only the nonminimal coupling of matter to the Ricci 
curvature; in \cite{Mukohyama:2003nw, 
Dolgov:2003fw}, where  the idea of making 
the kinetic term of a (minimally 
coupled) scalar field dependent on the curvature, while keeping 
$f_1(R)=R$ was exploited in attempts to 
cure the  cosmological constant problem; in 
\cite{Bertolami:2007vu} the consequences of such a theory for  
stellar equilibrium were studied; finally, 
 generalized constraints in order to   
avoid the instability  discussed in 
Sec.~\ref{sec:stabilitymetricsubsub} were derived 
in~\cite{Faraoni:2007sn}. The viability of theories with such 
couplings between $R$ and  matter is still under investigation. 
However, the case in which both $f_1$ and $f_2$ are linear has 
been shown to be non-viable~\cite{Sotiriou:2008dh} and, for the 
more general case of the action~(\ref{BBHLaction}), serious 
doubts have been expressed~\cite{Sotiriou:2008it} on whether extra 
forces are indeed  present in galactic environments and, 
therefore,  whether this theory can really account for the 
MOND-like phenomenology that initially motivated its use in 
\cite{Bertolami:2007gv} as a substitute for dark matter. 

One could also consider extensions of $f(R)$ gravity in which 
extra fields appearing in the action couple to 
different curvature 
invariants. A simple example with a scalar field is 
the action~(\ref{dSonly2}), which is sometimes dubbed {\em 
extended 
quintessence} \cite{Perrotta:1999am}, similarly to the extended 
inflation realized with Brans--Dicke theory. However,  
such generalizations lie beyond the scope of this review.

Finally, it is worth mentioning a different perspective on 
$f(R)$ gravity. It is common in the literature that we 
reviewed here to treat $f(R)$ gravity as an {\em exact} theory: 
the generalized action is used to derive field equations, the 
solutions of which describe the exact dynamics of the 
gravitational field (in spite of the fact that the action might 
be only an approximation and the theory merely a toy theory). A 
different  approach \cite{Bel:1985zz,Simon:1990ic} which was recently revived in 
\cite{DeDeo:2007yn} is 
that of treating metric $f(R)$ as an effective field theory. 
That is, to assume that the extra terms are an artifact of some 
expansion of which we are considering only the leading order 
terms. Now, when we consider a correction to the usual 
Einstein--Hilbert term, this correction has to be suppressed 
by some coefficient. This approach assumes that this coefficient 
controls the order of the expansion and, therefore, the field 
equations and their solutions are only to be trusted to the 
order with which that coefficient appears in the action (higher 
orders are to 
be discarded). Such an approach is based on two assumptions: 
first, some power (or function) of the coefficient of the 
correction considered should be present in all terms of the 
expansion; second, the extra degrees of freedom (which manifest 
themselves as higher order derivatives in metric $f(R)$ gravity) 
are actually an artifact of the expansion (and there would be a 
cancellation if all orders where considered). This way, one can 
do away with these extra degrees of freedom just by proper power 
counting. Since many of the viability issues troubling higher 
order actions are related to the presence of such degrees of 
freedom ({\em e.g.},~classical instabilities), removing  these 
degrees of freedom could certainly alleviate many problems 
\cite{DeDeo:2007yn}. However, the assumptions on which this 
approach is based should not be underestimated either. For 
instance, early results that showed renormalization of higher 
order actions were based on an {\em exact} treatment, {\em 
i.e.},~ it is fourth order gravity that is renormalizable 
\cite{Stelle:1976gc}. Even though, from one hand, the effective 
field theory approach seems very reasonable (these actions are 
regarded as low energy limits of a more fundamental theory 
anyway), there is no guarantee that extra degrees of freedom 
should indeed not be present in a non-perturbative regime.

\subsection{Concluding remarks}

Our goal was to present a comprehensive but still thorough 
review  of $f(R)$ gravity in order to provide a starting point 
for the 
reader less experienced in this field and a reference guide for the expert.  
However, even though we have attempted to cover 
all angles, no review can replace an actual study 
of the literature itself. It seems inevitable that certain 
aspects  of $f(R)$ might have been omitted, or analyzed less 
than rigorously and,  therefore, the reader is  urged to 
resort to the original sources.

Although many shortcomings of $f(R)$ gravity have been 
presented which may reduce the initial enthusiasm with which one 
might  have approached this field, the fact that such 
theories are mostly considered as toy theories should not be 
missed. The fast progress in this field, especially in 
the last five years, is probably obvious by now. And very useful 
lessons, which have helped significantly in the understanding of 
(classical) gravity, have been learned in the study of 
$f(R)$ gravity. In this sense, the statement made in the 
Introduction that $f(R)$ gravity  is a very useful toy theory  
seems to  be fully justified. Remarkably, there are still 
unexplored aspects of $f(R)$ theories or their extensions, such 
as those  mentioned in the previous section, which can  
turn out to be fruitful.


\section*{Acknowledgments}

It is a pleasure to acknowledge discussions with, and/or 
comments  by,  L. Amendola, E. Barausse, J. Barrow, S. 
Capozziello, P. Dunsby, F. 
Finelli,  A. Frolov, W. Hu,  T. Jacobson, S. Joras,  T. 
Koivisto, P. 
Labelle, N. 
Lanahan-Tremblay, S. Liberati, 
F. Lobo, J.  Miller, I. Navarro, S. Odintsov,  G. 
Olmo, A. Starobinsky, H. Stefan\u{c}i\'{c}, N. Straumann,  K. 
Van Acoleyen, M. Visser, and D. Vollick.
The work of T.~P.~S.~was supported by the National Science 
Foundation under grant PHYS-0601800. V.~F.~acknowledges  support by  
the Natural Sciences and  Engineering Research Council of Canada  
(NSERC) and by a Bishop's University Research Grant.

\bibliography{biblio}{}
\bibliographystyle{apsrmp}

\end{document}